\documentclass{aastex63}

\accepted{June 9, 2021}

\submitjournal{ApJ}
\shorttitle{Measuring and Replicating the Energy Distributions of the Coldest Brown Dwarfs}
\shortauthors{Leggett et al.}

\graphicspath{{./}{figures/}}
\begin{document}

\title{Measuring and Replicating the 1 -- 20~$\mu$m Energy Distributions of the  Coldest Brown Dwarfs: \\
Rotating, Turbulent and  Non-Adiabatic Atmospheres}

\correspondingauthor{Sandy Leggett}
\email{sandy.leggett@noirlab.edu}

\author[0000-0002-3681-2989]{S. K. Leggett}
\affiliation{Gemini Observatory/NSF’s NOIRLab, 670 N. A’ohoku Place, Hilo, Hawai’i, 96720, USA}

\author{Pascal Tremblin}
\affiliation{Université Paris-Saclay, UVSQ, CNRS, CEA, Maison de la Simulation, 91191, Gif-sur-Yvette, France}

\author{Mark W. Phillips}
\affiliation{University of Exeter, Stocker Road, Exeter EX4 4PY, United Kingdom}
%\affiliation{}
%\nocollaboration{1}

\author[0000-0001-9823-1445]{Trent J.~Dupuy}
%\affiliation{Gemini Observatory/NSF’s NOIRLab, 670 N. A’ohoku Place, Hilo, Hawai’i, 96720, USA}
\affiliation{Institute for Astronomy, University of Edinburgh, Royal Observatory, Blackford Hill, Edinburgh, EH9 3HJ, UK}

\author{Mark Marley}
\affiliation{NASA Ames Research Center, Moffett Field, CA 94035, USA}

\author{Caroline Morley}
\affiliation{Department of Astronomy, University of Texas, Austin, TX 78712, USA}

\author{Adam Schneider}
\affiliation{US Naval Observatory, Flagstaff Station, P.O. Box 1149, Flagstaff, AZ 86002, USA}
\affiliation{Department of Physics and Astronomy, George Mason University, MS3F3, 4400 University Drive, Fairfax, VA 22030, USA}

\author{Dan Caselden}
\affiliation{Backyard Worlds: Planet 9}

\author{Colin Guillaume}
\affiliation{Backyard Worlds: Planet 9}

\author{Sarah E. Logsdon}
\affiliation{Kitt Peak National Observatory/NSF’s NOIRLab, 950 North Cherry Avenue, Tucson, AZ 85719, USA}

\begin{abstract}
Cold, low-mass, field brown dwarfs are important for constraining the terminus of the stellar mass function, and also for optimizing atmospheric studies of exoplanets. In  2020 new model grids for such objects were made available: Sonora-Bobcat and ATMO 2020. Also, new candidate cold brown dwarfs were announced, and new spectroscopic observations at $\lambda \approx 4.8~\mu$m were published. In this paper we present new infrared photometry for some of the coldest  brown dwarfs,  and put the new data and  models together to explore the properties of these objects.  We reconfirm the importance of mixing in these atmospheres, which leads to CO and NH$_3$ abundances that differ by orders of magnitude from chemical equilibrium values. We also demonstrate  that the 
new models retain the known factor $\gtrsim 3$  discrepancy 
with observations at $2 \lesssim \lambda~\mu$m $ \lesssim 4$, for brown dwarfs cooler  than 600~K.  We show that the entire 
$1 \lesssim \lambda~\mu$m $ \lesssim 20$
energy distribution of six brown dwarfs with $260 \leq T_{\rm eff}$~K $\leq 475$ can be well  reproduced, for the first  time,  by  model  atmospheres which include dis-equilibrium chemistry as well as a photospheric temperature gradient which deviates from the standard  radiative/convective equilibrium value. This change to the pressure-temperature profile is not unexpected for rotating  and turbulent atmospheres which are subject to diabatic processes. A  limited grid of modified-adiabat model colors is generated, and used to estimate temperatures and metallicities for the currently known Y dwarfs. A compilation of the photometric data used here is given in the Appendix. 
\end{abstract}


\section{Introduction} 

Historically, the discovery of cooler main sequence stars has led to tension between the observations and the synthetic spectral energy distributions (SEDs) generated by model atmospheres. Major advances are made with every discovery, which resolves most discrepancies, until the next coolest type is found. Plane-parallel, radiative-convective atmospheres in local thermodynamic and hydrostatic equilibrium  did not reproduce observations of M dwarfs until more complete linelists of molecular transitions for hydrides and oxides were calculated \citep[e.g.][]{Allard_1995, Cushing_2003,Cushing_2005, Tennyson_2007}. Discovery of the very red L dwarfs led to the recognition of condensation and settling as important processes in cool atmospheres
\citep{Tsuji_1996, Ruiz_1997, Burrows_1999, Lodders_1999,  Ackerman_2001, Woitke_2003, Morley_2012, Morley_2014}.  
Infrared observations provided 
evidence of additional non-equilibrium processes, with more CO absorption at $\lambda \approx 4.5~\mu$m, and less NH$_3$ at    $\lambda \approx 1.5~\mu$m and  $\lambda \approx 11~\mu$m than would be present in an atmosphere
in chemical equilibrium \citep[e.g.][]{Saumon_2000,Saumon_2006, Leggett_2007}. Vertical transport of gas in the atmospheres of the solar system giant planets produces non-equilibrium chemical abundances 
\citep{Fegley_1985, Noll_1997} and this became recognised as an intrinsic feature of cool stellar and substellar atmospheres also.

In the last decade, cold substellar objects have been discovered which have even more in common with the giant planets. Substellar objects, or brown dwarfs, have insufficient mass for stable fusion and they cool with time
\citep{[e.g.][]Dantona_1985, Burrows_1993, Baraffe_1998, Saumon_2008, Phillips_2020}. These objects form the extended low-mass tail of the stellar mass function \citep[e.g.][]{Kirkpatrick_2019, Kirkpatrick_2020}, and brown dwarfs as low-mass as 4 Jupiter-masses have been found in young clusters and associations \citep{Best_2017, Esplin_2017, Luhman_2020, Lodieu_2021}. Older, free-floating and cold, very low-mass objects have also been found; the most extreme example is the few-Gyr-old WISE J085510.83$-$071442.5, hereafter J0855, which is a 260~K, $\sim$5 Jupiter-mass object, 2~pc from the Sun \citep{Luhman_2014, Luhman_2016, Leggett_2017}. The properties of giant planets and brown dwarfs overlap significantly \citep{Showman_2013, Morley_2014, Line_2015, Showman_2019}, and the difference between their formation mechanisms is an active research area \citep{Schlaufman2018, Nielsen_2019, Wagner_2019, Bowler_2020}. The coldest objects
have SEDs that are currently difficult to reproduce, and resolving this problem is important;  
characterization of the cold field brown dwarfs is vital for understanding both the  terminus of the   mass function and for optimizing studies of exoplanets. We tackle this problem here.

All but one of the known brown dwarf systems with effective temperature ($T_{\rm eff}$) $<$ 500~K 
were discovered by the mid-infrared all-sky survey executed by the {\it Wide-field Infrared Survey Explorer} \citep[WISE,][]{Wright_2010}. The additional cold brown dwarf, a distant companion to the white dwarf WD 0806$-$661
\citep{Luhman_2011},  was discovered in mid-infrared  images taken by the Infrared Array Camera \citep[IRAC,][]{Fazio_2004}
on board the {\it Spitzer Space Telescope} \citep{Werner_2004}. Some of these have been resolved into close similar-mass binary systems 
\citep[e.g.][]{Liu_2011, Liu_2012, Dupuy_2015},
while others appear super-luminous (compared to models) but have not been resolved in high spatial-resolution imaging \citep{Beichman_2013, Opitz_2016}.

Synthetic SEDs show that half of the energy emitted by a brown dwarf with $T_{\rm eff} < 600$~K is captured by the {\it WISE} W2 filter centered at $\lambda \approx 4.6~\mu$m (or the similar {\it Spitzer} [4.5] filter). In contrast, very little flux emerges through the W1 filter bandpass (or the  {\it Spitzer} [3.6] filter), which includes the strong 3.3~$\mu$m CH$_4$ absorption \citep[e.g.][]{Leggett_2017}. Hence cold brown dwarfs can be identified by very red W1 $-$ W2 (or [3.6] $-$ [4.5]) colors. Currently $\sim 50$ brown dwarfs with $T_{\rm eff} \lesssim 450$~K, classified as Y dwarfs, are known 
\citep{Cushing_2011, Luhman_2011, Kirkpatrick_2012, Tinney_2012, Kirkpatrick_2013, Cushing_2014, Luhman_2014, Pinfield_2014b, Schneider_2015, Martin_2018, Marocco_2019,
Bardalez_2020,
Meisner_2020a, Meisner_2020b}. Based on spectral analyses of an early subset of these objects, \citet{Leggett_2017}
found that most  are relatively young, lower-gravity, and lower-mass objects --— $\sim 1$--3 Gyr-old and $\sim 6$ Jupiter-mass  --- but there were also a few older, higher-gravity, and higher-mass objects --— $\sim 6$ Gyr-old and $\sim 14$ Jupiter-mass; a range of metallicity was also indicated. It is likely that the larger sample follows a similar distribution in  age and metallicity as these values  are typical of the low-mass solar neighborhood \citep{Dupuy_2017, Buder_2019}.

In the  year 2020, two new cold brown dwarf model grids  were made available. One of these is the Sonora-Bobcat grid
\footnote{\url{https://zenodo.org/record/1405206\#.XqoiBVNKiH4}}
of solar- and non-solar metallicity atmospheres, with the atmospheres in chemical equilibrium  \citep[][and submitted]{Marley_2017}. The other is the ATMO 2020 grid 
\footnote{\url{http://opendata.erc-atmo.eu}}
of solar-metallicity models both in chemical equilibrium and out of equilibrium with weak and strong mixing 
\citep{Phillips_2020}. 
Also in 2020, new candidate  $T_{\rm eff} <$ 400~K
brown dwarfs were announced \citep{Bardalez_2020, Kirkpatrick_2020, Meisner_2020a, Meisner_2020b}, and new ground-based spectroscopic observations at $\lambda \approx 4.8~\mu$m were published \citep{Miles_2020}. A study of the cold planet-like brown dwarfs which includes the mid-infrared, and uses state-of-the-art model atmospheres, is now possible. Such a study is  timely, given the scheduled 2021 launch of the  {\it James Webb Space Telescope} ({\it JWST}) for which such objects will be prime targets.

We present new infrared photometric measurements of cold brown dwarfs in Section 2. In Section 3 we compare the observed colors of late-T and Y-type brown dwarfs to the synthetic colors generated by the new atmospheric models. We show that, while the models can reproduce much of the SED, large discrepancies remain. In Section 4 we describe possible missing physics in the current models, 
which impacts the  pressure-temperature adiabatic profile of the atmospheres. We test adiabat-adjusted model atmospheres in Section 
5 by comparing synthetic spectra and photometry to observations of seven brown dwarfs, at wavelengths of 1 -- 20~$\mu$m. We show that a much improved fit can be obtained, and in Section 6 we use a  grid of the adiabat-adjusted models to explore the properties of a larger sample of Y dwarfs. Our Conclusions are given in Section 7. In the Appendix we illustrate trends with temperature for $JWST$ colors, 
provide a grid of colors generated by the adiabat-adjusted models, and give a compilation of the photometry used in this work.

\bigskip
\section{New Photometry} 

\smallskip
\subsection{Image Processing}

The DRAGONS software package \citep{Labrie_2019} was used to reduce all the new imaging data obtained at Gemini Observatory for this work. DRAGONS documentation is available at: https://dragons.readthedocs.io/en/stable/.

For Gemini's infrared cameras, DRAGONS performs these initial steps: the  non-linearity correction is applied; counts are converted from data numbers to electrons; bad pixel masks are applied; and the read and Poisson noise is added to the FITS extension which carries the variance information. Multiple dark observations are stacked to create a master dark. A master flat is created from multiple lamps-on and lamps-off observations; the flat is normalized and thresholded for out-of-range values. 

Science data is divided by the appropriate flat field for filter and read mode. The sky contribution is determined for each pointing using the images taken at other positions in the dither pattern. The sky is then subtracted from each science image. Point sources are detected in each image, and these are used to align and stack the data set for each object. Each sky-subtracted image in the stack is numerically scaled based on the background signal, by factors typically $< 5$\%, to produce a final image. For images obtained with the adaptive optics multi-detector imager GSAOI at Gemini South \citep{McGregor_2004}, an add-on package called Disco-Stu determines the astrometric transformations to perform the stacking and create the final image.

We used simple aperture photometry to measure magnitudes from processed images. The processed images either came from our new Gemini observations or from data archives, as we describe below. We used circular apertures with annular sky regions positioned to avoid nearby sources. The size of the target aperture was typically small, with diameters of 6 to 10 native pixels, 
in order to reduce noise and exclude potential nearby sources. 
We corrected for any loss of flux through the aperture by determining aperture corrections using bright isolated point sources in the science target image. Zeropoints for the processed images were determined from calibrators in the image or observed separately, or from the FITS header in the case of archival data. Extinction corrections were not applied to the ground-based data because the near-infrared extinction is small\footnote{\url{
https://www.gemini.edu/observing/telescopes-and-sites/sites}} and the targets were observed at airmasses $\lesssim 1.7$.

\smallskip
\subsection{Gemini Observatory J-band Photometry of Candidate Cold Brown Dwarfs}

To examine the nature of the candidate late-type brown dwarfs identified by \citet{Marocco_2019} and \citet{Meisner_2020a, Meisner_2020b}, we obtained $J$-band imaging at Gemini Observatory using the Near-InfraRed Imager (NIRI) at Gemini North \citep{Hodapp_2003} and FLAMINGOS-2 at Gemini South \citep{Eikenberry_2006}. 
Table 1 gives target names and Gemini program identifications; the targets were selected as those accessible at the time of the Observatory's Proposal Calls.

\setlength\tabcolsep{2pt}
\begin{deluxetable*}{cccccccccccccrc}[t!]
\tabletypesize{\scriptsize}
\tablecaption{New Near-Infrared Photometry and Estimates of $T_{\rm eff}$}
\tablewidth{0pt}
\tablehead{
\colhead{{\it WISE} Name} & \colhead{Disc.} &  \colhead{Spec.} & \colhead{Type} &
\colhead{Gemini} & \colhead{Obs. Date} & \colhead{Instrument} &   \colhead{On-Source} &   \multicolumn{5}{c}{Photometry, MKO mag} & \multicolumn{2}{c}{$T_{\rm eff}$~K}\\
\colhead{RA/Dec J} & \colhead{Ref.} &  \colhead{Type} & \colhead{Ref.} & \colhead{Program ID} &  \colhead{yyyymmdd} &  \colhead{Name} &  \colhead{Exp., hr} &  \colhead{$Y$} &
\colhead{$J$} & \colhead{$H$} & 
\colhead{$K_s$\tablenotemark{a}} & \colhead{$K$} & 
\colhead{Est.} &  \colhead{Ref.} 
}
\startdata
021243.55 & Me20a & Y1 & 1 & GS-2019B-DD-107 & 20191211 & FLAMINGOS-2 & 1.38 &  & 22.70    & &  & & 390 & 2\\
$+$053147.2  &   &     &    &                &          &             &      &  & $\pm$ 0.09 & & & && \\
030237.53   & Ti18 & Y0: &  Ti18 &  &   2013 & VIRCAM\tablenotemark{b} &  &  &  20.67 &  &  &&  460 & 2\\
$-$581740.3  &     &     &       &  &        &        &  &  &  $\pm$ 0.23 &   &   &   & & \\
032109.59 & Me20a & Y0.5 & Me20a  & GN-2020B-Q-321 & 20200930 & NIRI & 0.58 &  & 21.30 & & & & 415 & 2\\
$+$693204.5 &     &      &          &              &          &      &      & & $\pm$ 0.06 &&  & & &\\ 
033605.05 & Ma13b & Y0 & Ma18 & GN-2020B-ENG-1 & 20201001 &  NIRI & 0.57 & 21.02\tablenotemark{c}    & 21.26   & 21.59   & 21.4 & & 445 & 2\\ 
$-$014350.4 &    &    &     &                 &           &      &      &  $\pm$ 0.11 & $\pm$ 0.14 & $\pm$ 0.31 & $\pm$ 0.5 & & &\\
040235.55 & Me20a & Y1& Me20a & GS-2021A-FT-205 & 20210322 & FLAMINGOS-2 & 0.8 &  &  24.0 &  & & & 370 & 2 \\
$-$265145.4 &  &     &      &                    &          &             &       &    &  $\pm$ 0.5 & & & & & \\
050305.68  & Me20b & Y1  & Me20b & GS-2021A-FT-205 & 20210303 & FLAMINGOS-2 & 2.11 &    &   22.54 &  & & &  345 & 2 \\
$-$564834.0 &    &     &      &                    &          &             &       &    &  $\pm$ 0.09 &&  & & & \\
050615.56 & Me20b & T8 & PGpc  & GS-2013B-Q-16 & 20131224 & FLAMINGOS-2 & 0.16 &   & 20.31       & 20.89 & & & 600 & 1\\
$-$514521.3 &    &     &      &                    &          &             &       &    &  $\pm$ 0.05 & $\pm$ 0.14 & & && \\
064723.23 & Ki13 & Y1 & Ki13 & GS-2019B-Q-220 & 20121210,  & GSAOI & 1.30 &    &   &  & 23.03 & & 405 & 2\\
$-$623235.5 &    &    &     &                & 12, 13, 14 &       &      &    &   &   &   $\pm$ 0.15 & & &\\
085938.95 & Me20a & Y0 & Me20a & GN-2020B-Q-321 & 20201225 & NIRI & 0.13 &   & 21.39 & & & & 450 &  2\\
$+$534908.7 &     &     &      &                 &         &      &      &   &   $\pm$ 0.15 &  & &   &   &  \\
092503.2   & Ki20 & T8 & 1 &         &  2017     & VIRCAM\tablenotemark{d}     &      &     &   18.29  & &   & &  700  & 1 \\ 
$-$472013.8 &    &     &   &         &       &      &      &     &  $\pm$ 0.05 &   &    &   &  & \\
093852.89  & Me20a &  Y0 & Me20a &  GS-CAL20210429 & 20210429 & FLAMINGOS-2 & 0.44 &   & 21.08 & 21.49 & 21.11 & & 455 & 2 \\  
$+$063440.6 &      &     &       &                 &          &             &      &   & 0.10 & 0.21 & 0.23 &  && \\
094005.50 & Me20a & $\ge$Y1 & Me20a & GN-2020A-FT-205 & 20200310 & NIRI & 0.88 &  & 21.88     & & & & 410 & 2\\
$+$523359.2 &     &         &        &                &          &      &     &   & $\pm$ 0.11 &  & & & &\\
125721.01 & Me20b& Y1 & Me20b & GN-2021A-FT-206 & 20210409 & NIRI & 1.72 &   &  23.35 &   &  &  & 390 & 2\\
$+$715349.3  &    &           &       &                &             &           &    &  & $\pm$ 0.20 & & &  &&\\
144606.62 & Me20a & $\ge$Y1 & Me20a & GS-2020A-FT-204 & 20200305 & FLAMINGOS-2 & 4.32 &  & 23.20 & & && 350 & 2\\
$-$231717.8 &    &           &       &                &             &           &    &  & $\pm$ 0.14 & & & &&\\
193054.55  & Me20b & $\ge$Y1 & Me20b & GN-2020B-Q-321 & 20201001 & NIRI & 1.78 &  & 22.54 & & & & 365 & 2 \\
$-$205949.4 &        &       &       &                &          &      &      &   & $\pm$ 0.13 &  & & &    &\\ 
193518.58 & Ma19 & $\ge$Y1 & Me20a & GN-2020B-Q-321 & 20200823, & NIRI & 1.70 &   & 23.93 & & & & 365\tablenotemark{e} & 2\\
$-$154620.3 &    &          &      &                & 20200929, 30 &  &      &    & $\pm$ 0.33 &  & & & &\\ 
193656.08 & Me20a & Y0 & Me20a & GN-2020B-Q-321 & 20201001 & NIRI & 0.03 &  &20.16 & & & & 450 & 2\\
$+$040801.2 &     &     &       &               &          &      &      &   & $\pm$ 0.12 &  & &  & \\
200520.38 & Ma13a & sdT8 & Ma13a & GN-2021A-FT-206 & 20210511, & NIRI & 1.4 & 19.99\tablenotemark{f}  & 19.54 & 19.55 &  & 21.00 & 600 & 1 \\
$+$542433.9 &     &      &       &                 & 20210517  &       &    & 0.07 & 0.07 &  0.03 &  & 0.09 &
 &  \\
223022.60 & Me20a & $\ge$Y1 & Me20a & GN-2020B-Q-321 & 20201001, & NIRI & 1.30 &   & 22.99 & & & & 395 & 2\\
$+$254907.5 &      &         &       &                & 05        &      &      &    & $\pm$ 0.20   & &   &   \\
224319.56 & Me20b & Y0 & Me20b &  &  2013 &  VIRCAM\tablenotemark{b}     &      &  21.16   & 21.14 & & &  &  450 & 2 \\
$-$145857.3 &     &    &       &  &       &             &      &  $\pm$  0.34    &  $\pm$ 0.26  &   &  &  &   &  \\
224916.17& Me20a & T9.5 & Me20a & GN-2020B-Q-321 & 20200917 & NIRI & 0.28 &    &21.89 & & & & 460 & 2\\
$+$371551.4 &   &       &        &               &          &      &      &    &  $\pm$ 0.10 & & & & &\\
\enddata
%%\smallskip
\tablenotetext{a}{\citet{Leggett_2015}  measure  $K - K_s = 0.4 \pm 0.1$ for a T8 and a T9  dwarf using FLAMINGOS-2, implying 
$K = 21.8 \pm 0.5$ for J033605.05$-$014350.4,
$K = 23.43 \pm 0.18$ for J064723.23$-$623235.5, and 
$K = 21.51 \pm 0.25$ for J093852.89$+$063440.6.}
\vskip -0.1in
\tablenotetext{b}{Measured here using  VISTA VHS  imaging data.}
\vskip -0.1in
\tablenotetext{c}{In the native NIRI system $Y = 21.19 \pm 0.10$ for J033605.05$-$014350.4; we adopted ${Y}_{\mathrm{NIRI}}-{Y}_{\mathrm{MKO}}=0.17\pm 0.03$ as determined by
\citet{Liu_2012}  for late-T and Y dwarfs.}
\vskip -0.1in
\tablenotetext{d}{Measured here using  VVVX ESO Public Survey imaging data.}
\vskip -0.1in
\tablenotetext{e}{Assuming the system is an equal-mass binary, see Section 6.3.}
\vskip -0.1in
\tablenotetext{f}{In the native NIRI system $Y = 20.03 \pm 0.05$ for J200520.38$-$145857.3; we synthesized ${Y}_{\mathrm{NIRI}}-{Y}_{\mathrm{MKO}}$ for this object using the observed $Y$-band spectrum from \citet{Mace_2013b} and the filter profiles for 
NIRI\footnote{\url{https://www.gemini.edu/instrumentation/niri/components\#Filters}} and 
MKO\footnote{\url{https://http://svo2.cab.inta-csic.es/svo/theory/fps3/index.php?mode=browse&gname=UKIRT&gname2=UKIDSS&asttype=
%svo2.cab.inta-csic.es/theory/fps/index.php?id=UKIRT/UKIDSS.Y&&mode=browse&gname=UKIRT&gname2=UKIDSS#filter
}}.}
\vskip -0.05in
\tablerefs{(1) this work, type  ($\pm \approx 0.5$) based on the type-color, and $T_{\rm eff}$
 ($\pm \approx 50$~K) based on the
$T_{\rm eff}$-color, relationships of \citet{Kirkpatrick_2019, Kirkpatrick_2020}; 
(2) this work, $T_{\rm eff}$ ($\pm \approx 25$~K)
based on the $T_{\rm eff}$-color relationships determined in Section 6.2, with $T_{\rm eff}$ values rounded to 5~K;
Ki13 -- \citet{Kirkpatrick_2013}; 
Ki20 -- \citet{Kirkpatrick_2020};
Ma13a -- \citet{Mace_2013b}; 
Ma13b -- \citet{Mace_2013a}; 
Ma18 -- \citet{Martin_2018};
Ma19 -- \cite{Marocco_2019};
Me20a,b -- \citet{Meisner_2020a, Meisner_2020b};
PGpc -- Pinfield, P. and Gromadzki, M. private communication 2014;
Ti18 -- \citet{Tinney_2018}.
}
\end{deluxetable*}

The $J$ filter is defined by the Mauna Kea Observatories photometric system \citep{Tokunaga_2002}. 
The camera pixel scales are  $0\farcs 12$ for NIRI and  $0\farcs 18$ for FLAMINGOS-2.  Telescope dithers of 12 -- 15" were used, in the form of a 5- or 9-point grid. All nights were photometric and the targets were observed at airmasses 
of 1.1 -- 1.7. The delivered full width half maximum (FWHM) of the point spread function (PSF) was  $0\farcs4$ to $1\farcs0$.
The magnitude zeropoint  was determined from UKIDSS or VISTA sky survey photometry 
\citep{Lawrence_2007, McMahon_2013, Sutherland_2015, Dye_2018}
of stars in the field of view; typically four to eight such stars were available. 
In the case of J094005.50$+$523359.2, hereafter J0940,  only two survey stars were available and the zeropoint was determined by averaging the value implied by those stars plus a
measurement of a UKIRT Faint Standard \citep{Leggett_2006} executed immediately after the one-hour science observation and at a similar airmass (1.1 cf. 1.2 for the science); the three zeropoint measurements agreed to 10\%. Sky noise for these images was typically at the 5 -- 10\% value, and usually dominated the uncertainty.  Table 1 gives the final values.

One of the targets, CWISEP J021243.55$+$053147.2, hereafter J0212, was identified by \citet{Meisner_2020a} as having very red [3.6] $-$ [4.5] colors, but not having significant motion, and therefore not listed in their table of Y dwarf candidates. Subsequently, \citet{Kirkpatrick_2020} also determined a low-significance motion of $\mu_{\alpha} = -59.8 \pm 45.0$ and
$\mu_{\delta} = 57.0 \pm 27.4$~mas yr$^{-1}$, as well as a poor-quality parallax of $24.7 \pm 16.3$~mas; \citet{Kirkpatrick_2020} suggest that J0212 is a background source.
However, the extremely red $J -$ [4.5] color that we measure for this object, with very little flux at [3.6], implies that J0212 is cold and molecule-rich. 
The study of {\it WISE} colors by \citet{Nikutta_2014} shows that AGN and infrared luminous galaxies can be very red in W1 $-$ W2, however such objects are also red in W2 $-$ W3;  if J0212 falls into such a category it would be detected in W3, which it is not. A more plausible solution is that the object is a binary  and the actual parallax value is close to the upper limit of the current measurement;
we show below that the luminosity of J0212 is then consistent with the observed $J -$ [4.5] and [3.6] $-$ [4.5] colors. We therefore suggest that J0212 is a binary Y dwarf at a distance of $\sim$24~pc.

\smallskip
\subsection{Other New Near-Infrared Photometry}

As part of a project to measure photometric transformations between the UKIDSS and VISTA sky survey, NIRI, and FLAMINGOS-2 systems, a field containing the Y dwarf WISE J033605.05$-$014350.4
was observed at Gemini North at $YJHK_sK$ (the brown dwarf was not detected at $K$),
and a field containing the Y dwarf CWISEP J093852.89$+$063440.6  was observed at Gemini South  at $JHK_s$.  The data were reduced in the manner described in the previous section, and the results are given in Table 1.

Table 1 lists the $JH$ magnitudes for WISEA J050615.56$-$514521.3,  which was listed by \citet{Meisner_2020b} as a very late T dwarf candidate. This object was also targeted in the deep WISE search by \citet{Pinfield_2014a} and the unpublished photometry and spectral type (from their spectroscopy) is provided courtesy of a private communication with  that team.

We measured $YJHK$ magnitudes for the late-T subdwarf  WISE J200520.38$+$542433.9, also known as Wolf 1130C \citep{Mace_2013b}, in order to have a set of near-infrared colors for a known very metal-poor object with [m/H] $\approx -0.75$ \citep{Kessel_2019}. The data were obtained using NIRI at Gemini North and were reduced in the manner described in the previous section. The results are given in Table 1.

CWISE J092503.20$-$472013.8 was listed by \citet{Kirkpatrick_2020} as a candidate Y0 dwarf  based on its motion, and W1 $-$ W2 color (3.93 $\pm$ 0.38). We used VVVX ESO Public Survey data \footnote{\url{https://www.eso.org/sci/publications/announcements/sciann17186.html}} to determine the $J$ magnitude given in Table 1. The brown dwarf was not detected in the $K_s$ survey data. The $J -$ W2 color of the target (2.99 $\pm$ 0.06) provides an improved spectral type estimate of T8, based on Figures 13 and 14 of \citet{Kirkpatrick_2020}.

In addition, we searched for detections in the UKIDSS and VISTA  surveys' imaging data for Y dwarfs 
without near-infrared photometry. We determined magnitudes for two Y0 dwarfs from the 
VISTA Hemisphere Survey \citep[VHS,][]{McMahon_2013}: WISEA J030237.53$-$581740.3 ($J$), and WISEA J224319.56$-$145857.3 ($Y$, $J$). The results are given in Table 1.

Finally, to explore the known discrepancy between observations and models at $\lambda \approx 2~\mu$m \citep[e.g.][]{Leggett_2019}, we obtained $K$-band images
of the Y1 dwarf WISE J064723.23$-$623235.5 \citep{Kirkpatrick_2013}, hereafter J0647. This object was chosen in order to better measure the discrepancy for the coldest Y dwarfs, where little $K$-band imaging is available. 
Because of the faintness of the target,
we used the adaptive optics imager GSAOI \citep{McGregor_2004}
at Gemini South. Table 1 gives the program identification and the dates on which J0647 was observed. 
The imager has a pixel scale of $0\farcs 02$. 
The nights were photometric and the delivered FWHM  was $\sim 0\farcs 1$. Sixty-six 90~s observations were made, of which 52 with better seeing of  $\leq 0\farcs 095$ were used in the final image. 
J0647 was observed at an airmass of $\sim 1.2$, and the telescope was dithered by random 1 -- 4" offsets.  
Aperture photometry was carried out with apertures of diameter $0\farcs 12$ and  $0\farcs 20$, which gave consistent results after the application of the aperture corrections. 
The magnitude zeropoint  was determined using stars from the VISTA Hemisphere Survey \citep{McMahon_2013} which were in the GSAOI field of view.  Table 1 gives our derived $K_s$ for J0647.

\smallskip
\subsection{Mid-Infrared Photometry}

Our goal is to reproduce the SED of the coldest brown dwarfs over all wavelengths where significant flux is emitted. It is important therefore to include the mid-infrared region; furthermore, knowledge of the mid-infrared is crucial for planning observations with   {\it JWST}.

\begin{figure}
\plottwo{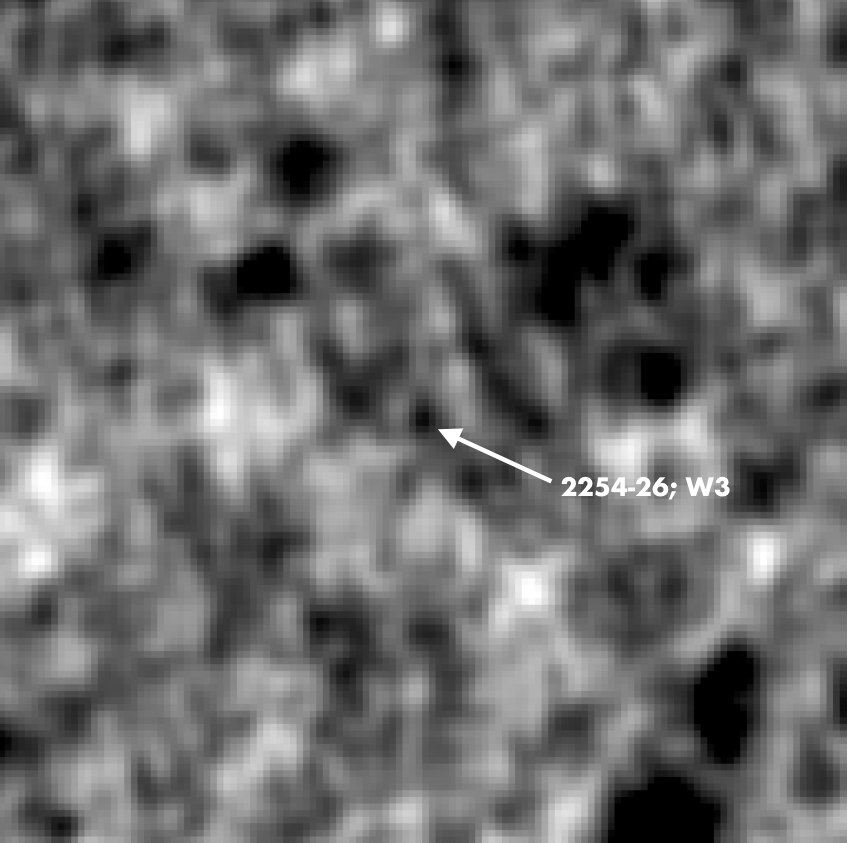}{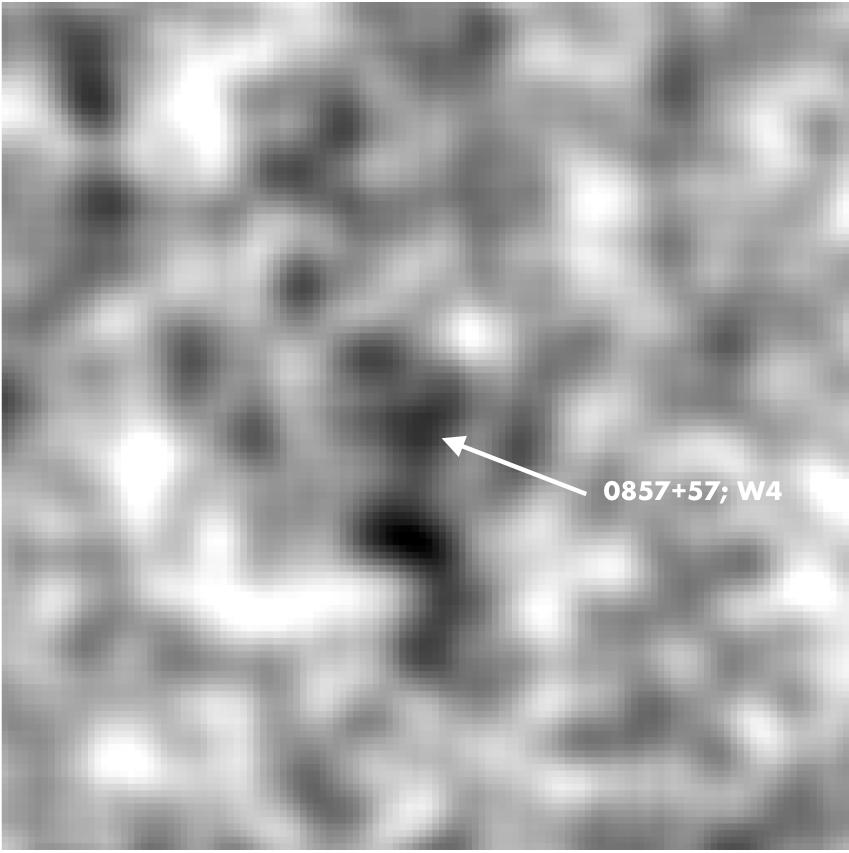}
%\vskip -0.25in
\caption{Examples of {\it WISE} images where faint brown dwarfs were previously not included in the ALLWISE catalog (left) or where the photometry was compromised by background sources (right). In the latter case, the smaller aperture used here allowed the brown dwarf to be better isolated, resulting in a W4 magnitude fainter than the catalog value by $\sim 2$ magnitudes.  
}
\end{figure}

\setlength\tabcolsep{3pt}
\begin{deluxetable*}{lcccRRRRRR}
\tabletypesize{\footnotesize}
\tablecaption{Revised and New  {\it WISE} and {\it Spitzer}  Photometry}
%\tablewidth{0pt}
\tablehead{
\colhead{{\it WISE} Name} &  \colhead{Disc.} &  \colhead{Spec.} &   \colhead{Type} &  \multicolumn{2}{c}{ALLWISE Catalog} &  \multicolumn{4}{c}{This Work}
\\
\colhead{RA/Dec J} &  \colhead{Ref.}     &  \colhead{Type} &  \colhead{Ref.}     &  \colhead{W3} & \colhead{W4} &
 \colhead{W3} & \colhead{W4} & \colhead{[3.6]} &  \colhead{[4.5]}
}
\startdata
001449.96$+$795116.1& Ba20  & T8 & Ba20 & &   & 13.69  $\pm$ 0.40 & & & \\
002810.59$+$521853.1 & Me20b & T7.5 & Me20b &  &   & 13.95 $\pm$	0.43 &  & &  \\
013217.78$-$581825.9 & Me20b & T9 & Me20b & &  & 14.10 $\pm$	0.41 &  & &  \\
014603.23$-$261908.7 & Me20b & T7.5 & Me20b & &  & 13.63 $\pm$	0.34 &   & & \\
081117.81$-$805141.3 & Ma13b & T9.5  & Ma13b &  12.64 $\pm$ 0.32 & 9.21 $\pm$ 0.38 & & 11.09 $\pm$ 0.65  & & \\
085510.83$-$071442.5\tablenotemark{a} & Lu14 & $>$Y4 & Ki19 & 11.14 \pm 0.13 &  & 11.51 \pm 0.06 & 10.56 \pm 0.50  & & \\
085757.95$+$570847.5 & Ge02 & L8 & Ge02 & 10.32 \pm 0.06 & 8.64 \pm 0.35 & & 10.48 \pm 0.50  & & \\
093735.63$+$293127.2 & Bu02 & T6pec & Bu06 & 10.70 \pm 0.10 & &  & 10.36 \pm 0.34 & &  \\
105349.41$-$460241.2 & Me20b & T8.5  & Me20b & &   &  14.13 \pm	0.40 &  & &  \\
112106.36$-$623221.5 & Ki20 & T7 & 1 &    &   &   &   &  16.47 \pm 0.10 & 15.13 \pm 0.04 \\
125721.01$+$715349.3 & Me20b& Y1 & Me20b & &  & 13.55 \pm	0.33 &  & &  \\
182831.08$+$265037.8 & Cu11 & $>$Y2 & Ki12 & 12.44 \pm 0.34 & & & 10.65 \pm 0.52  & & \\
193054.55$-$205949.4 & Me20b & Y1 & Me20b & &  & 14.44 \pm	0.58 &  & &  \\
214025.23$-$332707.4 & Me20b & T8.5 & Me20b &  &   &  13.32 \pm	0.32 &  & &  \\
225404.16$-$265257.5 & Me20b & T9.5 & Me20b  & &  & 13.29 \pm	0.29 &  & &  \\
\enddata
\tablenotetext{a}{\citet{Wright_2014} and \citet{Kirkpatrick_2019} demonstrate that the first epoch of {\it WISE} observations of J0855 are significantly contaminated at W1 by background sources. The W3 and W4 images date to the same epoch and the background sources will therefore be at the same location as J0855. \citet{Wright_2014} measure W1 $= 16.12$ and W1 $-$  W2 $=0.67 \pm 0.17$ for these sources from images where J0855 has moved away (post-cryo). \citet{Nikutta_2014} analyse {\it WISE} colors for large samples of Galactic sources; their Figure 6 (panel 3) shows that the W1 $-$  W2 color is likely to be on the bluer side of the \citet{Wright_2014} measurement, and the most likely values of  W2 $-$  W3 and 
W3 $-$  W4 are $\sim 0.8$ and $\sim 1.0$ respectively. Hence the background sources are 
expected to have  W3 $\approx$ 15 and W4 $\approx$ 14,
and so are not likely to significantly contaminate the J0855 W3 and W4 values in the Table. The successful model fits we show in Section 5.2 support this conclusion.}
\vskip -0.05in
\tablerefs{
1 -- this work;
Ba20 -- \citet{Bardalez_2020};
Bu02 -- \citet{Burgasser_2002};
Bu06 -- \citet{Burgasser_2006};
Cu11 -- \citet{Cushing_2011};
Ge02 -- \citet{Geballe_2002};
Ki12 -- \citet{Kirkpatrick_2012};
Ki19  -- \citet{Kirkpatrick_2019}; 
Ki20 -- \citet{Kirkpatrick_2020};
Lu14 -- \citet{Luhman_2014};
Ma13b -- \citet{Mace_2013a};
Me20b -- \citet{Meisner_2020b}.
}
\end{deluxetable*}

{\it WISE} catalog photometry 
\footnote{\url{https://irsa.ipac.caltech.edu/cgi-bin/Gator/nph-dd}}
of faint targets can be compromised by nearby objects, and fainter objects are sometimes omitted altogether.
The sensitivity limits for a signal-to-noise ratio (SNR) $= 5$ are $\sim$11.5 and 8.0 magnitudes for the
W3 ($\lambda \sim 14~\mu$m ) and W4 ($\lambda \sim 22~\mu$m ) filters,  respectively
\footnote{\url{https://https://wise2.ipac.caltech.edu/docs/release/allwise/expsup/sec2_3a.html}}.
We examined the W3 and W4 images provided by the {\it WISE} Image Service 
\footnote{\url{https://irsa.ipac.caltech.edu/applications/wise/}}
for the colder brown dwarfs, and determined new or revised values for the photometry based on this visual inspection.
We also looked for W3 and W4 data for warmer brown dwarfs to determine color trends.  We identified sources where a point source could be resolved by eye, at the correct location for the epoch of the W3 or W4 observation, allowing for the proper motion of the source. Figure 1 gives examples of sky regions where we obtained new or revised {\it WISE} magnitudes.

We carried out
aperture photometry on the {\it WISE} images using apertures of 3- or 5-pixel radii (4 or 7") and annular skies. These apertures are smaller than the predefined fitting radius used by the ALLWISE profile-fitting photometry routine, $r_{fit}$: 
$r_{fit} = 1.25 \times FWHM$ where $FWHM$ is 6" for bands 1 -- 3 and 12" for band 4\footnote{\url{https://wise2.ipac.caltech.edu/docs/release/allsky/expsup/sec4_4c.html\#wpro}}. The smaller aperture reduced the noise contribution from the background and improved exclusion of nearby sources. Aperture corrections were measured using isolated and brighter stars in the field. Zeropoints are taken from the {\it WISE} image header. Table 2 gives our new W3 and W4 measurements, as well as the ALLWISE Source Catalog values.
The uncertainties in the new measurements are due to background noise and are large in most cases, with SNRs of 2 or 3 only. Nevertheless significant differences exist between our values and those reported in the catalog (Table 2). These long-wavelength colors are useful for comparing to colors generated by current model atmospheres, and for planning {\it JWST} observations.

CWISE J112106.36$-$623221.5 was listed by \citet{Kirkpatrick_2020} as a candidate Y0 dwarf, based on its motion, W2 detection and W1 non-detection. {\it Spitzer} imaging data are available for the source via AORs r42735360 and r23699712 at the {\it Spitzer} Heritage Archive \footnote{\url{https://sha.ipac.caltech.edu/applications/Spitzer/SHA/}}. We carried out aperture photometry on these images using apertures of 3-pixel radii ($1\farcs8$) and annular skies. Aperture corrections were measured using isolated and brighter stars in the field, and the counts calibrated photometrically according to the {\it Spitzer} IRAC Manual\footnote{\url{https://irsa.ipac.caltech.edu/data/SPITZER/docs/irac/iracinstrumenthandbook/14/\#_Toc59022361}}. 
Table 2 gives the [3.6] and [4.5] magnitudes for the source, which was not detected at longer wavelengths in the earlier cryogenic observation. The two measurements of the source, taken four years apart, agree to 20\% at [3.6] and 2\% at [4.5]. The [3.6] $-$ [4.5] color of the target (1.34 $\pm$ 0.11) provides an improved spectral type estimate of T7, based on Figure 14 of \citet{Kirkpatrick_2020}.

\bigskip
\section{Observed and Modelled Colors of T and Y Dwarfs}

\begin{figure}
%\centering
%\includegraphics[width=6in]{col_col_noTune.pdf}
\plotone{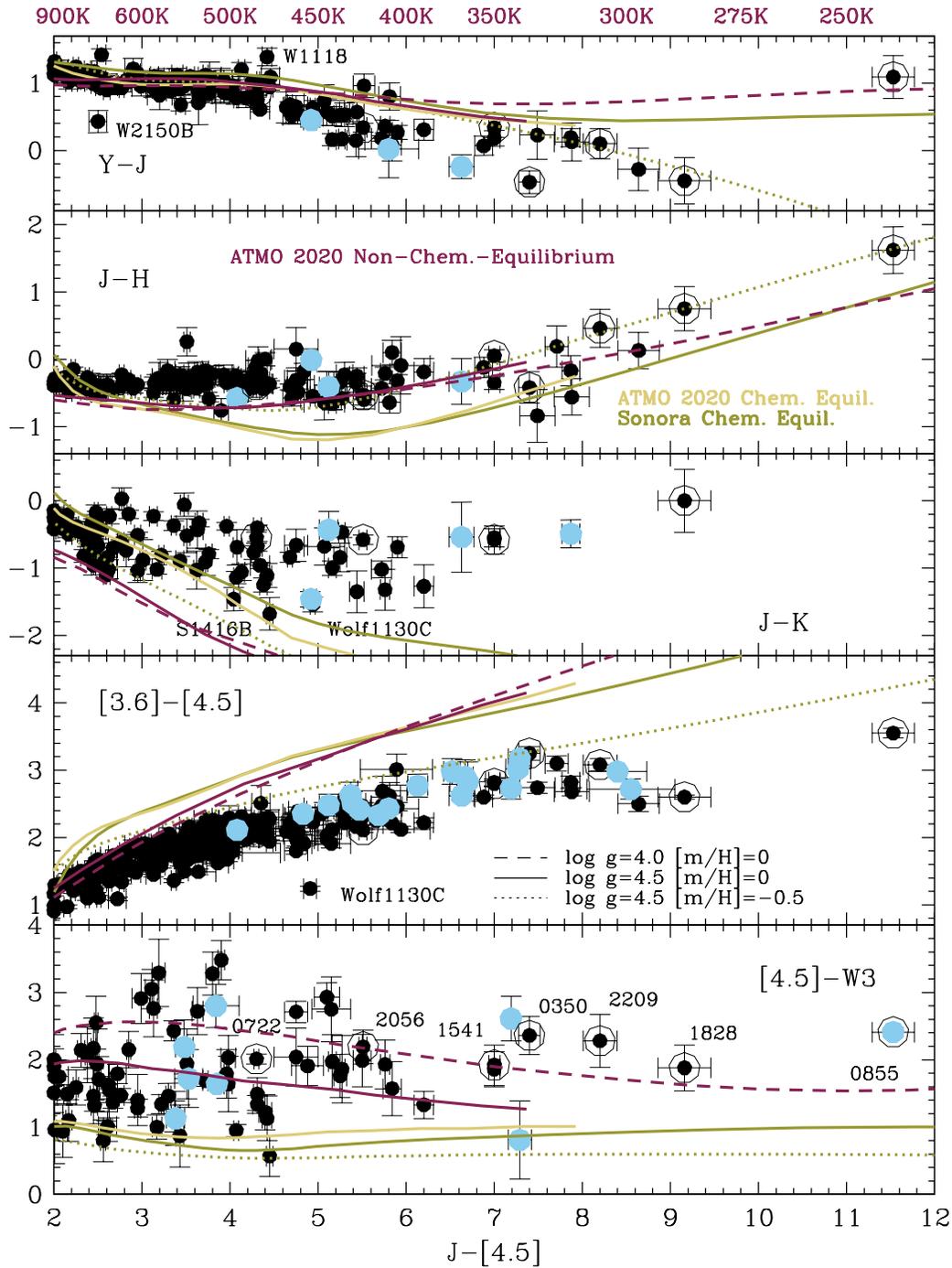}
\vskip -0.15in
\caption{Color-color diagrams for late T and Y dwarfs. Black dots are photometry from the literature; 
blue dots are new data presented here.  
Olive green lines are chemical equilibrium Sonora-Bobcat models, and 
yellow  lines are chemical equilibrium ATMO 2020 sequences for a mass of 0.015~$M_{\odot}$  ($\log~g \approx 4.5$). 
Dark red lines are chemical non-equilibrium ATMO  2020  sequences  
for  masses  of  0.015~$M_{\odot}$  ($\log~g \approx 4.5$) and  0.005~$M_{\odot}$  ($\log~g \approx 4.0$).
%The ATMO 2020 iso-mass sequences terminate at an age of 10~Gyr. 
Line types indicate gravity and metallicity as in the legend.
Approximate $T_{\rm eff}$ values  along the top axis are from the ATMO 2020 non-equilibrium chemistry models.
Circled points  indicate seven  dwarfs that we analyse in detail in Section 5, which are identified by short name in the bottom panel. Four color outliers are also identified:
ULAS J141623.94$+$134836.30 (``S1416B''),
WISE J111838.70$+$312537.9 (``W1118''),
WISEA J215018.25$-$752039.7B (``W2150B''), and Wolf 1130C. 
W1118 is a distant companion to a quadruple system composed of F and G stars \citep{Wright_2013}. S1416B and W2150B are companions to L dwarfs \citep{Burningham_2010a, Faherty_2020}. Wolf 1130C is a companion to an sdM and white dwarf binary \citep{Mace_2013b}. W1118, S1416B and Wolf 1130C  are members of  metal-poor systems with [m/H] $= -0.3$, [m/H] $\approx -0.3$  and [m/H] $\approx -0.75$, respectively \citep{Wright_2013, Gonzales_2020, Kessel_2019}.
}
\end{figure}

\begin{figure}
%\plotone{mc2_noTune.pdf}
\plotone{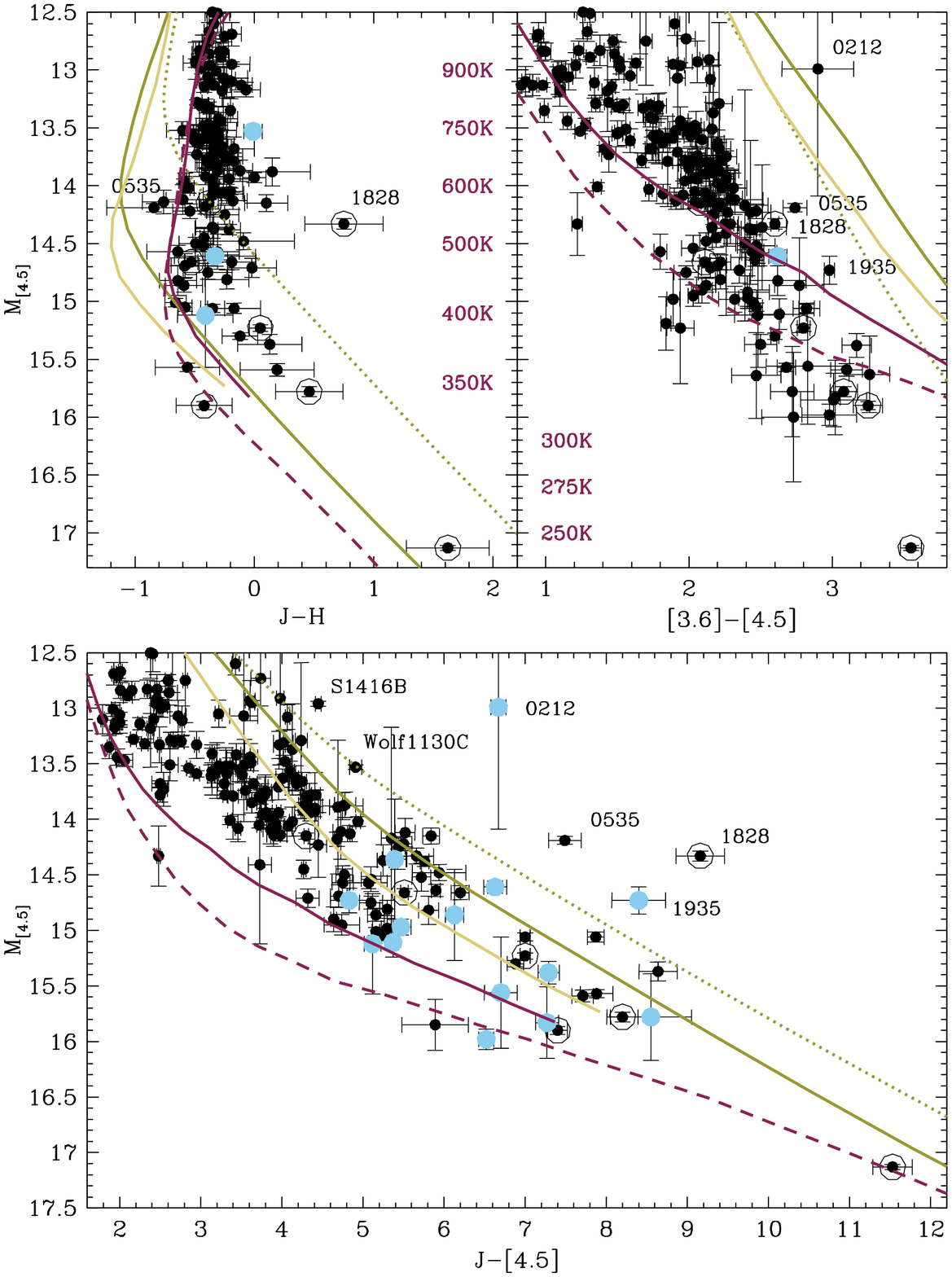}
\vskip -0.25in
\caption{Color-magnitude diagrams for late T and Y dwarfs. Symbols and lines are as in Figure 2. Approximate $T_{\rm eff}$ values  along the top middle axis are from the ATMO 2020 non-equilibrium chemistry models.
Over-luminous Y-dwarfs which are possibly unresolved binaries  are identified: 
CWISEP J021243.55$+$053147.2,
WISE J053516.80$-$750024.9, WISEPA J182831.08$+$265037.8, and
CWISEP J193518.59$-$154620.3. In the lower panel, the metal-poor T dwarfs S1416B and Wolf 1130C are identified (see also Figure 2).
%ULAS J141623.94$+$134836.30 \citep[``S1416B'',][]{Burningham_2010a} and Wolf 1130C \citep{Mace_2013b} are identified. These two objects are members of  metal-poor systems  with [m/H] $\approx -0.3$ and [m/H] $\approx -0.75$, respectively \citep{Gonzales_2020, Kessel_2019}.
%Apparently single
%objects which are likely to have $T_{\rm eff} \lesssim$ 400~K are identified by the first four digits of the {\it WISE} catalog Right Ascension, or their binary name in the case of the white dwarf companion. 
}
\end{figure}

Models of brown dwarf atmospheres are typically characterized by a set of physical and chemical parameters. The most fundamental is the total energy output, or luminosity ($L$) which is defined by Stefan's Law as $L = \sigma T_{\rm eff}^4 \times 4\pi R^2$, where $\sigma$ is the Stefan-Boltzmann constant, $R$ is the radius of the object, and $T_{\rm eff}$ the effective temperature. 
Another important parameter is the surface gravity $g$, which is defined as $g = GM/R^2$ where  $M$ is the mass and $G$ is the gravitational constant. The chemical composition of the atmosphere is usually described as the abundance of metals relative to hydrogen [m/H], normalized to the solar value. In addition, some models include cloud formation via a sedimentation parameter and a fractional cloud cover \citep[e.g.][]{Morley_2014}. Also, some models represent vertical transport of gas (which results in disequilibrium chemical abundances) as a diffusive process, via the vertical eddy diffusivity parameter  $K_{zz}$  \citep[cm$^2$ s$^{-1}$, e.g.][]{Saumon_2006}. The models we use here are parameterized by: $T_{\rm eff}$, $g$, [m/H] and $K_{zz}$. They are cloud-free and we discuss the possible impact of clouds later in this paper.

Figures 2 and 3 show color-color and color-magnitude diagrams for late-T and Y-type brown dwarfs. Observed colors are plotted, as well as sequences from the  Sonora-Bobcat models
\footnote{\url{https://zenodo.org/record/1405206\#.XqoiBVNKiH4}}
\citep[][and submitted]{Marley_2017} and the
ATMO 2020 models 
\footnote{\url{http://opendata.erc-atmo.eu}}
\citep{Phillips_2020}. 

Figure 2 shows various colors plotted against $J -$ [4.5], as a proxy for $T_{\rm eff}$. Note however that $J -$[4.5] is also sensitive to
gravity, metallicity, mixing, and clouds (e.g. Figure 3 bottom panel).
The photometry is taken from this work (Tables 1 and 2) and the literature \citep[][see also the photometry compilation in the Appendix]{Leggett_2017, Kirkpatrick_2019, Marocco_2019, 
Bardalez_2020,Faherty_2020, 
Kirkpatrick_2020,
Marocco_2020, Meisner_2020a, Meisner_2020b}. 
Figure 3 shows color-magnitude diagrams for late T and Y dwarfs with measured trigonometric parallaxes. Parallaxes are taken from \citet{Leggett_2017, Martin_2018, Kirkpatrick_2019,  Bardalez_2020, Kirkpatrick_2020, Marocco_2020}. The absolute [4.5] magnitude 
is shown as a function of the near-infrared color $J - H$, the mid-infrared color [3.6] $-$ [4.5], and the long-baseline color $J -$ [4.5]. The absolute [4.5] magnitude
can be used as a proxy for luminosity because $\sim$half of the total energy is captured by this filter for cold brown dwarfs. Luminosity in turn is strongly correlated with $T_{\rm eff}$ through the Stefan-Boltzmann law, because the radius of a brown dwarf does not change significantly after around 0.3~Gyr \citep[][see also Section 5.5]{Burrows_1997}. Note however that the [4.5] flux is also sensitive to gravity, metallicity, and mixing (e.g. Figure 3 bottom panel).

The new photometric measurements presented here (Tables 1 and 2) are represented by blue points in Figures 2 and 3. The new data support and build on the empirical sequence in each panel of Figure 2; the $K$-band datapoint for J0647 nicely fills in a gap in the $J - K$ sequence at $J -$ [4.5] $\approx 8$, and the new $J$-band data  improves the definition of the tight $J -$ [4.5]:[3.6] $-$ [4.5] observational sequence. For the 400 -- 600~K brown dwarfs, the $J - K$ and [4.5] $-$ W3 colors appear to have a large degree of intrinsic scatter; we discuss this further in Section 6.2.

\begin{figure*}[b]
\vskip -1.1in
\hskip -0.2in
\gridline{\fig{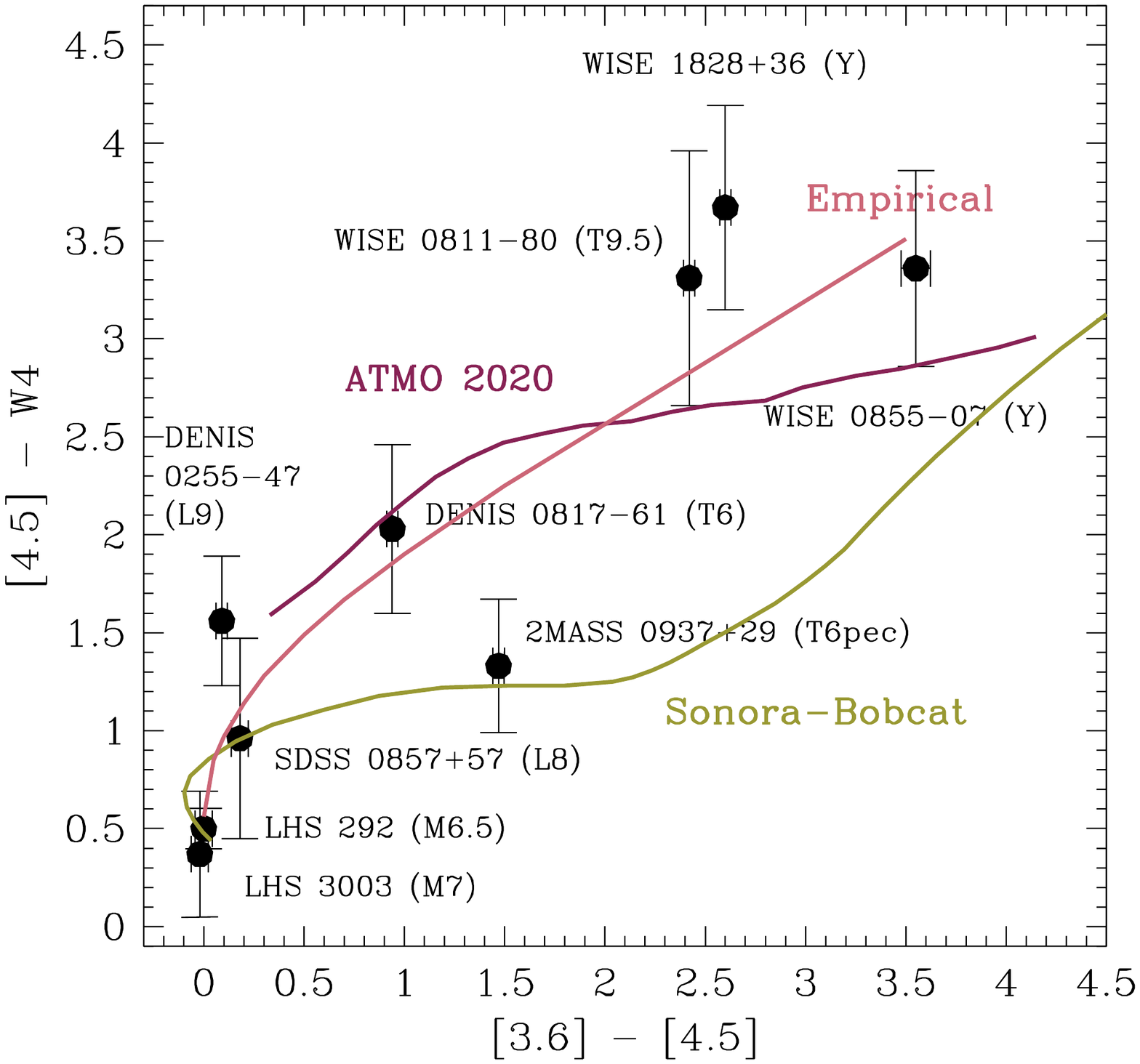}{0.53\textwidth}{}
\hskip -0.2in
        \fig{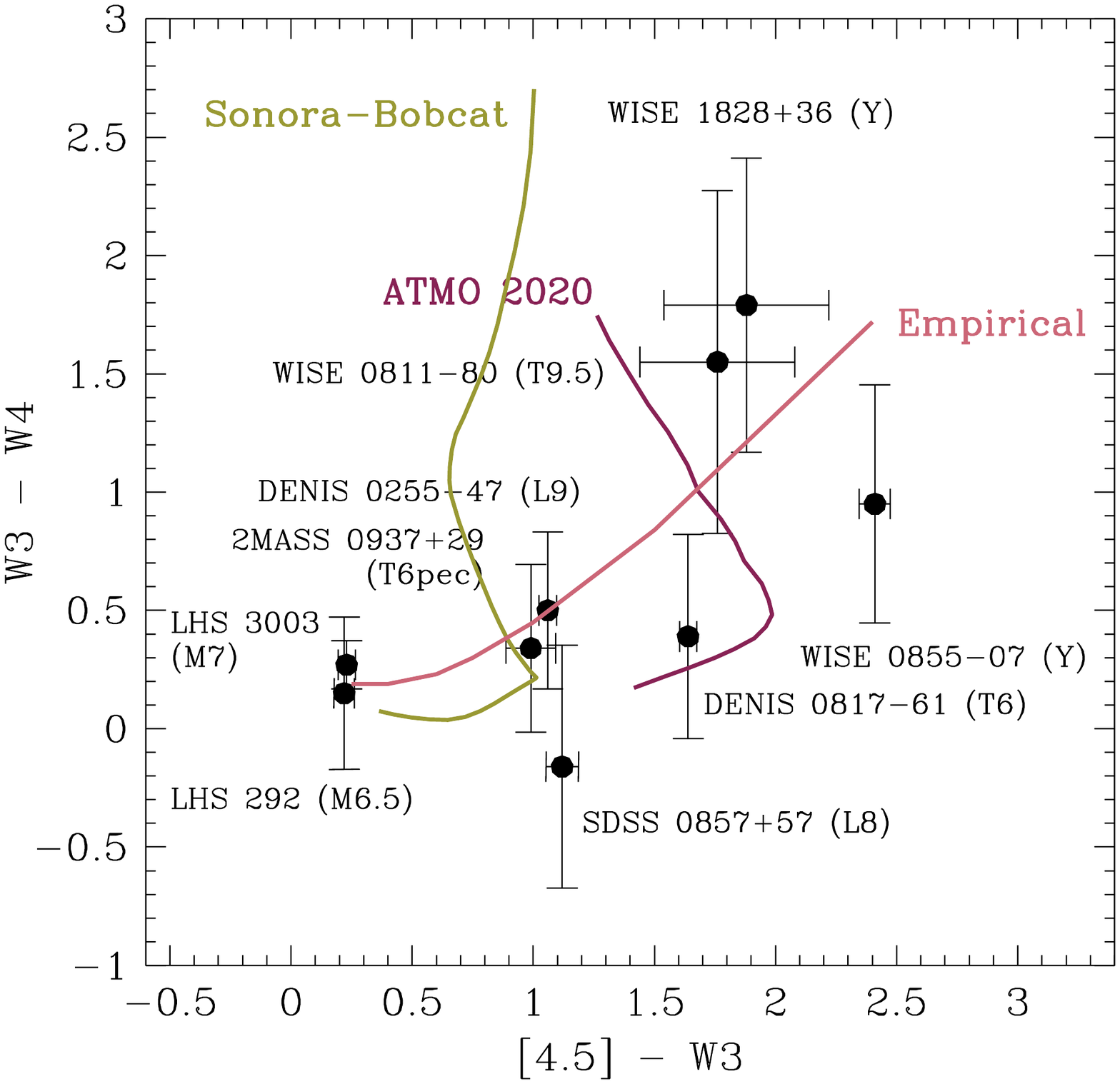}{0.53\textwidth}{}        
          }
\vskip -0.4in        
\caption{
Mid-infrared color-color diagram for M, L, T and Y dwarfs. Model sequences are  cloud-free,  with  solar-metallicity and $\log g = 4.5$. Olive green lines are Sonora-Bobcat chemical equilibrium sequences for  250 $\leq T_{\rm eff}$~K $\leq$ 2400, and dark red are ATMO 2020 non-equilibrium chemistry  sequences for 330 $\leq T_{\rm eff}$~K $\leq$ 1280.
The Sonora-Bobcat chemical equilibrium models reproduce the mid-infrared colors of late-M to late-L-type dwarfs, and the ATMO 2020 non-equilibrium chemistry models reproduce the colors of late-L to late-T dwarfs. 
An empirical by-eye sequence is shown which combines the two, and uses the observations of the Y dwarfs to anchor the red end of the sequence.
}
\end{figure*}

Figures 2 and 3 show that the most recent models at the time of writing, the ATMO 2020 and Sonora-Bobcat models, generate very similar colors for the same parameters. That is, the chemical equilibrium solar-metallicity cloud-free ATMO 2020 and Sonora-Bobcat model sequences (yellow and olive green solid lines in the figures) are very similar.  The models which include  vigorous mixing  (dark red lines) do a better job of reproducing the observed $J -$ [4.5]:$J - H$ and $J -$ [4.5]:[4.5] $-$ W3  sequences in Figure 2, and the $J - H$:$M_{[4.5]}$ and [3.6] $-$[4.5]:$M_{[4.5]}$ sequences in Figure 3. This is because 
mixing in these cool atmospheres has the net result of decreasing the NH$_3$ abundance and increasing N$_2$, and increasing CO at the expense of CH$_4$ \citep[e.g.][]{Noll_1997, Saumon_2006, Saumon_2007, Visscher_2011,Zahnle_2014, Leggett_2015, Tremblin_2015, Phillips_2020}.
The $H$- and W3-bands brighten when the NH$_3$ absorption decreases, and [4.5] becomes fainter due to increased CO.  For a representative 400~K brown dwarf with $\log g = 4.5$, the ATMO 2020 models  with no mixing and with strong mixing ($\log K_{zz} = 6$) give $\delta H = -0.7$, $\delta$W3 $= -0.2$, and $\delta$[4.5] $= +0.3$.

However, although the non-equilibrium chemistry models reproduce much of the data in Figures 2 and 3,  Figure 2 shows that all models diverge from the observed $J - K$ and [3.6] $-$ [4.5] colors for  $T_{\rm eff} \lesssim 600$~K. Discrepancies between observations and synthetic colors are also apparent in the $J -$ [4.5]:$M_{[4.5]}$ plot in Figure 3.

Figure 4 shows observed mid-infrared colors for M, L,  T and Y dwarfs which
can be used to estimate 5 --  20~$\mu$m colors of cool dwarfs, for example for {\it JWST} observations. If used for this purpose, the reader should note that
the uncertainties are large and exposure estimates should therefore be conservative. 
We include a by-eye empirical sequence which can be used for interpolation. 
It is important to note that  {\it chemical equilibrium models will underestimate the [4.5] $-$ W3 and  [4.5] $-$ W4 colors of T and Y dwarfs by 
$\sim 1$~magnitude.}

\bigskip
\section{Modifications to Brown Dwarf Model Atmospheres}

Given the discrepancies between observations and models for brown dwarfs with $T_{\rm eff} < 600$~K (Figures 2 and 3), we explored modifications to the model structure. We used the ATMO 2020 models which include strong mixing as the starting point, as overall they reproduce the observations better than the chemical equilibrium models.
%Brown dwarfs have a metallic hydrogen/helium core which is supported by electron degeneracy, and insulated by a molecule-rich atmosphere through which heat escapes  over $\sim$Gyr timescales \citep[e.g.][]{Burrows_2001}. 

Energy transport in a cool dwarf atmosphere is predominantly convective, with radiative cooling becoming important  high in the atmosphere  where the pressure is too low for convection to be efficient. 
Convection is treated as an adiabatic process  where pressure $P$ and temperature $T$ are defined by 
${P}^{(1-\gamma )}{T}^{\gamma }=\mathrm{constant}$. For an ideal gas, $\gamma$ is the ratio of specific heats at constant pressure and volume and, for a gas composed entirely of molecular hydrogen, $\gamma = 1.4$. 
The reader is referred to
\citet{Marley_2015} and \citet{Zhang_2020} for  reviews of the important processes in model atmospheres.  

One-dimensional models, such as the ATMO and Sonora-Bobcat models, represent the atmosphere as a  $P-T$ profile which maps the cooling from the core out to the surface, and by a chemical abundance profile which maps the chemical changes that occur through the atmosphere as $P$ and $T$ change. The $P-T$ profile can be thought of as a slice through the atmosphere, where both temperature and pressure decrease with increasing altitude. 

Of course, an actual brown dwarf atmosphere is more complex. These objects rotate rapidly  with periods of a few hours, similar to the solar system giant planets \citep{Zapatero_2006,Cushing_2016, Esplin_2016, Leggett_2016b,Scholz_2018, Vos_2020, Tannock_2021}. They also have a radius approximately equal to Jupiter's \citep[e.g.][]{Burrows_1997}.
The atmospheres are turbulent, and are likely to have 
planetary-like features such as zones, spots and planetary-scale waves \citep{Apai_2017, Showman_2019}. 
\citet{Showman_2013} simulate the dynamics of a brown dwarf atmosphere  and demonstrate that for a rotation period of a few hours, large-scale, organized horizontal wind speeds of tens of m s$^{-1}$ are plausible, and coherent vertical  circulation moves air parcels over a scale height ($\sim 7$~km) in $\sim 10^5$~seconds. These motions translate into a diffusion parameter $K_{zz}\sim 10^6$~cm$^2$ s$^{-1}$, typical of the values used in the ATMO 2020 non-equilibrium chemistry models \citep[][their Figure 1]{Phillips_2020}. The coefficient is higher
in the atmospheres of Jupiter and Saturn where
$K_{zz}\approx 10^8$~cm$^2$ s$^{-1}$ \citep{Wang_2016}. The $\lambda \sim 5~\mu$m spectrum of the very cold brown dwarf J0855 is also best fit with a high mixing coefficient of $K_{zz}\approx 10^{8.5}$~cm$^2$ s$^{-1}$ \citep[][and Section 5.2]{Miles_2020}.

\citet{Augustson_2019}, and references therein, describe how convection in a 
rotating stellar or planetary atmosphere can change the chemical composition and thermodynamic properties of the gas and therefore impact the differential rotation, opacity, and thermodynamic gradients of the atmosphere. 
The model developed by \citet{Augustson_2019} connects the rotation rate and vertical diffusion coefficient
to the velocity of the gas motion, the divergence from adiabacity,
 and characteristic scale lengths.  
The damping effect of rotation can decrease the size of the convection zone, leading to sharper thermodynamic and chemical gradients than would otherwise be present. Furthermore, both superadiabatic and subadiabatic temperature gradients can exist in the atmosphere.

The atmospheres of the solar system giant planets are not perfectly adiabatic  \citep[e.g.][]{Guillot_1994, Guillot_2005, Vazan_2020} and
various mechanisms can produce a non-adiabatic cooling curve in giant planet
and brown dwarf atmospheres. 
%Discrete radiative zones can form within the convective zone \citep{[][their Figure 2]Marley_2015}, or 
These include
compositional changes  such as those due to condensation \citep[e.g.][]{Robinson_2012}, or the CO $\Leftrightarrow$ CH$_4$ 
%or  N$_2$ $\Leftrightarrow$ NH$_3$
changes at the L- to T-type spectral transition  \citep{Tremblin_2015, Tremblin_2019}. The upper atmosphere  can be heated by a cloud deck, or by breaking gravity waves  \citep[e.g.][]{Schubert_2003, O'Donoghue_2016}.  Further evidence in support of non-adiabatic $P-T$ profiles in brown dwarf atmospheres comes from 
retrieval analyses.  \citet{Line_2015, Line_2017} and \citet{Zalesky_2019} reproduce near-infrared observations of T and Y dwarfs with non-adiabatic 
$P-T$ curves, and \citet{Piette_2020} show that a parametric $P-T$ profile can be used to determine accurate atmospheric parameters from a high precision spectrum of a T dwarf.

In summary, there is significant evidence that the $P-T$ curve of a brown dwarf atmosphere does not, and should not be expected to, follow the standard adiabat. In this work we treat the adiabatic parameter $\gamma$ as a variable, along with $T_{\rm eff}$, $g$, [m/H] and $K_{zz}$, and generate a small number of models to compare to observations of a sample of cold brown dwarfs. 
In the ATMO 2020 models, the initial value of $\gamma$ is determined for each atmospheric layer using the equation of state tables from \citet{Saumon_1995}; for our tuned models we force $\gamma$ to be constant in the upper atmosphere. The tuning process is described in the next section.

The models are cloud-free, and clouds are not expected to be significant in the photospheres of 400 -- 600~K brown dwarfs \citep{Morley_2012, Morley_2014}. However, for the warmest atmospheres in our sample there may be chloride and sulfide clouds in deep regions that can contribute flux at wavelengths where the atmosphere is clear. For the coldest  objects, water clouds may form high in the atmosphere, and these would impact the SED at wavelengths where the the atmosphere is opaque. We discuss this further in Section 5.3.

The ATMO 2020 models we use here have a fixed potassium abundance.
\citet{Phillips_2020} show that different treatments of potassium broadening produce large variations  in the shape of the blue wing of the $Y$-band  flux peak in brown dwarfs. Those  authors note that an order of magnitude reduction  in  the K abundance improves  the agreement between models and observations, and suggest that current modelling of the potassium chemistry, including its condensation into KCl,  is slightly incorrect. In this work we adopt a  K abundance of 4$\times 10^{-9}$ for the late-type T dwarf we use as a proof-of-concept, UGPS J072227.51$-$054031.2 (hereafter J0722). For the cooler Y dwarfs, we adopt  a  K abundance of $1\times 10^{-9}$. We return to the issue of potassium and the $Y$-band in Section 6.1. 

The analysis presented here  is a first step towards including processes currently missing in all brown dwarf models. We simplify the complex three-dimensional turbulent atmospheres by parameterizing the $P-T$ profile in a one-dimensional model. We show below that this simple approach significantly improves the agreement with observations.

\bigskip
\section{Tuning the  Pressure-Temperature Profile}

\medskip
\subsection{Proof of Concept: the 500~K Brown Dwarf UGPS J072227.51-054031.2}

\begin{figure}
\plotone{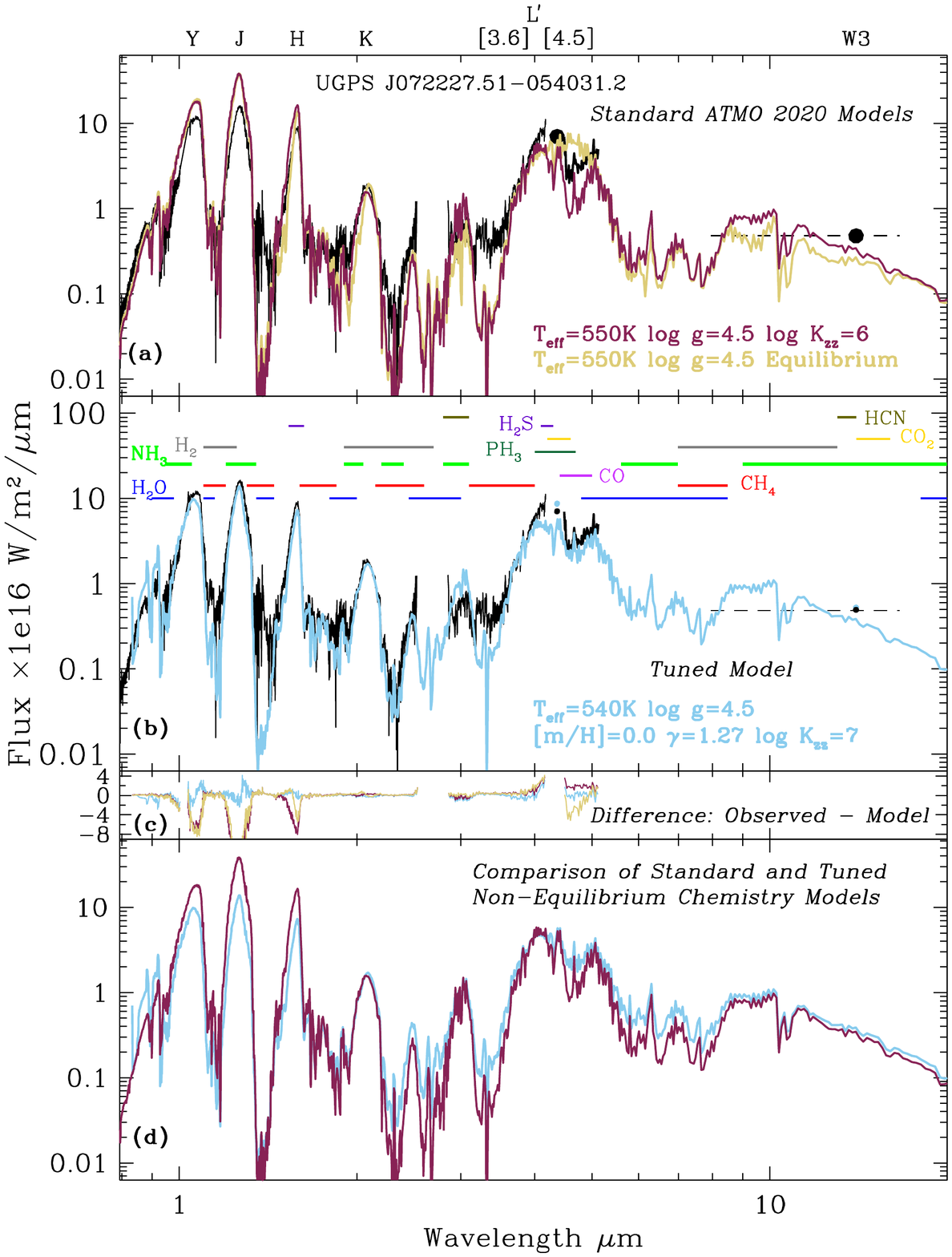}
\vskip -0.35in
\caption{The black line is the flux-calibrated spectrum of  UGPS J072227.51$-$054031.2 \citep{Lucas_2010, Leggett_2012, Miles_2020}. The black circles are the observed {\it Spitzer} [4.5] and {\it WISE} W3  photometric datapoints, and the dashed line indicates the width of the W3 filter which peaks at $\lambda \sim 14~\mu$m. 
The colored  lines are ATMO 2020  models with  parameters given in the legends. Significant absorbers are identified at the top of panel (b).  Also in panel (b), the small blue and black datapoints  demonstrate  the good agreement between the  observed and tuned-model photometry in the mid-infrared. Panel (c) demonstrates the improvement in fit provided by the tuned model (note the linear $y$-axis). Panel (d) compares the standard and tuned ATMO 2020 non-equilibrium models.
Note that the model fluxes are not scaled to match the data, they are scaled by the measured distance and by the brown dwarf radius calculated by ATMO 2020 evolutionary models. 
}
\end{figure}

We use observations of the bright late-type T-dwarf  J0722 for our initial  test. This brown dwarf has $T_{\rm eff} \approx 500$~K and has extensive observational data, including spectra at $\lambda \sim 3.5~\mu$m and $\lambda \sim 4.8~\mu$m \citep{Lucas_2010, Leggett_2012, Miles_2020}. Table 3 lists previous determinations of the atmospheric parameters of J0722. \citet{Leggett_2012} compare the observed near-infrared and $\lambda \sim 3.5~\mu$m spectra of J0722, and mid-infrared photometry, to chemical non-equilibrium cloud-free \citet{Saumon_2012} models. Constrained by luminosity, they find a range in the [$T_{\rm eff},\log g$] parameters  of [492,3.5] to [550,5.0]. The mid-infrared observations pushed the parameter selection to the lower temperatures and gravities, while the near-infrared was better fit by the higher temperature and gravity solution.  \citet{Filippazzo_2015} and \citet{Dupuy_2013} also use luminosity-based arguments to determine the parameters given in Table 3, while \citet{Miles_2020} use the Sonora-Bobcat model grid and near- and mid-infrared photometry to constrain $T_{\rm eff}$, evolutionary models to constrain $g$, and the $4.8~\mu$m spectrum to constrain $K_{zz}$.

The top panel of Figure 5 shows SEDs generated by standard models with parameters typical of those found for J0722. As found in earlier analyses \citep[e.g.][]{Leggett_2012, Miles_2020}, the fit is quite good, especially for the non-equilibrium chemistry models. However
the calculated $YJ$ fluxes are higher than observed and the $2 \lesssim \lambda~\mu$m $\lesssim 4$ flux is lower than observed. The direction of these offsets is consistent with the systematic discrepancies seen in the colors of the cooler brown dwarfs in Figures 2 and 3: 
\begin{itemize}
\item the modelled $J - H$ and $J -$ [4.5] are too blue because  $J$ (in particular, see Figure 5) is too bright,
\item the modelled $J - K$ is much too blue because $J$ is too bright {\it and} $K$ too faint, 
\item and the modelled [3.6] $-$ [4.5]  is too red because [3.6] is too faint. 
\end{itemize}
In other words, {\it the discrepancies demonstrated in the top panel of Figure 5, for standard models, apply to all brown dwarfs with $T_{\rm eff} < 600$~K}.

Panel (b) of Figure 5 shows a model we have tuned to better fit the observations.  The tuning is done manually, iterating over ages of 100~Myr, 1~Gyr, 5~Gyr, and 10~Gyr, and metallicities of $-0.5$, 0 and $+0.3$ dex. The steps involved are:
\begin{itemize}
    \item assume {\it a priori} that $\log K_{zz} = 7$
    \item select $\log g$ and radius based on evolutionary models for the selected age
    \item select $T_{\rm eff}$ to reproduce the observed flux at [4.5]
    \item decrease $\gamma$ to reduce the $YJHK$ flux
    \item let $\gamma$ increase to the standard value at a depth in the atmosphere defined by pressure  $P(\gamma, max)$, and deeper, to increase the $YJ$ flux as necessary 
    \item adjust $\log K_{zz}$ if the other adjustments have changed the [4.5] flux 
\end{itemize}
The fits are also constrained by ensuring that the scaling used to transform the model surface flux to that detected at Earth, which depends on the distance to the dwarf and its radius, is consistent with  the evolutionary models. 
Once a reasonable fit is obtained, judged by eye for this preliminary analysis, selection between any similar quality fits is done by choosing the fit that best agrees with the observed  [3.6] and W3 photometry.

\begin{figure}
\gridline{
\fig{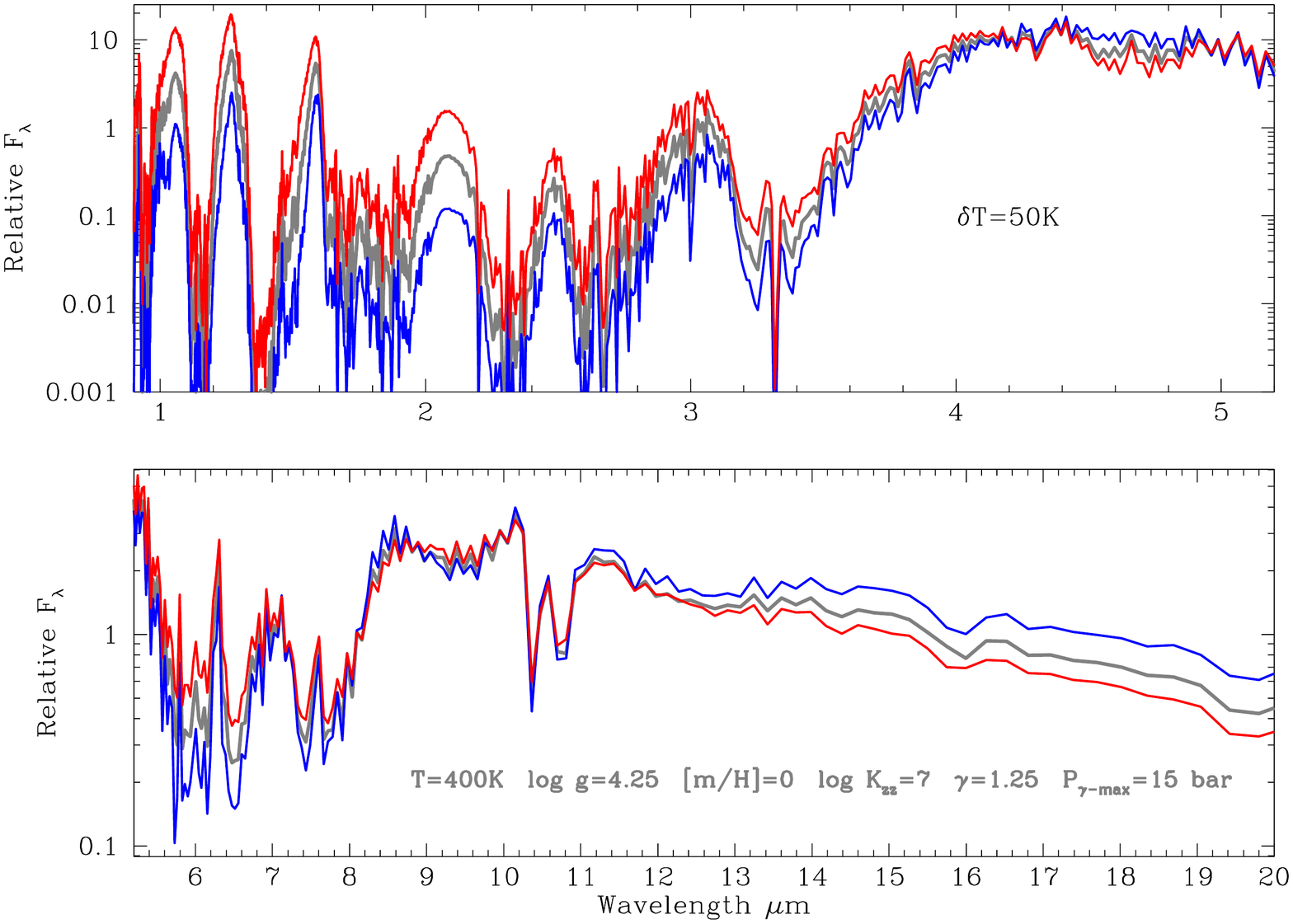}{0.5\textwidth}{}
        \fig{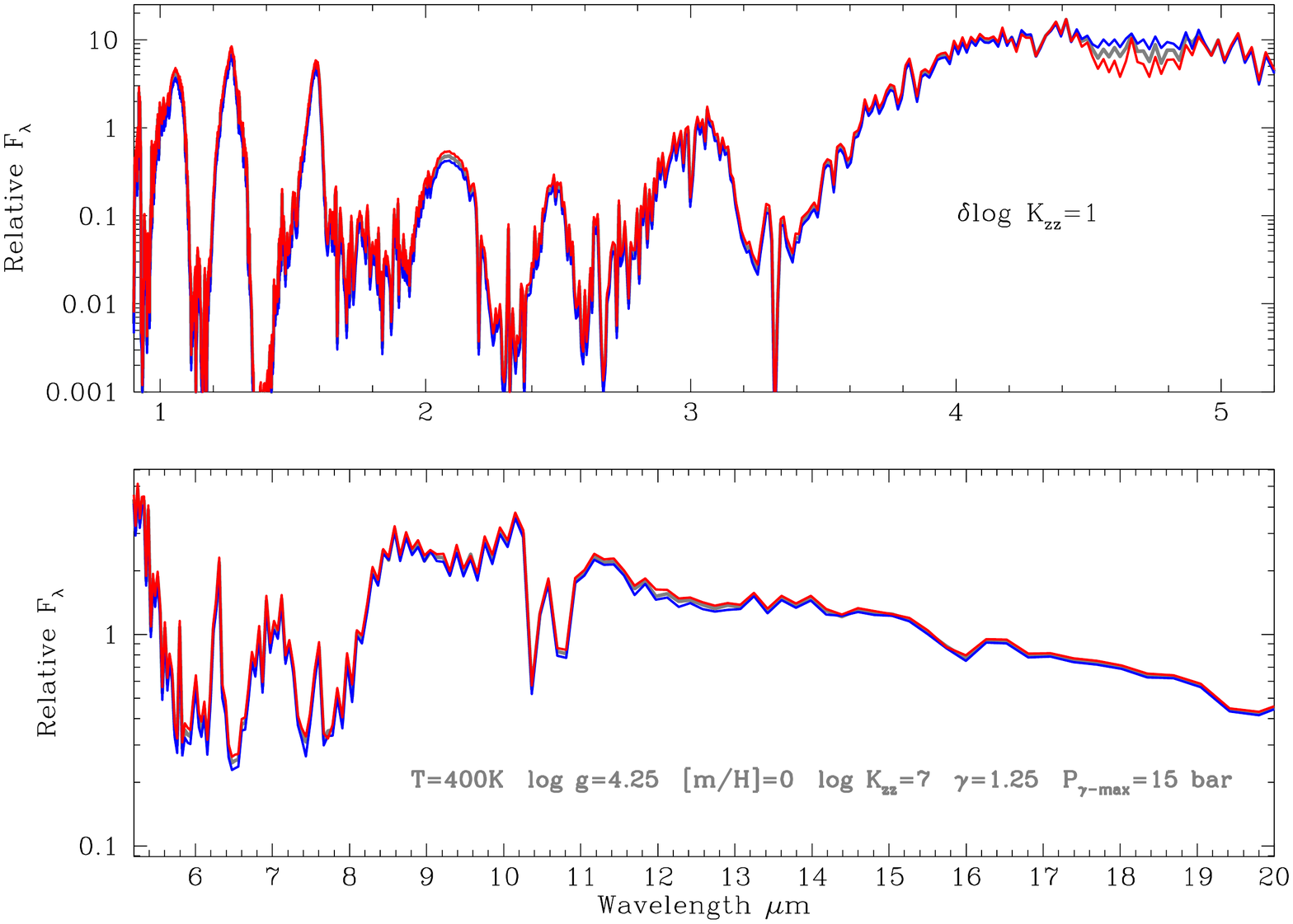}{0.5\textwidth}{}
         }
          \vskip -0.6in
\gridline{
\fig{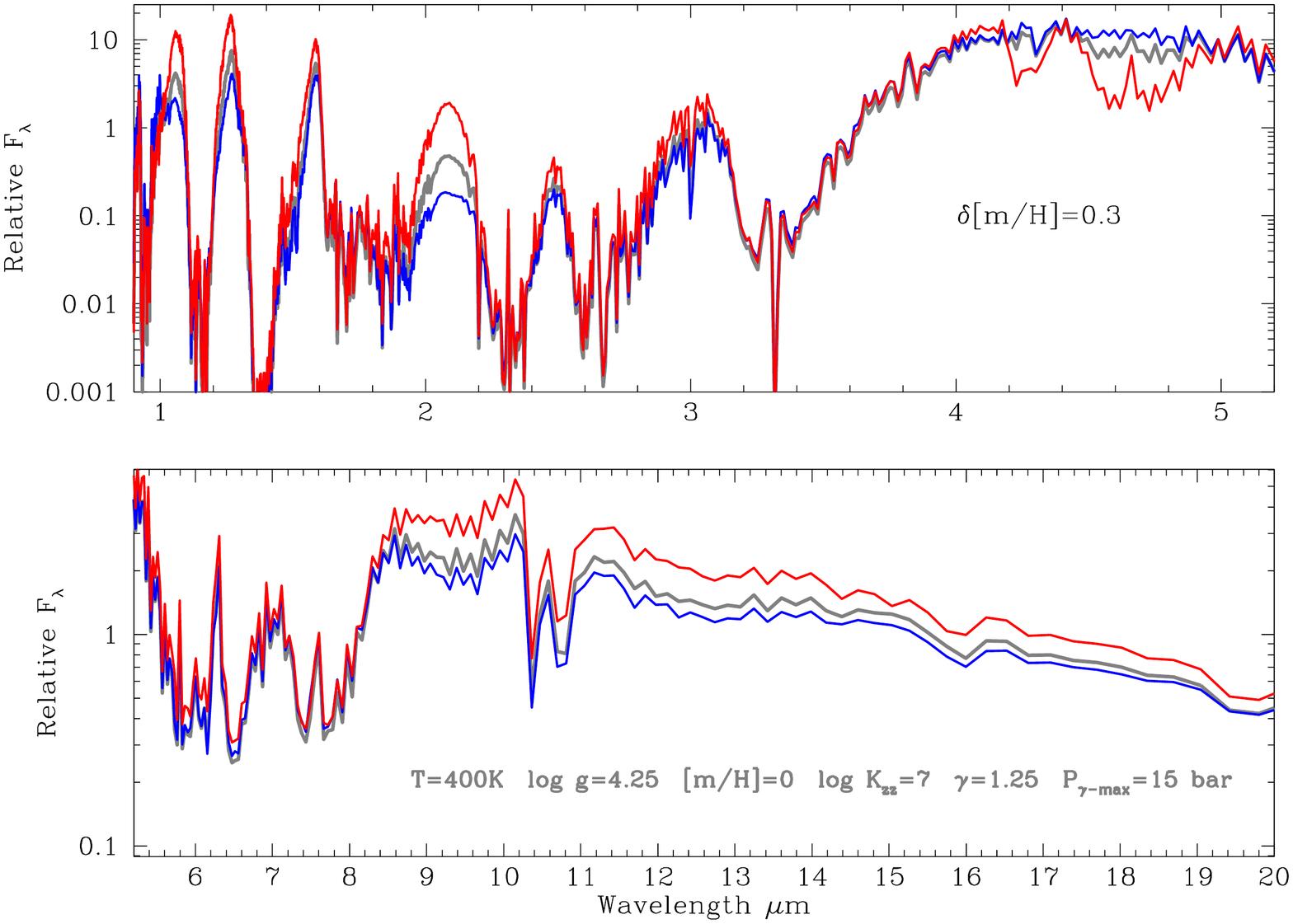}{0.5\textwidth}{}
        \fig{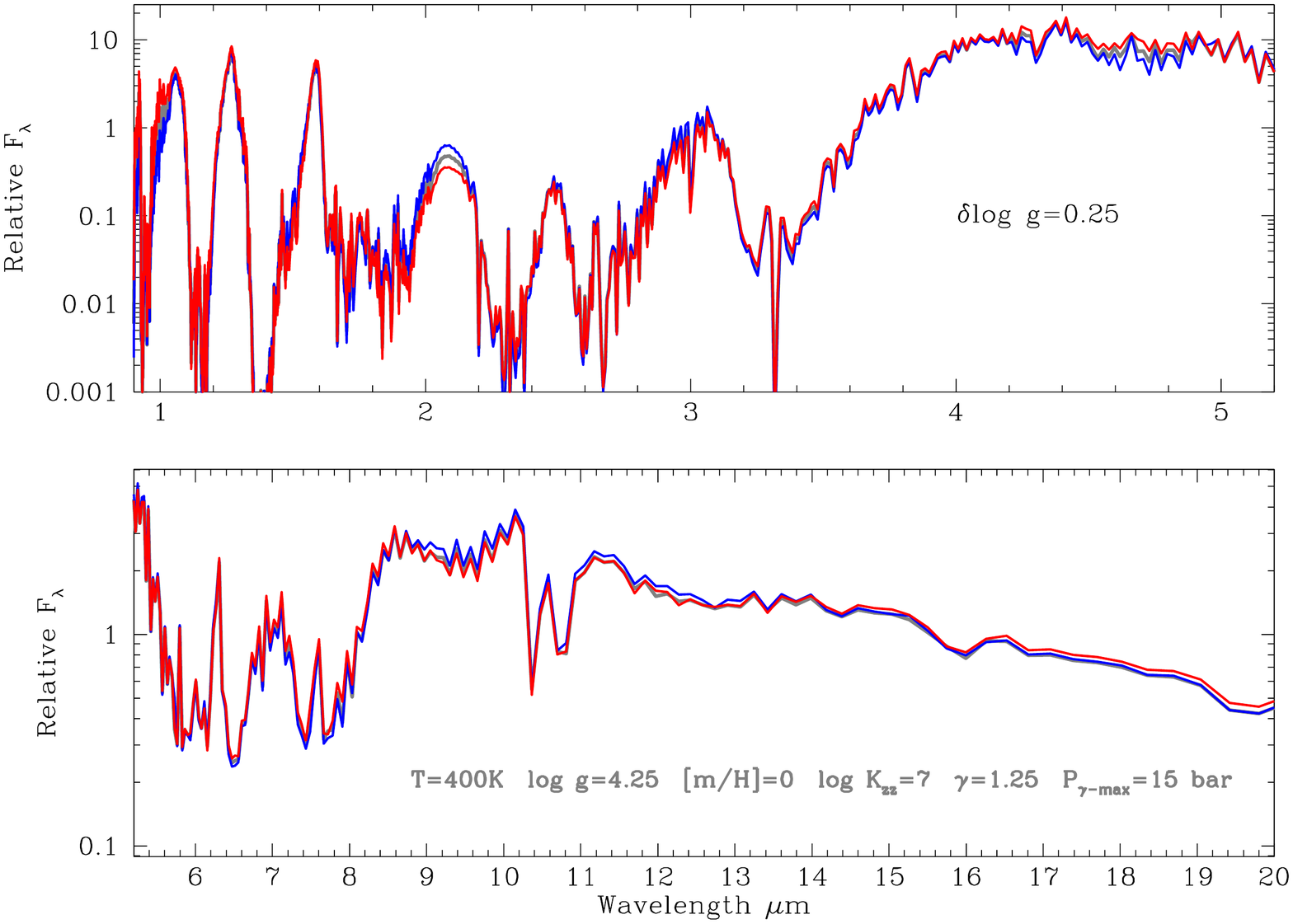}{0.5\textwidth}{}
         }
          \vskip -0.6in     
        \gridline{
\fig{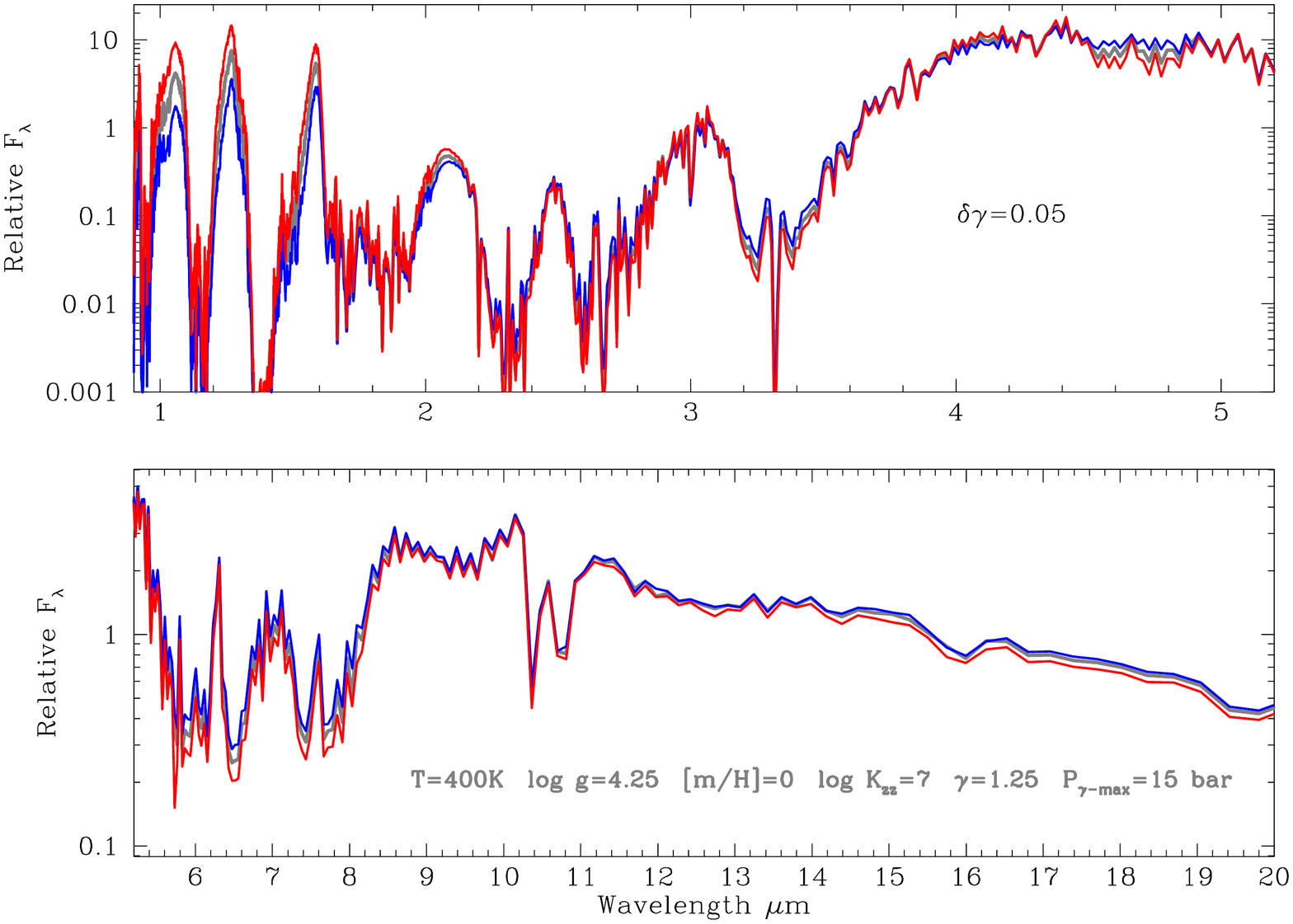}{0.5\textwidth}{}
        \fig{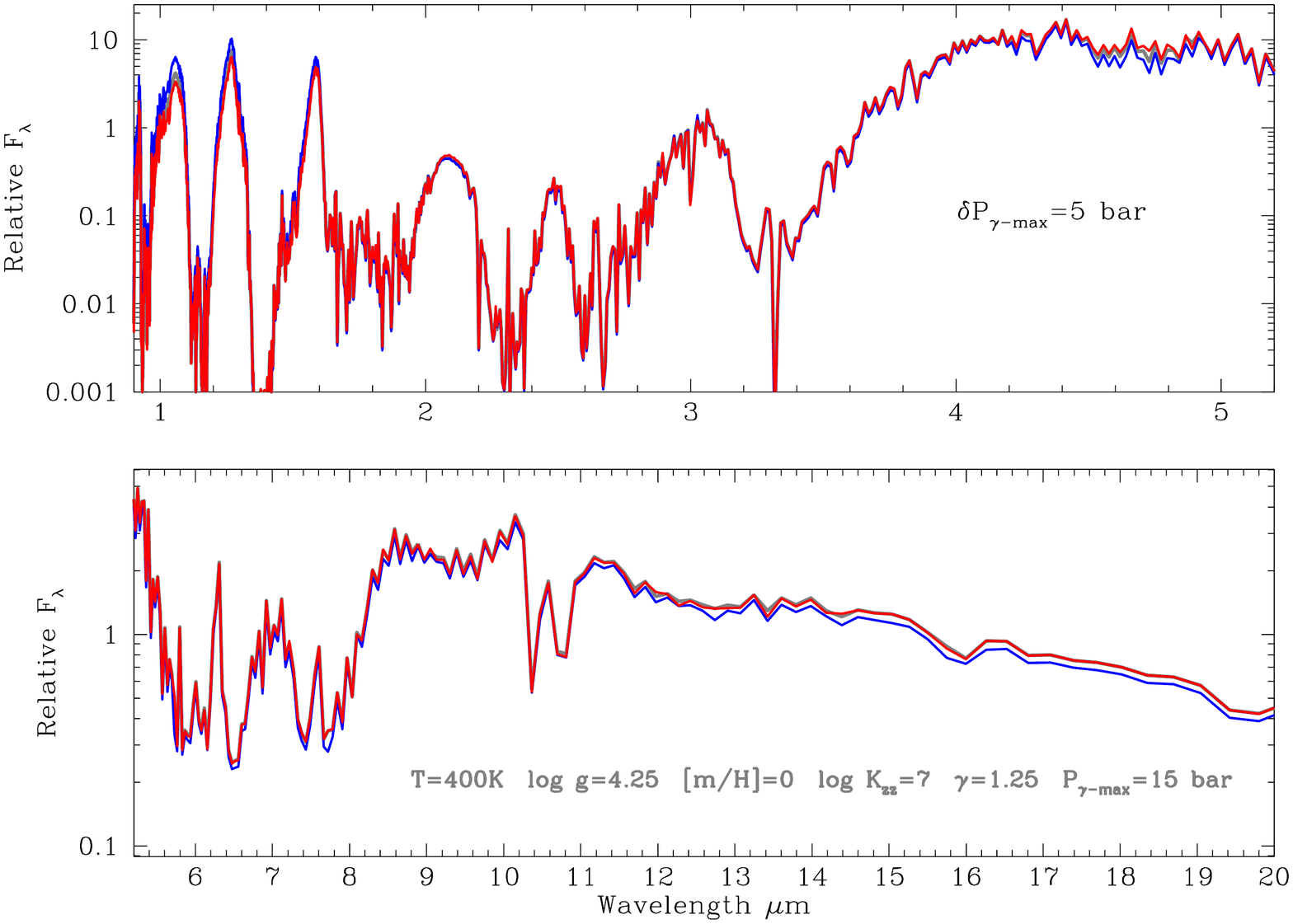}{0.5\textwidth}{}
         }
            \vskip -0.4in
\caption{Comparison of the effects of varying the model parameters in our tuning process, for a representative $T_{\rm eff}=400$~K model.  Each variation of the six parameters is displayed as a plot pair, with the  near-infrared region in the upper plot and the mid-infrared in the lower.
The grey line shows the SED for the model with parameters as in the legend. Red and blue lines show the SEDs generated by increasing or decreasing the parameter, respectively. The parameter that is being varied is shown in the upper panel. The spectra are normalized to a value of 10 at $\lambda = 4.98~\mu$m (a local flux maximum).
} 
\end{figure}

\begin{figure}
\gridline{
\fig{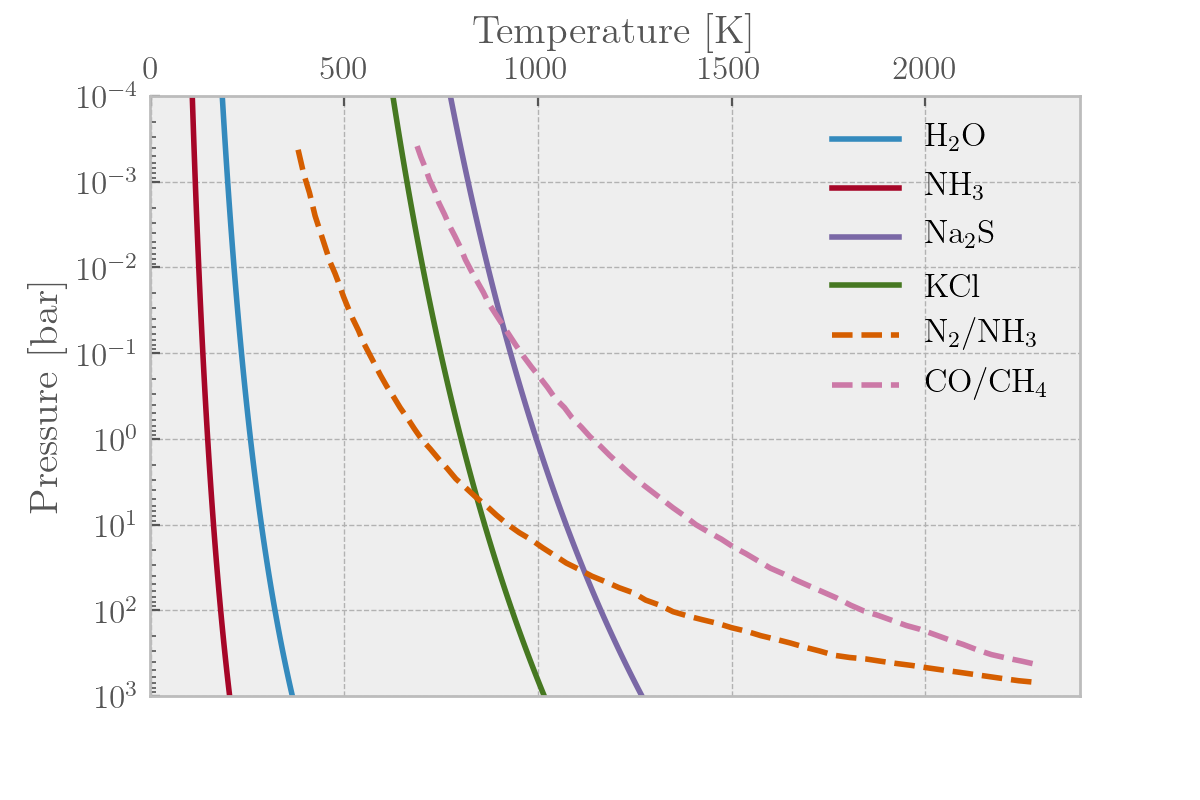}{0.47\textwidth}{}
          \fig{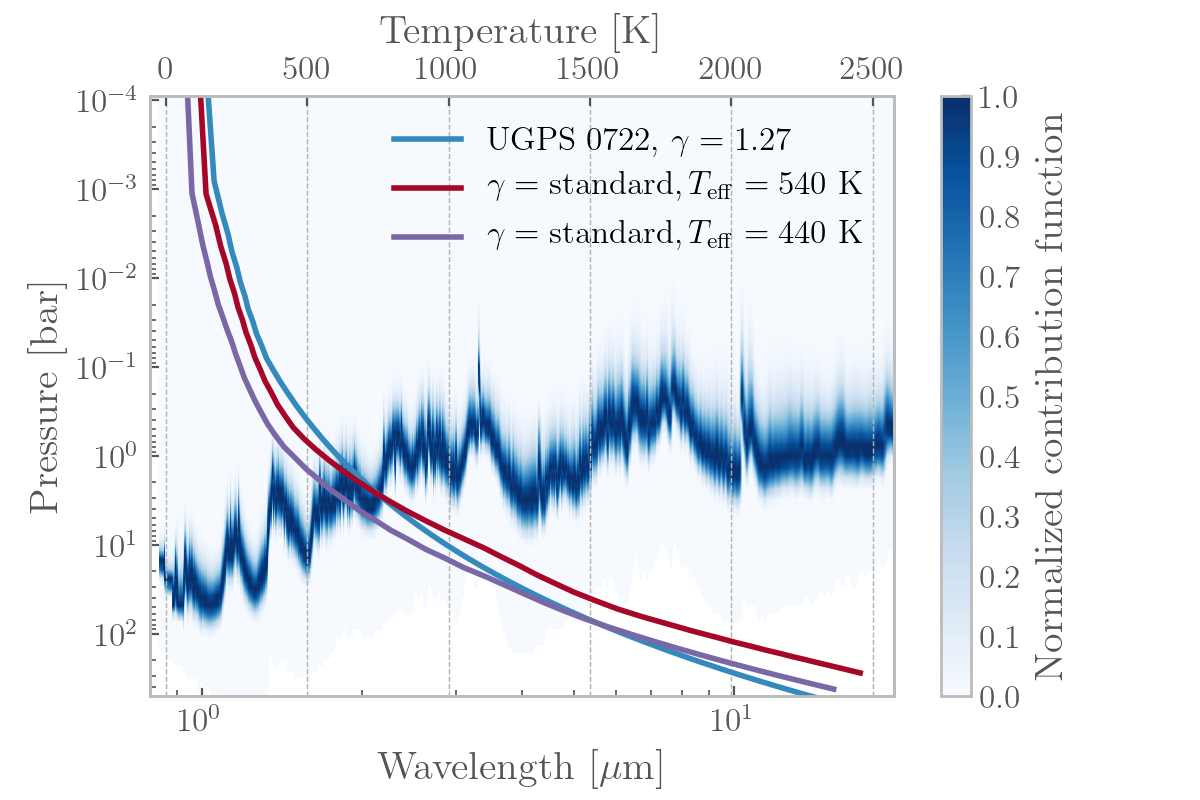}{0.47\textwidth}{}
          }
          \vskip -0.4in
\gridline{
\fig{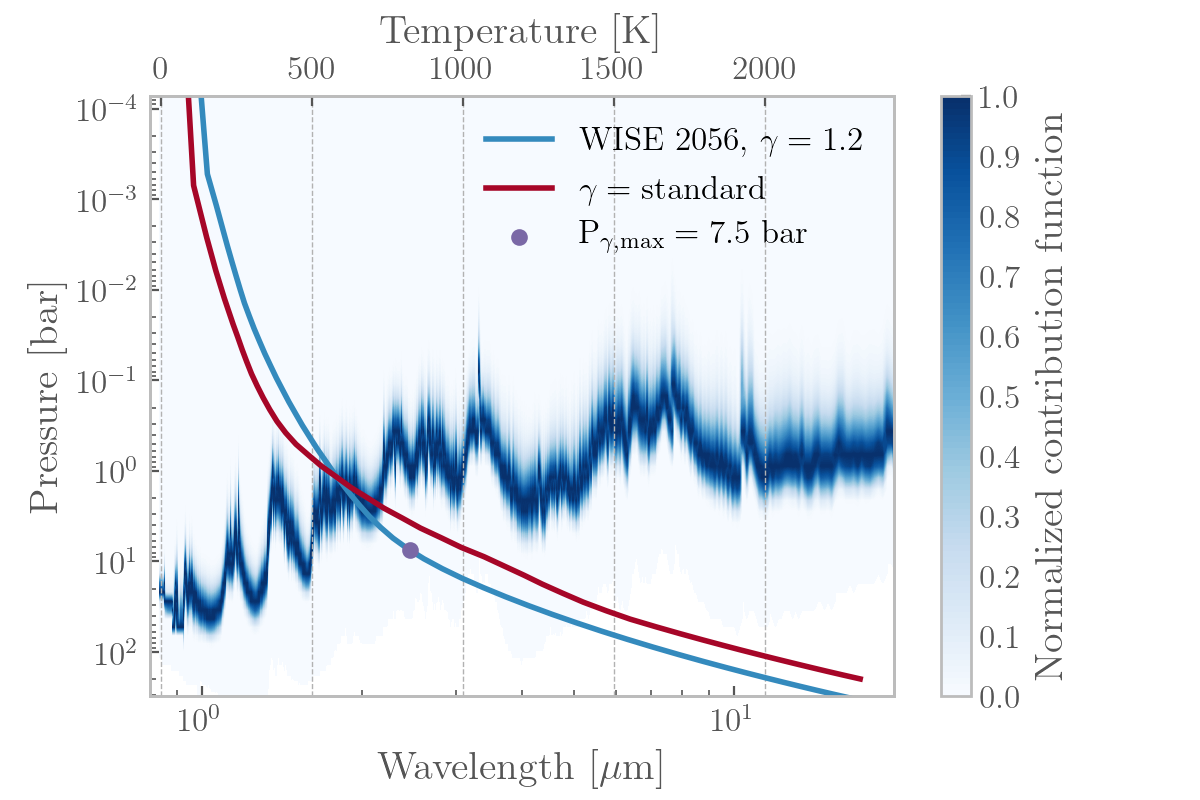}{0.47\textwidth}{}
  \fig{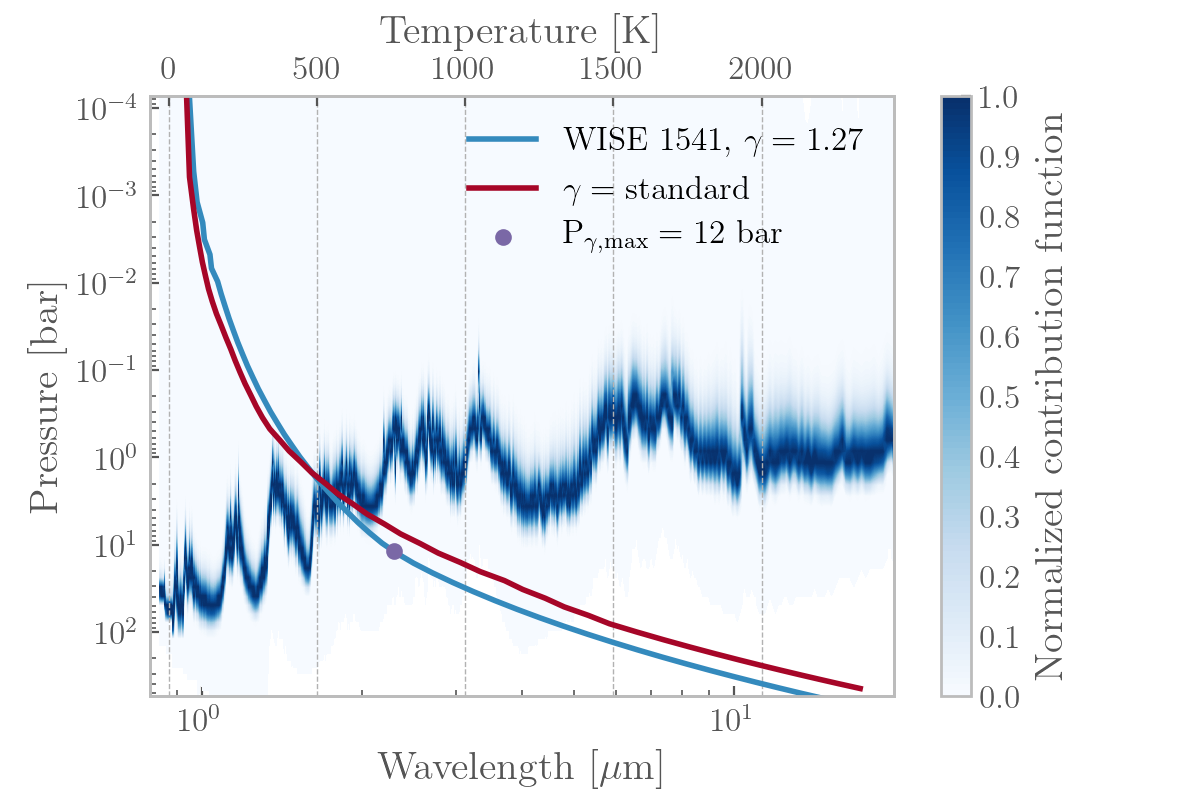}{0.47\textwidth}{}
  }
            \vskip -0.4in
\gridline{
\fig{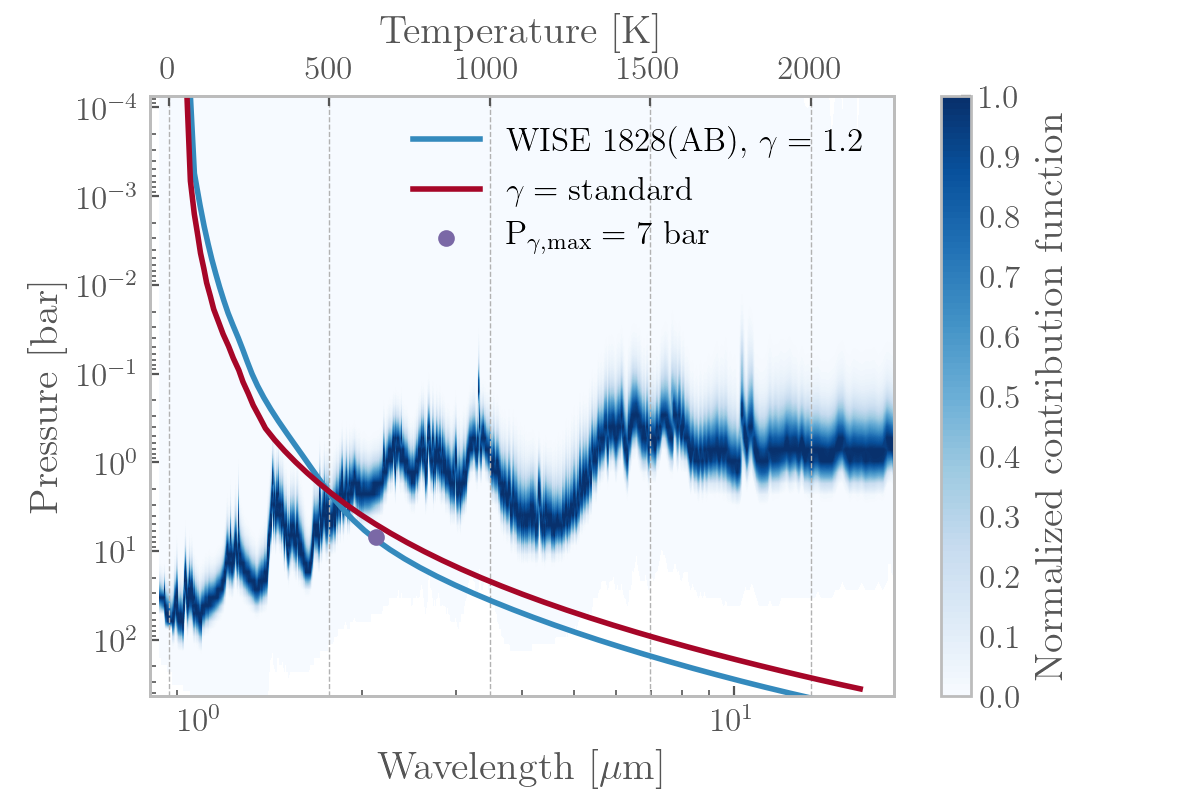}{0.47\textwidth}{}
\fig{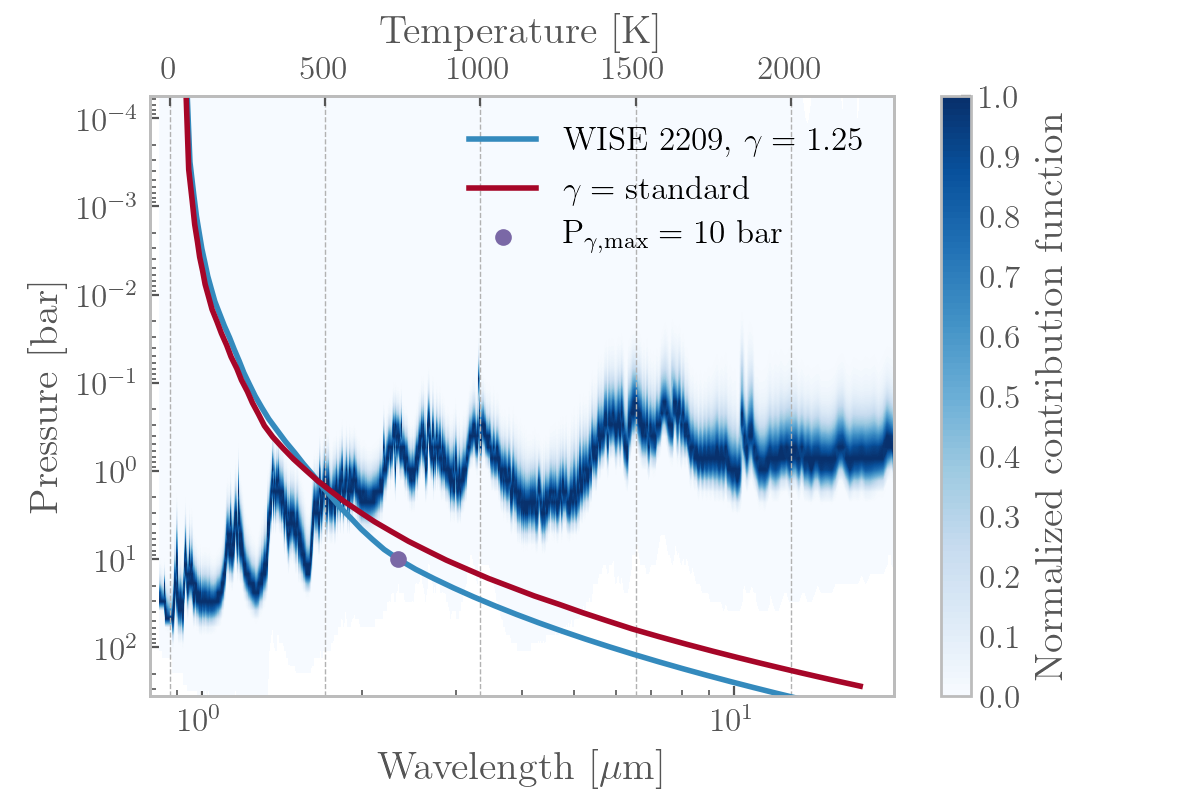}{0.47\textwidth}{}
}
          \vskip -0.4in
  \gridline{
  \fig{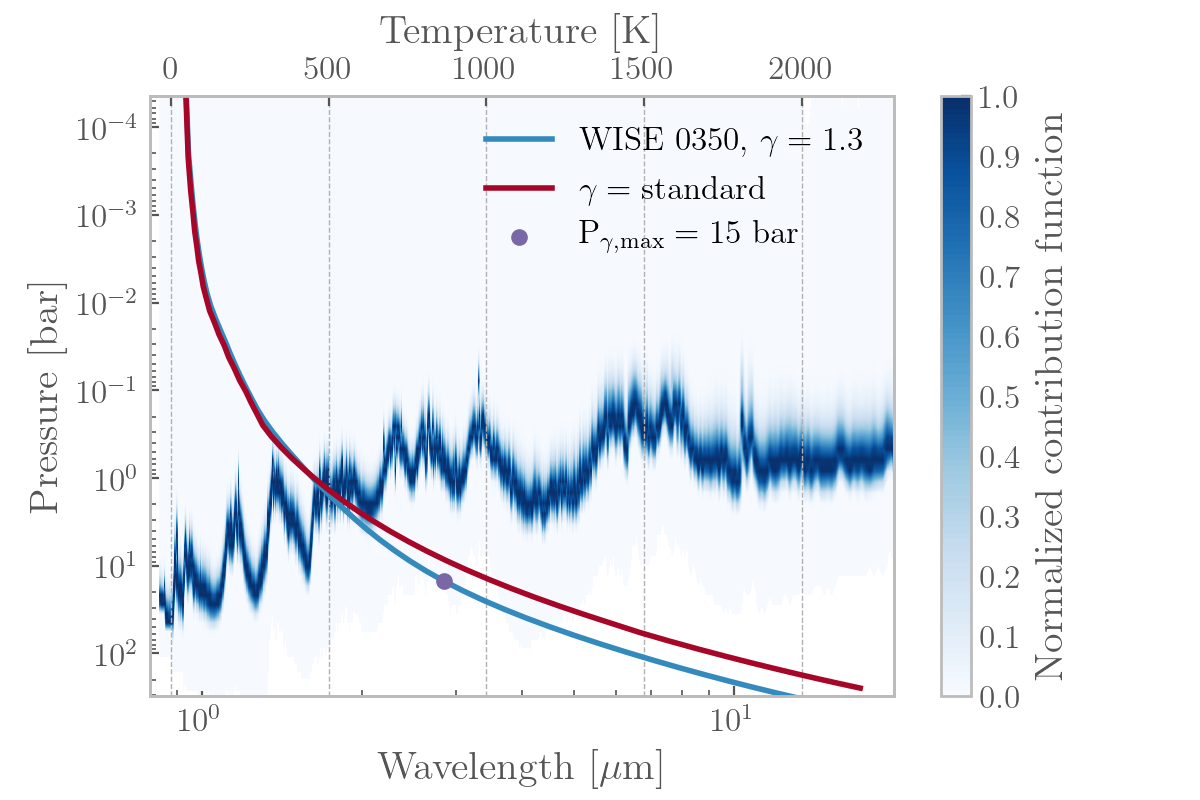}{0.47\textwidth}{}
  \fig{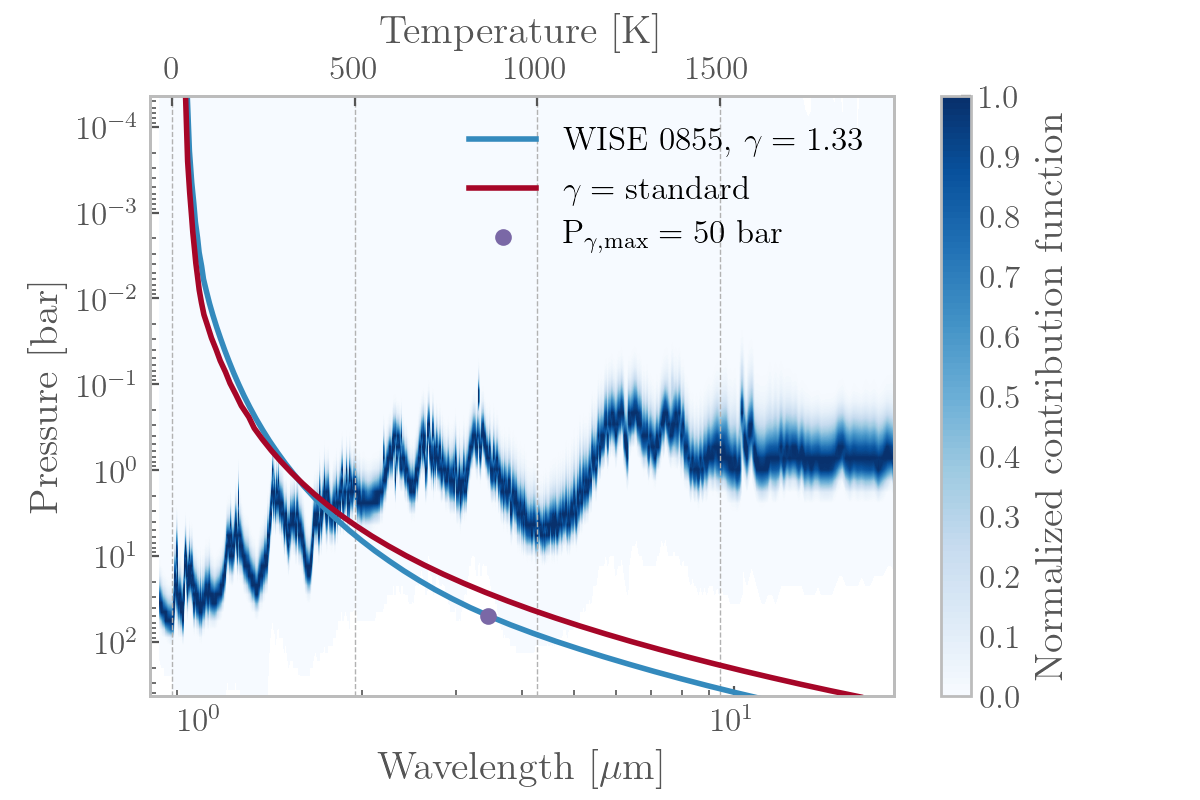}{0.47\textwidth}{}
  }
               \vskip -0.3in
\caption{
The top left panel shows condensation curves for elements in equilibrium (see text for discussion). The other panels are pressure-temperature profiles (left and upper axes) and flux contribution functions (thick blue line, left and lower axes), for the  brown dwarfs in our tuning sample. Blue $P-T$ profiles are tuned to fit the data by reducing the adiabat $\gamma$ at $P < P _{\gamma, \rm max}$; red lines have a standard radiative-convective profile.
} 
\end{figure}

\setlength\tabcolsep{3pt}
\begin{deluxetable*}{crcccclcccccccc}[t]
\tabletypesize{\scriptsize}
\tablecaption{Atmospheric Parameters for the $P$-$T$ Tuned Brown Dwarf Sample}
\tablehead{ 
\\
&
&  \multicolumn{5}{c}{Previous Work}
&  \multicolumn{6}{c}{This Work}
&  \multicolumn{2}{c}{Evol. Model} \\
{Name} 
& {$V_{tan}$}
&{$T_{\rm eff}$}  
& {$\log g$}  
& {[m/H]}  
& {$\log K_{zz}$ } 
& {Ref.} 
& {$T_{\rm eff}$}  
& {$\log g$}
& {[m/H]} 
& {$\log K_{zz}$} 
& {$\gamma$} 
& $\Delta\gamma$(P,T)
& {$\sim$Mass}
& {$\sim$Age}
\\
& {km~s$^{-1}$}
&{K}  
& {cm~s$^{-2}$} 
&   
& {cm$^2$s$^{-1}$} 
& 
& {K}  
& {cm~s$^{-2}$}
& 
& {cm$^2$s$^{-1}$} 
&
& bar, K
& {$M_{Jup}$}
& {Gyr}
}
\startdata
{\it WISEA} & 16.8 & 310 -- 340 & 3.75 -- 4.25 & $\gg 0$  & 6.0 &  Le17 &  325   &  4.0     &  +0.3  & 6.0  &  1.30     &  (15,860)   &  5  &   1.0  \\
      J035000.31         &      $\pm$ 0.3          & 300 -- 350  &$\sim 5.00$   &    &   & Sc15 & &   &  &  &  &  &   \\
          $-$565830.5                          &   & 294 -- 341  & 3.92 -- 4.47   &   &  & Du13\tablenotemark{a} &   &  &  &  &  &  &  \\
\tableline
UGPS & 18.9 & 522 -- 558 & 3.70 -- 4.40 &  & 4.4 & Mi20 & 540 & 4.50  & 0.0  & 7.0 & 1.27 & \nodata   & 15 & 1.5  \\
 J072227.51                         &  $\pm$ 0.2       & 524 -- 614 & 4.15 -- 5.21 & $\sim$0 & & Fi15 &     &       &      &      &     &    &     \\
 $-$054031.2 &  & 493 -- 551 & 4.38 -- 4.92 & & & Du13\tablenotemark{a}  &  &   &   &  &  &  &    \\
 &   & 490 -- 520 & 3.50 -- 4.50 & $\sim$0 & 5.5 & Le12 &  &   &   &  &  &  &    \\
\tableline
{\it WISE} &  88.0
& 249 -- 260 & 3.50 -- 4.50 &   & 8.5    & Mi20 & 260 & 4.00  & 0.0  & 8.7 & 1.33 & (50,870) & 5  & 3.0    \\
 J085510.83& $\pm$ 0.6 & 240 -- 260 & 3.50 -- 4.30 &  &  6.0   & Le17 & &  &  &  &  &  &      \\
                   $-$071442.5 &                 & $\sim 240$ & $\sim 4.00$  &   &    & Lu16 &     &       &      &      &     &  & \\
\tableline
{\it WISEPA} & 25.7
& 396 -- 434 & 4.30 -- 4.90 &  & 6.0 & Mi20 & 375 & 4.50  & +0.3 & 6.0 & 1.27 & (12,760) & 12 & 3.0  \\
  J154151.66      &    $\pm$ 0.4                       & 302 -- 474 & 3.72 -- 4.24 &  &     & Za19 &     &       &      &      &     &    &     \\
$-$225025.2 & & 360 -- 390 & 4.25 -- 4.75 & $>0$  & 6.0 &  Le17 &  &   &  &  &  &   & \\
              &                  & $\approx$400  & 4.00 -- 4.50 &    &   & Sc15 &     &       &      &      &     &    &     \\ 
                   &                  & 335 -- 367  & 4.03 -- 4.54 &    &   & Du13\tablenotemark{a} &     &       &      &      &     &    &     \\          
\tableline
{\it WISEPA} & 48.6 &   310 -- 340 & 3.75 -- 4.25 & $\ll 0$  & 6.0 &  Le17 
&  375   &  4.0     &  -0.5    & 7.0 & 1.20     &  (7,640)    &  5  &   0.5  \\
J182831.08 & $\pm$ 1.1 & 421 -- 470 &   4.24 -- 4.78 &  &    &  Du13\tablenotemark{a}   &       &      &      &     &    &     & \\ 
$+$265037.8AB\tablenotemark{b}\tablenotemark{c} & &  &    &  &    &      &      &      &     &    &     & \\ 
\tableline
{\it WISEPC} & 33.6
&  471 - 522 & 4.40 -- 5.00 &   & 5.3 & Mi20 & 475 & 4.25  & 0.0  & 7.0 & 1.20 &(7.5,820)  & 8 & 0.5  \\
    J205628.90       &      $\pm$0.5                   & 447 -- 523 & 4.64 -- 5.18 &    &   & Za19 &     &       &      &      &     &    &      \\
$+$145953.3  && 410 -- 440 & 4.25 -- 4.75 &  $>0$ & 6.0 &  Le17 &  &   &  &  &  &  &   \\
                        &           & 400 -- 450 & 4.00 -- 4.50 &   &    & Sc15 &     &       &      &      &     &    &      \\ 
                         &           & 414 -- 460 & 4.23 -- 4.76 &   &    & Du13\tablenotemark{a} &     &       &      &      &     &    &      \\            
\tableline
{\it WISEA} &  53.2  & 310 -- 340 & 4.27 -- 4.75 &  $>0$ & 6.0 & Le17 & 350 & 4.00  & 0.0 & 7.0 & 1.25 & (10,740) & 5  & 0.5 \\
J220905.75 &$\pm$0.8 & 500 -- 550 & 4.00 -- 4.50 &   &   & Sc15\tablenotemark{d} &     &       &      &      &     &    &      \\
$+$271143.6 & &  &  &   &   &  &     &       &      &      &     &    &      \\
\enddata
\smallskip
\tablecomments{
Excluding any systematic errors, we estimate the uncertainties in our derived parameters to be $\pm$20~K in $T_{\rm eff}$, $\pm$0.25~dex in $\log g$, $\pm$0.3 dex in [m/H], $\pm$1 dex in $\log K_{zz}$, $\pm$0.1 in $\gamma$, and $\pm$10~bar in $P_{\gamma -max}$ (Figure 6).  These uncertainties lead to an uncertainty  in mass and age of a factor of $\sim$2 and  $\sim$3, respectively (Section 5.5).}
\vskip -0.05in
\tablenotetext{a}{The  \citet{Dupuy_2013} $T_{\rm eff}$ and $\log g$ values quoted in the Table use  the bolometric luminosities given in that paper combined with the more recent measurements of parallaxes used here.}
\vskip -0.1in
\tablenotetext{b}{J1828 could not be fit by us as a single star. The parameters given here and the fits shown in Figure 8 assume it is an equal-mass binary system.}
\vskip -0.1in
\tablenotetext{c}{The  \citet{Dupuy_2013} higher temperature for J1828 is based on the assumption that it is a single object.}
\vskip -0.1in
\tablenotetext{d}{A value as high as 500~K for $T_{\rm eff}$ is not plausible for J2209, as also pointed out by \cite{Martin_2018}. We suspect the noisy near-infrared spectrum skewed the model fit by \citet{Schneider_2015}.}
\vskip -0.05in
\tablerefs{Du13 - \citet{Dupuy_2013}, 
%Cu20 - Cushing et al. in prep., 
Fi15 - \citet{Filippazzo_2015}, Le12 - \citet{Leggett_2012}, Le17 - \citet{Leggett_2017}, Lu16 - \citet{Luhman_2016}, Mi20 - \citet{Miles_2020},
Sc15 - \citet{Schneider_2015},
Za19 - \citet[][constrained]{Zalesky_2019}. Tangential velocities are from \citet{Kirkpatrick_2019}.
}
%\vskip -0.5in
\end{deluxetable*}
\setlength\tabcolsep{6pt}

Figure 6 illustrates the sensitivity of the synthetic 0.9 -- 20~$\mu$m spectrum to the parameters (for these models, longer wavelengths of 20 -- 30~$\mu$m do not show significant sensitivity).  The shape of the SED is very sensitive to temperature, and also to metallicity. $\gamma$ impacts the slope from the near- to the mid-infrared, as well as the depth of the strong absorption bands. Gravity signatures are more subtle, and somewhat degenerate with metallicity. However gravity is also constrained by the flux scaling to Earth, via the mass-radius relationship used by the evolutionary models. We discuss this further in Section 5.5.

 The SED generated by the tuned model provides a significantly improved fit to observations of J0722. The agreement with the near-infrared spectrum and the $4 \leq \lambda~\mu$m $\leq 5$ spectrum is now excellent, instead of being a factor of $\sim 3$ discrepant. Also, the discrepancy at the bottom of the strong  3.3~$\mu$m CH$_4$ band is reduced to a factor of $\sim$ 2 from a factor of $\sim 10$. Apart from the reduced adiabat, the other atmospheric parameters --- $T_{\rm eff}$, $g$, [m/H] and $K_{zz}$ --- are consistent with previous determinations  (Table 3). Panel (d) of Figure 5 compares the spectra generated for J0722 by the standard and tuned non-equilibrium chemistry ATMO 2020 models. The difference in the near-infrared region is clear, as is that in the 2 -- 4~$\mu$m region. {\it JWST} spectra at 5 -- 9~$\mu$m, impossible to obtain from the ground, will provide an additional check on this approach.

Figure 7, top right panel, shows the standard and tuned $P-T$ diagram for J0722, as well as the contribution function -- the pressure or atmospheric layer from which flux at a certain wavelength arises. Standard curves for a $T_{\rm eff}$ value equal to that determined from the fit, and a temperature 100~K cooler, are shown; these demonstrate that the tuned model has an interior (where the $\lambda \sim 1~\mu$m flux originates) similar to the cooler standard model, and an upper atmosphere similar to the warmer standard model.  The fact that the 3.3~$\mu$m feature is still somewhat deeper than observed suggests that the revised $P-T$ profile  may not be warm enough where this flux originates, in the upper atmosphere at pressures $\sim$ 0.1~bar.  
Interestingly, the need for upper atmosphere heating has also been identified in retrieval analyses of L dwarf atmospheres \citep{Burningham_2017}.
We discuss this further in Section 6.

\medskip
\subsection{Tuned-Model Fits to 250 -- 500~K Brown Dwarfs}

\begin{figure}[b]
%\plotone{400spec.pdf}
\plotone{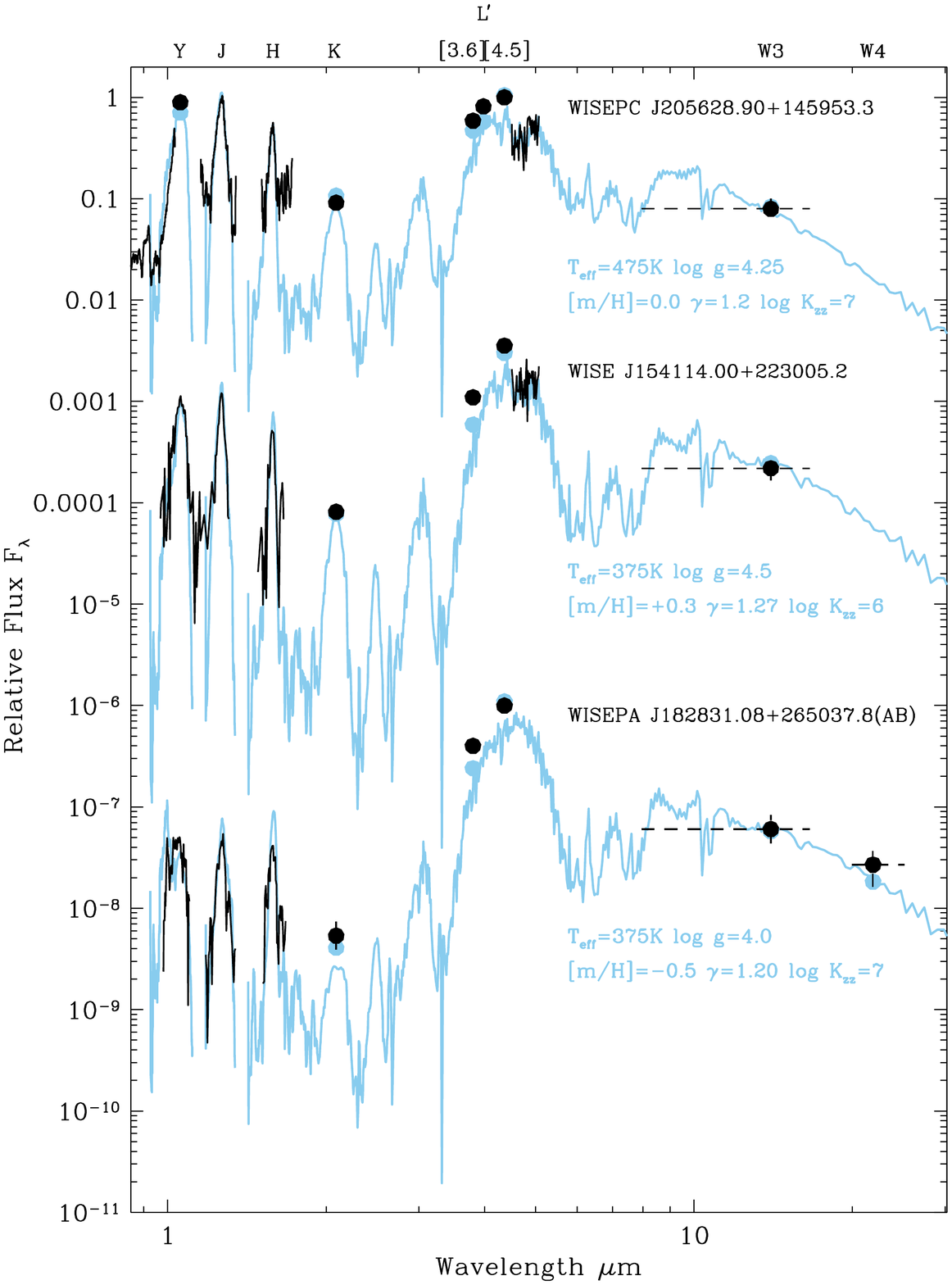}
\vskip -0.35in
\caption{Adiabat-tuned fits for three $T_{\rm eff}\sim 400$~K brown dwarfs, identified in the legends. Solid black lines are observed spectra \citep[][and Cushing et al. 2021 submitted]{Cushing_2011, Leggett_2013, Schneider_2015, Miles_2020}, and the black points are observed photometric data, with vertical error bars where these are larger than the symbol. Uncertainties in the observed J2056 spectra are negligible, in the J1541 and J1828(AB) spectra they are 10 -- 20\% in regions where there is significant flux.
Dashed black lines indicate the passbands of the broad W3 and W4 filters. Blue lines are synthetic spectra generated by the tuned models with parameters given in the legends, and blue points are the synthetic photometry.  
}
\end{figure}

\begin{figure}
%\plotone{300spec.pdf}
\plotone{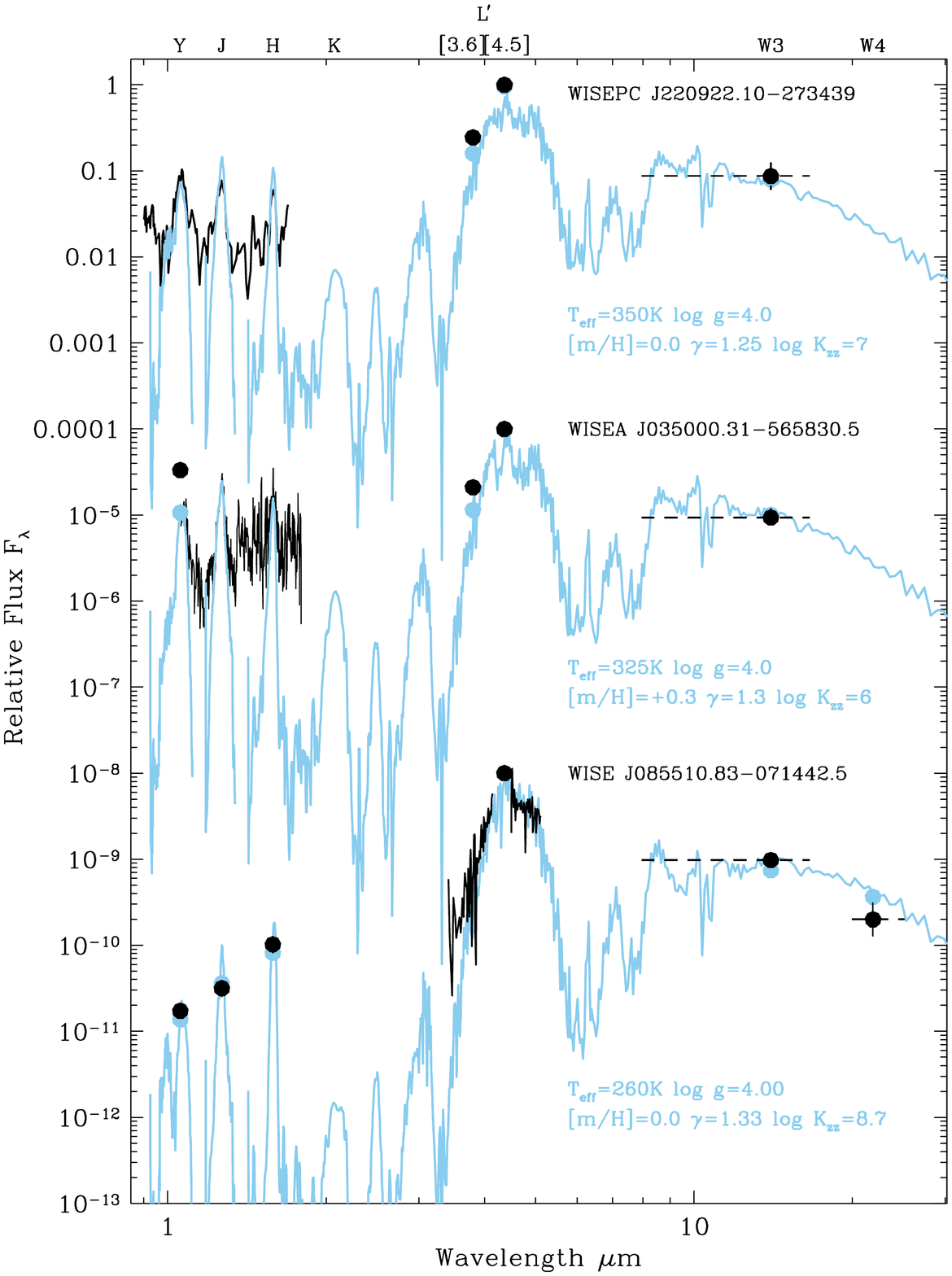}
\vskip -0.35in
\caption{Adiabat-tuned fits for three $T_{\rm eff}\sim 300$~K brown dwarfs, identified in the legends. Solid black lines are observed spectra
 \citep{Schneider_2015, Leggett_2016a, Morley_2018, Miles_2020}, and the black points are observed photometric data, with vertical error bars where these are larger than the symbol.
Uncertainties in the observed near-infrared spectra, in regions where there is significant flux,
are 20 -- 50\% for J2209 and J0350.
Uncertainties in the observed  spectra for J0855 
are 10 -- 30\% for the $L$-band and 5 -- 20\% for the $M$-band.}
\end{figure}

We extended the approach described above to colder brown dwarfs. The sample consists of 
three cold brown dwarfs for which 
\citet{Miles_2020} provide $\lambda \approx 4.8~\mu$m spectra, because this region is  sensitive to mixing of CH$_4$ and CO (Figures 5 and 6): J0855, {\it WISEPA} J154151.66$-$225025.2 (hereafter J1541), and {\it WISEPC} J205628.90$+$145953.3 (hereafter J2056). We added three other brown dwarfs with $J -$[4.5] colors between those of J2056 and J1541, and the extreme dwarf J0855, all  of which have W3 photometry available ---  {\it WISE}
J035000.31$-$565830.5
(hereafter J0350), {\it WISEPA} J182831.08+265037.8 (hereafter J1828), and {\it WISE} J220905.73+271143.9 (hereafter J2209).

Figures 2 and 3 identify the target objects in the color-color diagrams, and  Table 3 lists the six objects, with atmospheric parameters determined here and previously. We could not fit the absolute flux level of J1828 as a single object, but we did find a satisfactory fit assuming it is an equal-mass binary. We refer to J1828 from here on as J1828(AB) to clarify that the estimated properties assume binarity.

Figures 8 and 9 show the SEDs of the six Y dwarfs in our sample --- observational data as well as the best by-eye tuned model spectrum --- in order of decreasing $T_{\rm eff}$. The W4 photometric point is included for J1828(AB) and J0855 in the Figures; it was not used when judging fit quality as the uncertainty is large (Table 2), but the observed and modelled photometry agree within the uncertainties.

Note the increasing dominance of the mid-infrared region and the steady reddening of the [3.6] $-$ [4.5] color with decreasing temperature in Figures 8 and 9. Note also the pronounced difference between J1541 and J1828(AB) in Figure 8 although they have the same $T_{\rm eff}$ --- the lower metallicity and $\gamma$ of J1828(AB) suppress the $YJHK$ flux and broaden the $Y$ band peak. These changes in the SEDs are also demonstrated in Figure 6.
As $T_{\rm eff}$ drops to 260~K there is a loss of flux at $\lambda \sim 1~\mu$m.

The fits shown in Figures 8 and 9 are generally very good across the entire SED. The height and width of the near-infrared flux peaks are well reproduced, with the exception of J1828(AB) where the $H$-band peak is a factor of $\sim$2 too bright, and the $Y$-band peak for J0350, where the model is a factor of three too faint. The $Y$-band discrepancy suggests that a large amount  of flux at $\lambda \lesssim 1.0~\mu$m is missing from the models, as the red wing of the flux peak is well matched.
Both these systems are challenging --- J1828(AB) is a very metal-poor likely-multiple system, and J0350 is  a cold metal-rich brown dwarf.

The model flux at $\lambda \approx 3.3~\mu$m is low, as also seen for J0722 in Figure 5 (although the agreement is improved by a factor of $\sim 5$ compared to standard-adiabat models). This leads to [3.6] magnitudes that are a few-tenths to a magnitude fainter than observed. The spectrum of J0855 in Figure 9 suggests that the loss occurs only at the blue end of the $3.13 \lesssim \lambda ~\mu$m $\lesssim 3.92$ [3.6] filter bandpass.  
The mid-infrared fluxes are otherwise well matched. The coldest object, J0855, 
is very well matched 
--- the observed and synthetic photometry generated by the tuned model agree within the measurement uncertainties at all passbands apart from [3.6], and the 3.5 -- 4.1~$\mu$m and 4.5 -- 5.1~$\mu$m spectra are well reproduced (see also Section 5.4).

The agreement between these tuned non-equilibrium chemistry models and observations is better than has been possible in the past.    Previous efforts to fit the mid-infrared spectroscopy and photometry of WISE 0855 by \citet{Morley_2018}
found that models with lower CH$_4$ abundances could adequately fit the data, including models with sub-solar metallicity and C/O ratios
(see the low-metallicity sequence in the [3.6] $-$ [4.5] panel in Figure 2).
Low-metallicity models that adequately match the mid-infrared photometry are too bright at near-infrared wavelengths, but a deep continuum opacity source (e.g. clouds) could readily decrease the near-infrared flux to match the observed photometry. Those authors found that upper-atmosphere heating could not be invoked to fit the observed properties, but did not explore changes to the deep adiabatic structure. In other recent work,  the model comparisons to J0722, J2056, J1541 and J0855 by \citet[their Figure 3]{Miles_2020} show large discrepancies (factors of 2 -- 3) at most wavelengths.

 Table 3 gives our derived model parameters and compares these to previously determined values. Excluding any systematic errors, we estimate the uncertainties in our derived atmospheric parameters, based on the full fit to the SED, to be $\pm$20~K in $T_{\rm eff}$, $\pm$0.25~dex in $\log g$, $\pm$0.3 dex in [m/H], $\pm$1 dex in $\log K_{zz}$, $\pm$0.1 in $\gamma$, and $\pm$10~bar in $P_{\gamma max}$.  This is based on the sensitivity of the SED to the parameters (Figure 6); gravity is constrained by  both the SED and the mass-radius relationship of the ATMO 2020 evolutionary models \citep{Phillips_2020}. The absolute uncertainty in the parameter estimates is difficult to assess but is unlikely to be more than twice these values, given the agreement between the estimates for individual objects in Table 3, which were arrived at using different models and different methods. Furthermore, the  ATMO 2020 evolutionary models have been tested against a small sample of brown dwarfs with dynamically determined masses,  and the ages derived are appropriate for the solar neighborhood  \citep{Dupuy_2017, Buder_2019}. The evolutionary models also produce cooling curves very similar to earlier models, while using  a more recent equation of state for H–He
mixtures \citep{Chabrier_2019}.

The atmospheric parameters determined for J1541 and J2056 by \citet{Zalesky_2019} in Table 3 are of particular relevance, as those authors use a retrieval method to adjust the atmosphere properties in order to reproduce observations, somewhat similar to (but more complex than) our $P-T$ tuning technique \citep[see][]{Line_2015}. \citet{Zalesky_2019} constrain their fits using {\it HST} near-infrared spectra while we use longer-baseline observations, which allows us to probe the higher and cooler regions of the atmosphere (compare our Figure 7 to Zalesky et al. Figure 2). The shape of the profile we determine for J2056 is similar to that found by \citet{Zalesky_2019}, with the atmosphere cooler at deeper layers and warmer in the upper layers compared to the grid models. However the difference between the tuned and standard temperatures are larger at deeper layers in our models, for example at 100~bar we find $\delta T \approx 500$~K compared to 100~K for \citet{Zalesky_2019}.  For J1541, the deviation of the shape of the profile from the standard model is larger in the Zalesky et al. analysis than in our analysis. \citet{Zalesky_2019} find that both the upper and lower regions of the atmosphere are warmer by $\sim$500~K, while we find that the deeper layers are cooler with only small differences from standard in the upper regions. Nevertheless both analyses indicate that the
$P-T$ profile deviates from the standard form, typically with cooler regions in the deeper layers of late-T and Y dwarf atmospheres, from which the near-infrared radiation emerges. 

The parameters determined by \citet{Miles_2020} are also of interest, as the 4.8~$\mu$m spectra presented by those authors provides a constraint on $K_{zz}$. Both this work and Miles et al. find a very high $K_{zz}$ for the extremely cold J0855. We are in agreement for J1541, however Miles et al. find a lower value than ours for J0722 and J2056 (Table 3). 
We suggest that our estimates are more robust as they are based on broader wavelength coverage.

Our tuning sample of six Y dwarfs has a relatively small range in the photospheric adiabatic parameter $\gamma$ (typically 1.2 -- 1.3), and in the diffusion coefficient $\log K_{zz}$ (typically 6 -- 7), but some variation in these values for a larger sample would not be surprising. 
The global properties of a brown dwarf atmosphere are likely to vary with inclination to the line of sight. For example, models of turbulent convection in rapidly rotating atmospheres, including the solar system gas giants, calculate that $K_{zz}$ is latitude-dependent, decreasing from the equator to the poles \citep[e.g.][]{Flasar_1978, Visscher_2010, Wang_2016}.  Measurements of variability are also likely to be inclination-dependent
\citep{Vos_2017}.

\medskip
\subsection{Clouds, Chemical Changes and the Disruption of Convection in Y Dwarfs}

Figure 7 shows the standard and modified $P-T$ profile for the six Y dwarfs in our sample. Also shown is the contribution function, which indicates the pressure layer from which the near- to mid-infrared flux emerges in the tuned model. From the coldest to the warmest object, the $1~\mu$m light emerges from regions where $P \sim$ 10 -- 100~bar and temperatures are 900 -- 1500~K, while the $10~\mu$m light emerges from regions where $P \sim 1$~bar and temperatures are 250 -- 500~K. Where the atmosphere is more opaque, such as at $\lambda \sim$~3, 6 or 8~$\mu$m, the light emerges from high and cold regions where $P \sim 0.1$~bar and $T \sim$ 150 -- 350~K.

\begin{figure*}[t]
\gridline{\fig{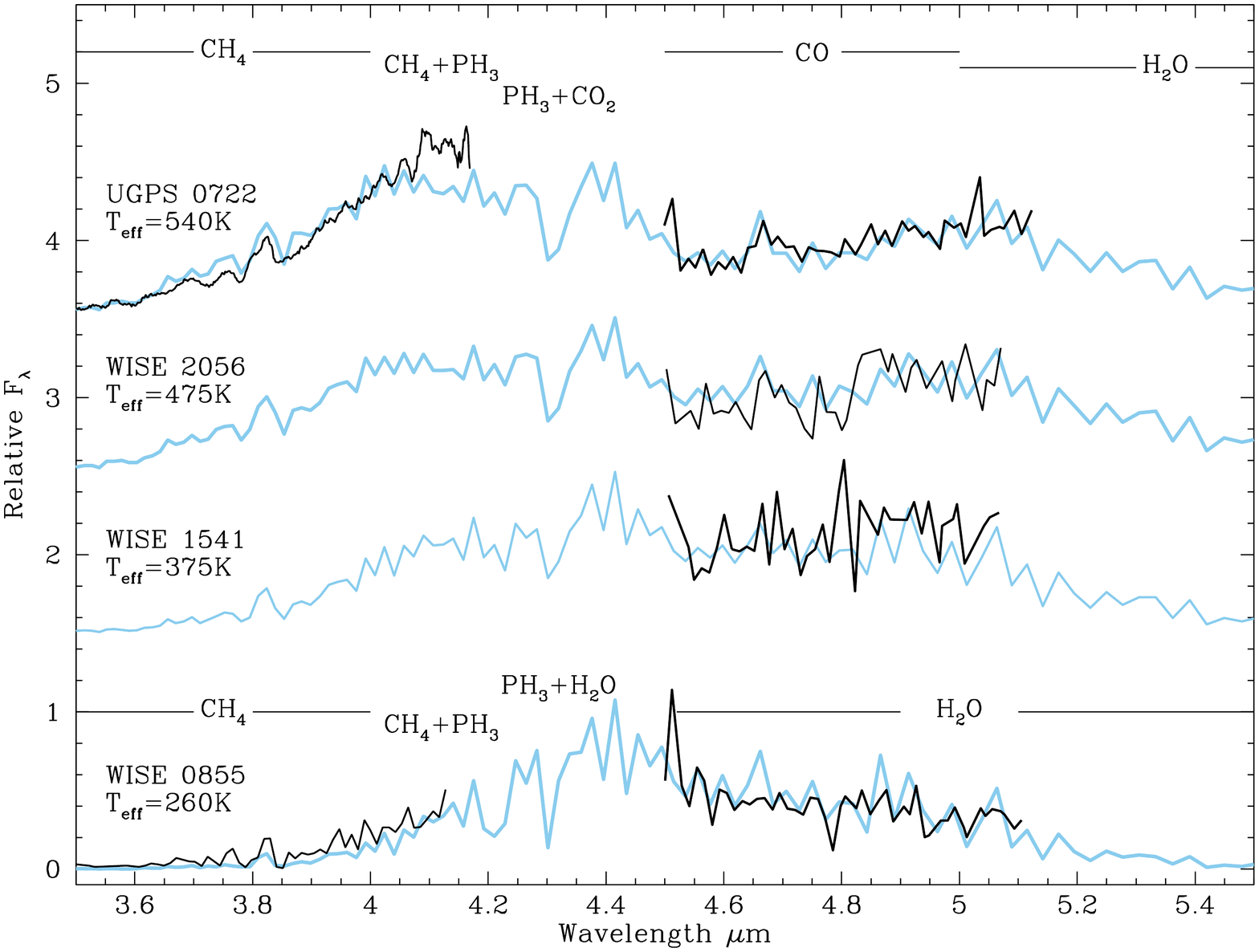}{1.0\textwidth}{\vskip -0.4in (a)}}
\vskip -0.4in
\gridline{\fig{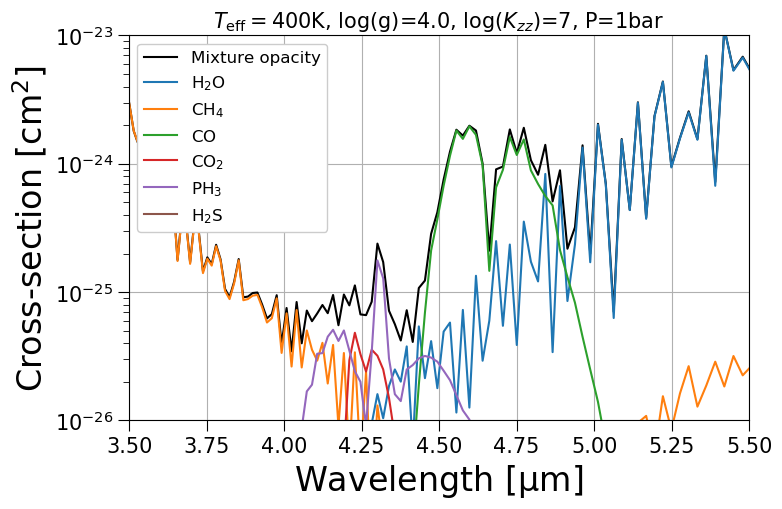}{0.5\textwidth}{\vskip -0.1in (b)}
        \fig{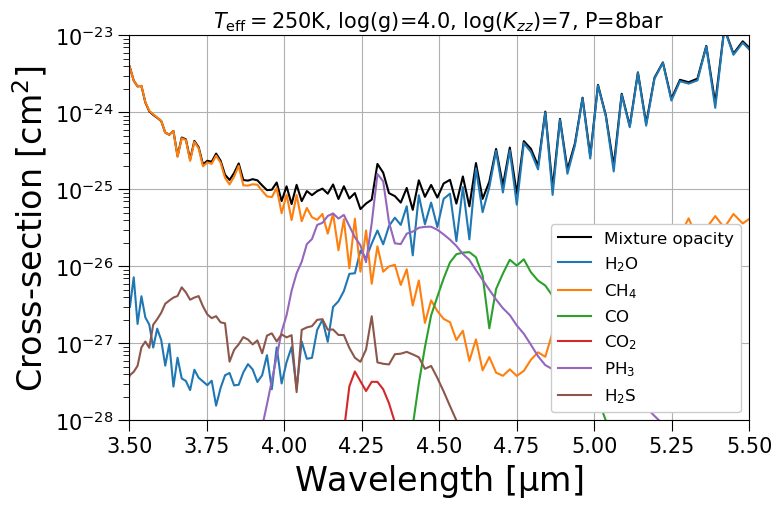}{0.5\textwidth}{\vskip -0.1in (c)}       
          }
\caption{The upper panel (a) shows observations (black line) and the  tuned-adiabat model spectra (blue line) from Figures 8 and 9, for 
$3.5 \leq \lambda~\mu$m $\leq 5.5$. The lower panels (b)(c) show 
ATMO 2020 opacity calculations for these wavelengths, for two representative values of $T_{\rm eff}$.}
\end{figure*}

The condensation curves in the top left panel of Figure 7 suggest that water clouds would be expected in the upper layers of the atmosphere of J0855, and possibly in the very upper atmosphere of J0350 and J2209 
\citep[see also][their Figure 6]{Morley_2014}. These could produce the heating in the upper atmosphere needed to increase the model flux at $\lambda \approx 3.3~\mu$m, although this  could also be accomplished by breaking gravity waves as is likely in the solar system giant planet atmospheres above the 1-bar pressure surface \citep[e.g.][]{Schubert_2003, O'Donoghue_2016}.

The condensation curves also indicate that KCl and Na$_2$S clouds would be important in the regions where  the near-infrared flux originates, for our sample, i.e.  $P \sim$ 10~bar and $T \sim$ 1000~K 
\citep[see also][their Figure 4] {Morley_2012}. The 10~bar/1000~K level also corresponds to where nitrogen moves  into the NH$_3$ form from N$_2$, in equilibrium conditions (Figure 7). It is interesting to note that our fits indicate that the $P-T$ curve reverts to the standard adiabat at pressures around 10~bar  for the 325 -- 475~K Y dwarfs in our sample, and 50~bar 
for the 260~K J0855; all at temperatures of 750 -- 870~K (the very metal-poor J1828(AB) system appears to transition at a slightly cooler 640~K).  This may indicate that  convection in Y dwarf atmospheres is disrupted once the atmosphere cools to $\sim$800~K, by  the change in nitrogen chemistry and/or the condensation of chlorides and sulfides. We find that for the warmer T9 dwarf J0722 any increase in $\gamma$ occurs at higher pressures which are not sampled by the emergent SED, suggesting different physics is at play for T dwarfs.

\medskip
\subsection{$5~\mu$m Spectra of Brown Dwarfs and the Detection of Phosphine}

Phosphine is a non-equilibrium species that is seen in the $5~\mu$m spectra of Saturn and Jupiter; it is a useful species which  can be used to study both atmospheric dynamics and the effect of photochemistry on planetary atmospheres \citep[e.g.][]{Fletcher_2009}.  PH$_3$  is not detected in ground-based spectra of J0855 and other cold brown dwarfs, although it is expected to be abundant \citep{Skemer_2016, Morley_2018, Miles_2020}. Because of the potential diagnostic value of the species, we explore what the new tuned models indicate for its detectability.

Figure 10(a) shows  3.5 -- 5.5~$\mu$m spectra of the four brown dwarfs in our tuning sample with such data. We also show the derived adiabat-tuned fit for each object, which reproduces the observations well. Figure 10(b)(c) show the opacity contributions from various species at these wavelengths, for representative temperatures. These opacity contributions are taken from the ATMO 2020 models with vertical mixing, which only consider the non-equilibrium abundances of the major carbon- and nitrogen- bearing species, and thus do not take into account the mixing of $\rm{PH_3}$ \citep{Phillips_2020}.

The spectral regions that can be observed from the ground, the $L$ and $M$ bands, are dominated by CH$_4$ and CO absorption bands, respectively, for the 400~K and warmer brown dwarfs. For the 260~K J0855,  H$_2$O becomes the dominant opacity source in the $M$-band. Although \citet[][their Figure 19]{Morley_2018} find that the red edge of the $L$-band and the blue edge of the $M$-band in J0855 should show  PH$_3$ absorption,  at the enhanced  abundance brought about by mixing, these are difficult wavelengths to work at from ground-based observatories. 
We calculate that there is a strong feature due to PH$_3$ at 4.30~$\mu$m in the spectra of cold brown dwarfs, even when assuming $\rm{PH_3}$ is in chemical equilibrium. 
Hence {\it JWST} observations should finally confirm the presence of PH$_3$ in brown dwarf atmospheres.

\medskip
\subsection{Estimating Masses and Ages for the Six Y Dwarfs}

\begin{figure}
\vskip -1.5in
\includegraphics[width = 7.5 in]{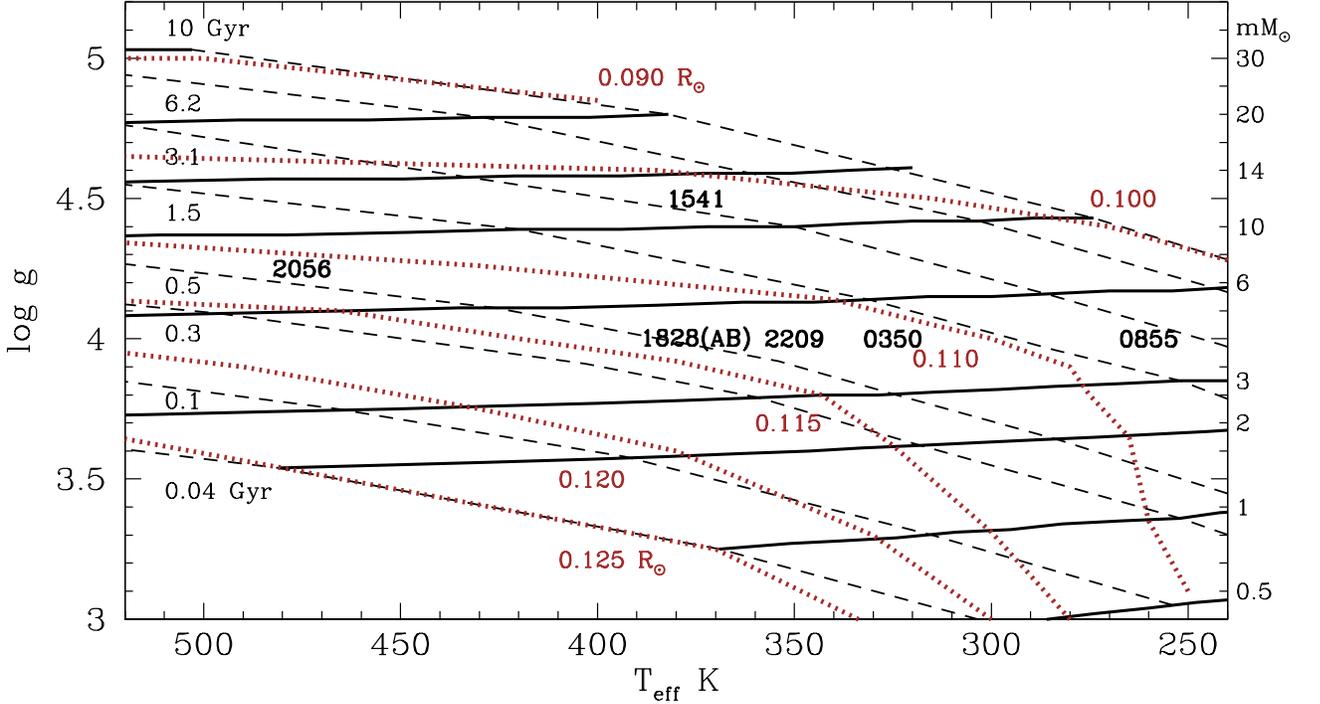}
\vskip -0.3in
\caption{Evolutionary curves from ATMO 2020 models (\url{http://opendata.erc-atmo.eu}). Solid black lines are iso-mass sequences  for objects with mass shown along the right axis. Evolution proceeds from left to right. Dashed lines are isochrones for the ages indicated, and dotted brown  lines are lines of constant radii, for the values indicated. The location of the six Y dwarfs in our tuning sample are shown by short name.
}
\end{figure}

Figure 11 shows the evolution of cold brown dwarfs in a $T_{\rm eff}$:gravity diagram.   The luminosity, or absolute brightness, of a brown dwarf, as measured at the Earth,  is determined by $T_{\rm eff}$, radius and distance. The uncertainty in distance is very small for these nearby objects, and the SED is very sensitive to temperature (Figure 6), with the net result that the absolute flux level constrains  radius to  $\sim 10\%$ (Figures 8 and 9).  Figure 11 shows that $\log g$ can then be constrained to $\pm 0.3$~dex, mass can be constrained to a factor of $\sim$2, and age to a factor of $\sim$3, for a notional 400~K brown dwarf.

Table 3 gives the atmospheric and evolutionary parameters we derived here from the  $T_{\rm eff}$ and gravity of each tuned-adiabat model fit. 
For our tuning sample of six Y dwarfs, the evolutionary models give ages of between approximately 0.5 and 3~Gyr (Table 3, see also Figure 11). These values agree, within the uncertainties, with what would be expected for a local sample ---  1 -- 3~Gyr  \citep{Dupuy_2017, Buder_2019}. Weak support for relative youth is provided by the tangential velocities  which suggest  thin disk membership \citep{Dupuy_2012} and so an age younger than 8~Gyr \citep{Kilic_2017}. The estimated masses for the six Y dwarfs are very low for this cold sample --- between 5 and 12 Jupiters.

\bigskip
\section{Application to the Larger Y Dwarf Sample}

\subsection{Color Trends}

\begin{figure}
\centering
\includegraphics[width=6in]{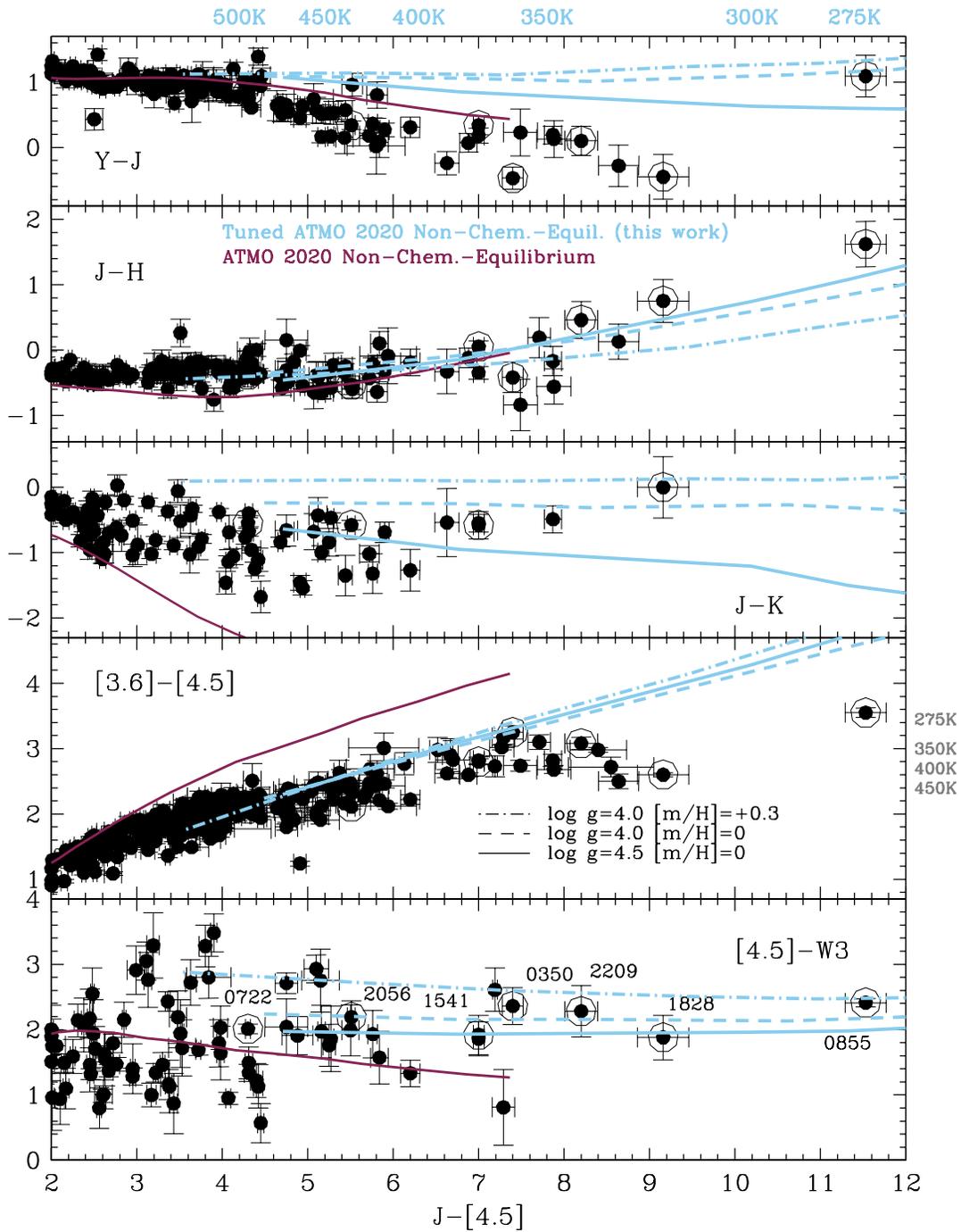}
\vskip -0.2in
\caption{Color-color diagrams for late T and Y dwarfs. Symbols and lines are as in Figure 2, with the addition of modified-adiabat model sequences shown in blue. The 
model has $K_{zz} = 10^7$ and $\gamma = 1.25$ at pressures of 15 bar and lower. Values of $T_{\rm eff}$ from this model are shown along the top axis. For the  frequently used [3.6] $-$ [4.5] color diagnostic, the  model deviates from observations for the coldest objects, and semi-empirical values of  $T_{\rm eff}$ are shown in grey  along the right axis (see Section 6.2).
}
\end{figure}

\begin{figure}
%\plotone{mc2_Tune.pdf}
\plotone{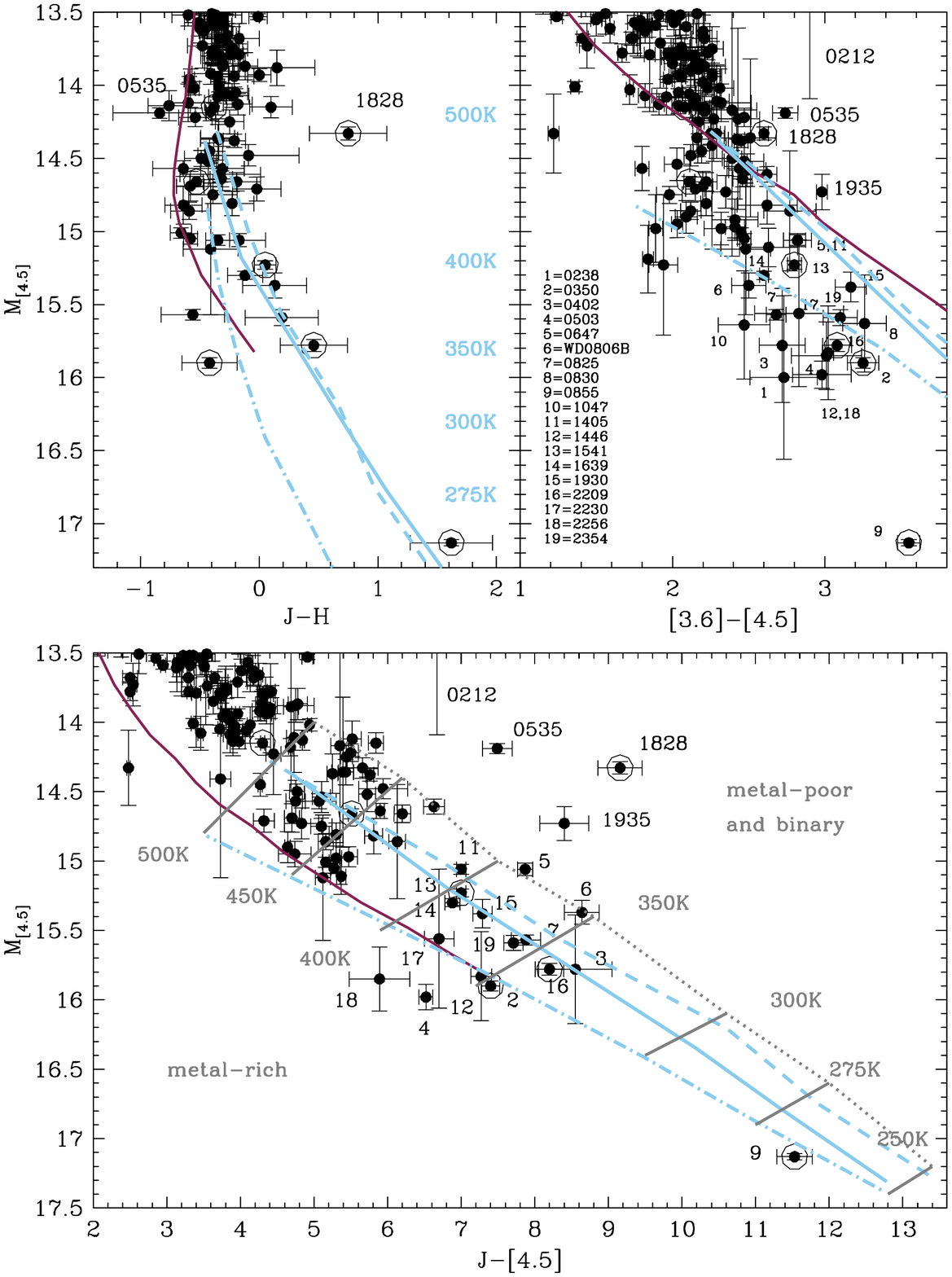}
\vskip -0.3in
\caption{Color-magnitude diagrams for late T and Y dwarfs. Sequences are as in Figure 12. Grey $\sim$diagonal lines in the bottom panel indicate constant $T_{\rm eff}$, as labelled, for metallicities ranging from approximately $+0.3$ on the left, to $-0.5$ on the right. 
The location of the metal-poor envelope edge in the bottom panel is consistent with the 
low-metallicity Sonora-Bobcat models (Figure 3), and with the 
observed population.
%Wolf 1130C has [Fe/H] $\approx -0.75$ \citep{Kessel_2019} and 
Our SED analysis of J1828 indicates that it is an equal-mass binary with [m/H] $\sim -0.5$ (Figure 8).
J0212, J0535, and J1935 are also likely to be similar-mass binary systems. Notionally single Y dwarfs which are estimated to have $T_{\rm eff} \lesssim$ 400~K are identified in the legend by  the first four digits of the {\it WISE} catalog Right Ascension, or their binary name in the case of the white dwarf companion. 
 }
\end{figure}

To check how the modified-adiabat non-equilibrium chemistry models perform for a larger sample, Figures 12 and 13 repeat the color-color and color-magnitude diagrams of Figures 2 and 3, but this time they include a model sequence  generated by a small grid of the $P-T$-modified models. For this grid we adopt $K_{zz} = 10^7$~cm$^2$s$^{-1}$, $\gamma = 1.25$ and $P_{\gamma -max} = 15$~bar. Colors are calculated for two gravities, $\log g = 4.0$ and 4.5, and two metallicities, [m/H] $=$ 0.0 and $+0.3$. A sequence generated by the standard non-equilibrium chemistry model is also shown  for comparison.

The top panel of Figure 12, $J -$ [4.5]:$Y - J$, shows that there is a systematic issue in the $Y$-band for the 325 -- 450~K brown dwarfs, as  the  models are fainter at $Y$ than observed, by a few-tenths to one magnitude. The spectral fits in Figures 8 and 9 suggest that the problem is too little flux in the models at the blue wing of the $Y$-band, suggesting in turn that
a more rigorous approach to the treatment of the strong $0.8~\mu$m K~I line is called for (see 
Section 4). The models are likely to have issues with two important chemical changes at 325 -- 450~K, exploration  of which are beyond the scope of this paper:  collisions with H$_2$ affect the shape of the wings of the alkali resonance lines \citep{Allard_2016}, and neutral K gas transitions to  KCl gas and then to  KCl solid \citep[e.g.][]{Lodders_1999}.

The fits to the other colors and magnitudes in Figures 12 and 13 are good to excellent. The agreement between the models and observations at $J - K$ and [3.6] $-$ [4.5] is greatly improved. The previous $\gtrsim 1$~magnitude discrepancy for these colors is now $\approx 0$ for $J - K$ and reduced to 
a few tenths of a magnitude for [3.6] $-$ [4.5]. In the color-magnitude diagram, Figure 13, the previous $\approx$ 0.4 mag discrepancy in $J - H$ is resolved, as is the $\approx$ 1.0 mag discrepancy in $J -$ [4.5].

\medskip
\subsection{$T_{\rm eff}$ and Metallicity  Estimates for Y Dwarfs}

Figures 12 and 13 show that, as well as temperature, the metallicity and gravity of the atmosphere can impact the colors of a brown dwarf. We find that temperature and metallicity have the largest impact, as also indicated by the synthetic spectra shown in Figure 6. 

Figure 14 is a plot of $T_{\rm eff}$ against [3.6] $-$ [4.5], $J -$ [4.5], and $M_{[4.5]}$, which are the most commonly available photometric measurements for Y dwarfs, currently. The relationships in Figure 14 are determined from the modified-adiabat  model grid, which spans $250 \leq T_{\rm eff}$~K $\leq 500$. These models have $K_{zz} = 10^7$~cm$^2$s$^{-1}$, $\gamma = 1.25$ and $P_{\gamma -max} = 15$~bar. 
The relationship for the [3.6] $-$ [4.5] color includes an empirical correction based on observations of the Y dwarfs for which we do a full SED fit in Section 5.2.
%The colors generated by the grid are given in the Appendix (Table 8).
The Figure suggests that the $J -$ [4.5] color is particularly sensitive to metallicity.
%If an object 
%has all three measurements available, and the  $T_{\rm eff}$ implied by  $J -$ [4.5] is discrepant, then this may be an indicator of non-solar metallicity. 
Table 4 gives polynomial fits to the solar metallicity relationships shown in Figure 14. We estimate the uncertainty in a color-derived $T_{\rm eff}$ to be $\pm 25$~K, based on the scatter seen when determining  $T_{\rm eff}$ from different colors, and comparing the SED-determined  $T_{\rm eff}$ to the color value.

\begin{figure}[b]
\vskip -1.0in
%\plotone{Teff_color.pdf}
\plotone{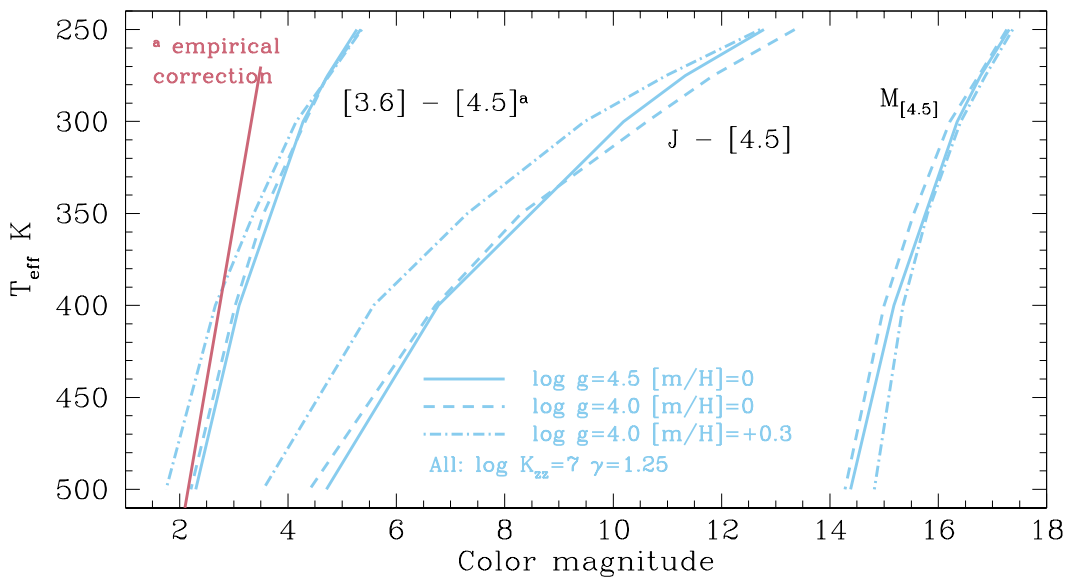}
\vskip -0.3in
\caption{Synthetic colors from the modified-adiabat model grid, see text. 
%The models underestimate [3.6] $-$ [4.5] for   [3.6] $-$ [4.5] $\gtrsim 2.6$, or $T_{\rm eff} \lesssim 400$~K, by 0.2 --  1.0  magnitude (Figure 10). An empirical correction is shown based  on estimates of $T_{\rm  eff}$ for  cold Y dwarfs  with additional data. 
Table 4 gives polynomial fits to the solar metallicity relationships shown in the Figure.  
}
\end{figure}

\begin{deluxetable*}{lccc}[t]
\tabletypesize{\normalsize}
\tablecaption{Polynomial Relationships\\for Estimating $T_{\rm eff}$ from Color}
\tablehead{
\colhead{Color} & \colhead{$a_0$} & \colhead{$a_1$} & \colhead{$a_2$} 
}
\startdata
[3.6] $-$ [4.5]$^a$ & 850 & $-$166.7 &  \\
$J -$ [4.5] & 816 & $-$81.64 & 2.9572 \\
$M_{[4.5]}$ & 5331 & $-$544.5 & 14.4990 \\
\enddata
\tablecomments{$T_{\rm eff}$ is estimated using: ~
$T_{\rm eff} = a_0 + a_1\times{\rm Color} + a_2\times{\rm Color}^2$\\
Relationships are valid for $250 \leq T_{\rm eff}$~K $\leq 500$. Excluding any systematic errors, the uncertainty in   $T_{\rm eff}$ is $\pm 25$~K.
Solar metallicity is assumed; metal-rich objects will be cooler, and metal-poor object warmer, for a given $J -$ [4.5] (see Table 5 and Figure 14).\\
$^a$ Semi-empirical.
}
\end{deluxetable*}

\begin{deluxetable*}{cccc}[b]
\tabletypesize{\normalsize}
%\tablewidth{0pt}
\tablecaption{Estimate of Color Sensitivity to Metallicity and Gravity for $T_{\rm eff} = 400$~K}
\tablehead{
\colhead{Color} & \multicolumn{2}{c}{
$\delta$ mag}  & \colhead{Important} \\
\colhead{} & \colhead{$\delta\log g = +0.5$} & \colhead{$\delta$[m/H] $= +0.3$} & \colhead{Chemistry} 
}
\startdata
$\delta(J - H)$ & $-0.1$ & $-0.3$ & H$_2$  at $J$ \\
$\delta(J - K)$ & $-0.7$ & $+0.4$ & (stronger) H$_2$ at $K$ \\
$\delta(J -$ [4.5]) &  $+0.1$ & $-1.1$ & H$_2$ at $J$,  CO at [4.5]\\
$\delta$([3.6] $-$ [4.5]) & $-0.2$ & $+0.4$ & CH$_4$ at [3.6], CO at [4.5] \\
$\delta$([4.5] $-$ W3) & $-0.2$ & $+0.6$ & CO at [4.5], H$_2$ at W3 \\
\enddata
\tablecomments{Generated by a $P - T$ modified-adiabat model with $K_{zz} = 10^7$~cm$^2$s$^{-1}$, $\gamma = 1.25$ and $P_{\gamma -max} = 15$~bar, except for the [3.6] $-$ [4.5] color which is empirical. See also Figure 5 for opacity identification and Figure 6 for SED sensitivity to gravity and metallicity.} 
%\vskip -0.1in
%\tablenotetext{a}{Empirical.}
\end{deluxetable*}

\begin{figure}[t]
\plotone{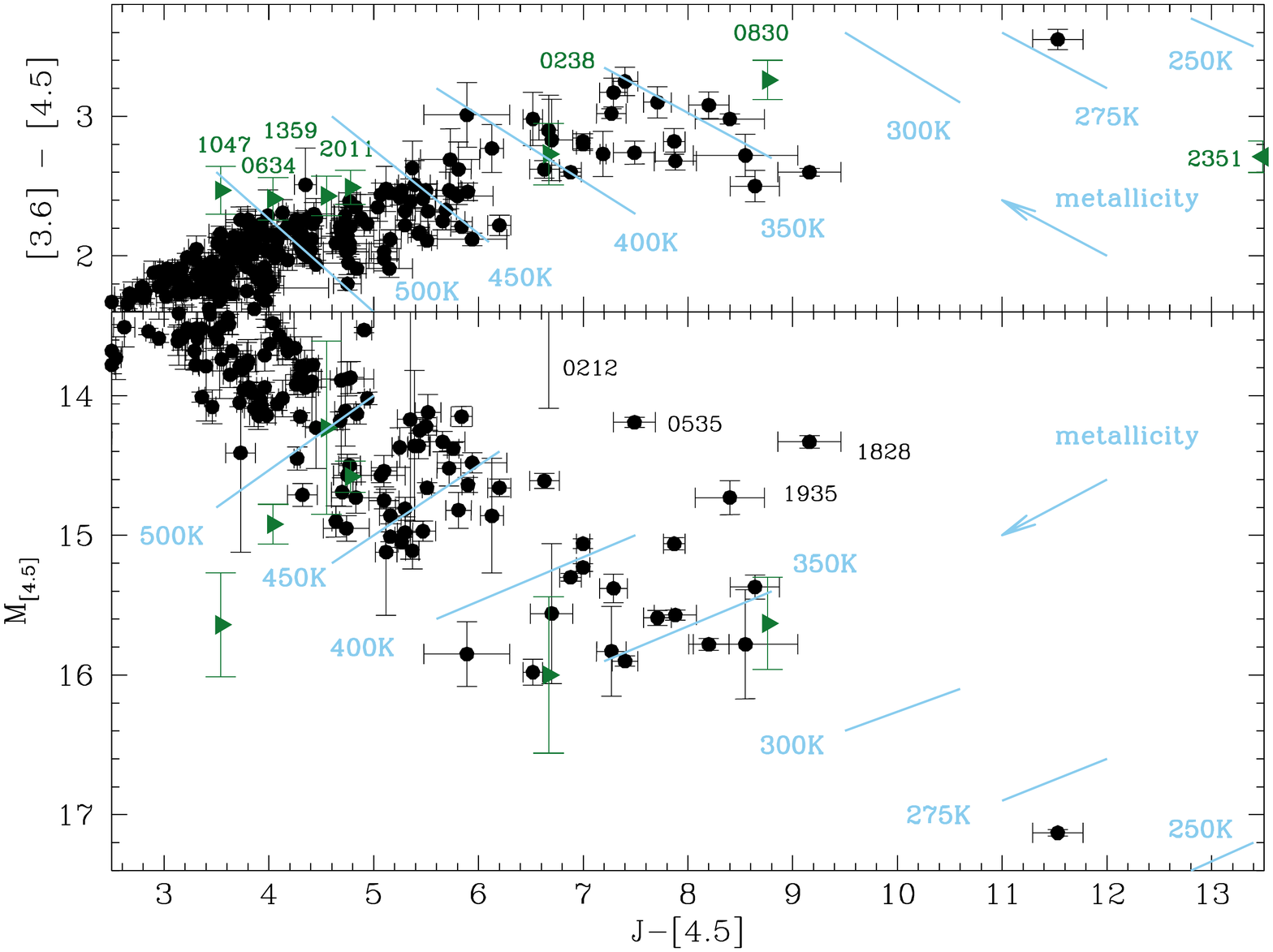}
\vskip -0.1in
\caption{Color-color plots for estimating $T_{\rm eff}$ and metallicity for Y dwarfs. Blue lines are isotherms with metallicity ranging from approximately $+0.3$ to $-0.5$ from left to right. Green triangles, identified in the upper panel with the first four digits of the object's RA, correspond to candidate Y dwarfs with lower limits only on  $J$, or no constraint on $J$ in the case of J2351. In the lower panel, 
the possible equal-mass binaries J0212, J0535, J1828, and J1935 are identified. Of the seven $J$-limit objects, J2351 is not in the lower panel as there is no parallax available. 
}
\end{figure}

Table 5 summarizes the dependencies of various colors on metallicity and gravity for a representative 400~K brown dwarf. All color changes are calculated by a  modified-adiabat  model, except for [3.6] $-$ [4.5] which is estimated from the two 375~K brown dwarfs analysed in Section 5, which differ in metallicity and gravity.
%The $Y$-band is omitted due to the significant discrepancy seen with the models (Figure 12). 
By referencing the  SED parameter dependence shown in Figure 6, and the opacity identifications shown in Figure 5, we find that there are two opacities which drive the pressure (gravity) and metallicity sensitivity in the models: the CO absorption at $\lambda \approx 4.6~\mu$m (Figure 6), and collision-induced H$_2$ opacity \citep[e.g.][]{Allard_1995, Burgasser_2002,Knapp_2004, Saumon_2012}.
The  H$_2$ opacity at these low temperatures has two broad peaks with similar absorption coefficients, at $\lambda \approx 2.2~\mu$m ($K$) and $\lambda \approx 11.1~\mu$m (W3); there is a weaker absorption peak at $\lambda \approx 1.2~\mu$m  \citep[$YJ$, ][their Figure 1]{Saumon_2012}.

\setlength\tabcolsep{3pt}
\begin{deluxetable*}{lcccrrclcccrr}[t]
\tabletypesize{\scriptsize}
\tablecaption{Estimates of $T_{\rm eff}$ and Metallicity for Candidate and Confirmed Y Dwarfs}
%%\tablewidth{0pt}
\tablehead{
\colhead{{\it WISE} Name} &   \colhead{Disc.} &  \colhead{Spec.} & \colhead{Type} & \colhead{$T_{\rm eff}$ K} &  
 \colhead{[m/H]} & &
 \colhead{{\it WISE} Name} &   \colhead{Disc.} &  \colhead{Spec.} & \colhead{Type} & \colhead{$T_{\rm eff}$ K} &  
 \colhead{[m/H]} 
 \\
     &  \colhead{Ref.} & \colhead{Type} &  \colhead{Ref.} &  &  & &
     &  \colhead{Ref.} & \colhead{Type} &  \colhead{Ref.} &  &
}
\startdata
\\
014656.66$+$423410.0B\tablenotemark{a} & Ki12 & Y0 & Du15 & 435   & $\sim 0$       &  & 120604.38$+$840110.6 & Sc15 & Y0 & Sc15 & 475 & $\sim$0 \\
021243.55$+$053147.2(AB)\tablenotemark{b}\tablenotemark{c} & Me20a & Y1  & 1    & 400   &     & & 121756.91$+$162640.2B & Ki11 & Y0 & Le14 & 460 & $\sim$0 \\
023842.60$-$133210.7 & Me20a   & Y1 & Me20a   & 400 & & & 125721.01$+$715349.3  & Me20b & Y1 & Me20b & 390 &  $\sim$0 \\
030237.53$-$581740.3 & Ti18 & Y0: &  Ti18 & 460 & $> 0$ & & 135937.65$-$435226.9\tablenotemark{c}  & Me20a & Y0 & Me20a & 455 & \\
030449.03$-$270508.3 & Pi14b  &  Y0pec & Pi14a & 465 &   $\sim$0 & & 140518.40$+$553421.4 & Cu11 & Y0.5 & Cu16 & 400  & $\sim$0 \\
032109.59$+$693204.5 & Me20a & Y0.5 & Me20a  & 415 &  $> 0$ & & 144606.62$-$231717.8 & Me20a &Y1 & Me20a  &  350  &      $> 0$ \\
033605.05$-$014350.4 & Ma13b & Y0 & Ma18 & 445 &   $<$ 0 & & 154151.66$-$225025.2\tablenotemark{d} & Cu11 & Y1 & Sc15 & 375 & $+0.3$  \\
035000.32$-$565830.2\tablenotemark{d} & Ki12   & Y1  & Ki12   & 325 & $+0.3$  & & 163940.86$-$684744.6 & Ti12 & Y0pec & Sc15 & 405 & $\sim$0 \\
035934.06$-$540154.6 & Ki12 & Y0 & Ki12   & 475 & $< 0$ & & 173835.53$+$273258.9 & Cu11 & Y0 & Cu11 & 450  &      $> 0$ \\
040235.55$-$265145.4 & Me20a & Y1 & Me20a & 370 & $< 0$ & & 182831.08$+$265037.8(AB)\tablenotemark{d}\tablenotemark{e} & Cu11 & $\ge$Y2 & Ki12 & 375 & $-0.5$  \\
041022.71$+$150248.5 & Cu11 & Y0  & Cu11 & 435 & $> 0$ & & 193054.55$-$205949.4 & Me20b &  Y1 &  Me20b &  365 & $\sim$0  \\
050305.68$-$564834.0 & Me20b & Y1 & Me20b &   345  & $> 0$   & & 193518.59$-$154620.3(AB)\tablenotemark{e} & Me20a & Y1  &  Me20a &  365 & $< 0$ \\
053516.80$-$750024.9(AB)\tablenotemark{e} & Ki12  &  $\ge$Y1:  & Ki13 & 415 &   $< 0$ & & 193656.08$+$040801.2 & Me20a & Y0 & Me20a & 450 & $\sim$0 \\
063428.10$+$504925.9 & Me20a &  Y0 & Me20a & 445 & & & 201146.45$-$481259.7 & Me20a & Y0 & Me20a & 465 & \\
064723.23$-$623235.5 & Ki13 &  Y1 & Ki13 & 405 & $< 0$ & & 205628.90$+$145953.3\tablenotemark{d} & Cu11  & Y0 & Cu11  & 475 & 0.0  \\
071322.55$-$291751.9 & Ki12 &  Y0 & Ki12 & 465 &   $\sim$0  & & 220905.73$+$271143.9\tablenotemark{d} & Cu14 & Y0: & Cu14 & 350 & 0.0  \\
073444.02$-$715744.0 & Ki12 & Y1  & Ki12 & 470 &   $\sim$0 & & 222055.31$-$362817.4 & Ki12 & Y0 & Ki12  & 450  & $\sim$0 \\
080714.68$-$661848.7 & Lu11 & Y1 & Ki19 & 415 & $< 0$ & & 223022.60$+$254907.5\tablenotemark{c}  & Me20a & Y1 &   Me20a     & 395 &      \\
082507.35$+$280548.5 & Sc15 & Y0.5 & Sc15  &  380  & $\sim$0 & & 224319.56$-$145857.3 & Me20b & Y0 & Me20b & 450 & \\
083011.95$+$283716.0 & Ba20 & Y1  & Ba20 &  335  &  & & 225628.97$+$400227.3 & Me20a & Y1 & Me20a & 345   & $> 0$\\
085510.83$-$071442.5\tablenotemark{d}  & Lu14 &  $\ge$Y4 & Ki19 & 260 & 0.0  & & 235120.62$-$700025.8 & Me20b & Y0.5 & 1 & 405 & \\
085938.95$+$534908.7\tablenotemark{c}  & Me20a  & Y0 & Me20a  & 450 & & & 235402.79$+$024014.1 & Sc15 & Y0 &  Sc15 & 355 &   $\sim$0 \\ 
093852.89$+$063440.6\tablenotemark{c}  & Me20a & Y0 & Me20a  & 455 & & & 235547.99$+$380438.9 & Me20a   &  Y0 & Me20a   & 480 & \\
094005.50$+$523359.2\tablenotemark{c}  & Me20a & $\ge$Y1 & Me20a & 410 & & & 235644.78$-$481456.3 & Me20a & Y0.5 & Me20a   & 425 & \\
104756.81$+$545741.6 & Me20a & Y0 & Me20a  & 400 &  & & & & & & &\\
114156.67$-$332635.5 & Ti14 & Y0 & Ti18 & 485 & $\sim$0  & & & & & & &\\
\enddata
\tablenotetext{a}{No measured resolved $5~\mu$m photometry is published for the close binary. For this work we deconvolve the {\it Spitzer} photometry \citep{Kirkpatrick_2019} using spectral types of T9 and Y0 for the components  \citep{Dupuy_2015}, and adopting $\delta$[3.6] $= 1.00 \pm 0.15$ and $\delta$[4.5]$= 0.7 \pm 0.10$ \citep[][their Figure 14]{Kirkpatrick_2020}. }
%\citet{Leggett_2017} estimate $T_{\rm eff} \approx 415$~K and [m/H] $\approx 0$ based on resolved near-infrared photometry and a deconvolution of the unresolved near-infrared spectrum.}
\vskip -0.1in
\tablenotetext{b}{$T_{\rm eff}$ is estimated from [3.6] $-$ [4.5] and $J -$ [4.5]; the value is consistent with the $M_{[4.5]}$-implied value if the system is an equal mass binary {\it and} the true parallax is close to the upper limit on the current uncertain measurement.}
\vskip -0.1in
\tablenotetext{c}{The $M_{[4.5]}$ magnitude was ignored in the estimate due to the large uncertainty in the distance modulus ($> 0.4$~mag).}
\vskip -0.1in
\tablenotetext{d}{The parameter estimates are based on the full SED fits described in Section 5.2.}
\vskip -0.1in
\tablenotetext{e}{The parameter estimates assume the system is an equal mass binary.}
\vskip -0.1in
%\tablenotetext{f}{The $J -$ [4.5] color was ignored in the estimate due to the large uncertainty in the $J$ magnitude.}
%\vskip -0.05in
\tablerefs{(1) this work, type  ($\pm \approx 0.5$) based on the type-color relationship of \citet{Kirkpatrick_2019}; Ba20 -- \citet{Bardalez_2020};
Cu11 -- \citet{Cushing_2011};
Du15 -- \citet{Dupuy_2015};
Ki11, 12, 13, 19  -- \citet{Kirkpatrick_2011, Kirkpatrick_2012, Kirkpatrick_2013, Kirkpatrick_2019};
Le14 -- \citet{Leggett_2014};
Lu11, 14 -- \citet{Luhman_2011, Luhman_2014};
Ma13b -- \citet{Mace_2013b}; 
Ma18 -- \citet{Martin_2018};
Ma19 -- \cite{Marocco_2019};
Me20a,b -- \citet{Meisner_2020a, Meisner_2020b};
Pi14a -- \citet{Pinfield_2014b};
Sc15 -- \citet{Schneider_2015};
Ti12, 14, 18 -- \citet{Tinney_2012, Tinney_2014, Tinney_2018}.
}
\end{deluxetable*}

Figure 15 shows late-T and  Y dwarf candidates in color-color diagrams which  take advantage of the metallicity-sensitivity of the $J -$ [4.5] color, which becomes redder with decreasing metallicity.
For warmer brown dwarfs, \citet[][their Figure 3]{Schneider_2020} show that metal-poor T-type (sub)dwarfs  are also red in $J -$ [4.5] for their W1 $-$ W2 color, which is similar to the [3.6] $-$ [4.5] color. 
Note that the observationally-defined metal-poor population edge, in the lower panels of Figures 13 and 15, is consistent with the location of the metal-poor chemical-equilibrium Sonora-Bobcat sequence shown in Figure 3. This would be expected, as metal-paucity reduces the size of the chemical changes brought about by mixing \citep[][their Figures 4 and 10]{Zahnle_2014}.

The  commonly available colors for Y dwarfs are shown in Figure 15: [3.6] $-$ [4.5] and $M_{[4.5]}$ as a function of $J -$ [4.5].
Observations, together with the modified-adiabat disequilibrium chemistry models (with an empirical correction to [3.6] $-$ [4.5]), show that 
$T_{\rm eff}$ and metallicity can be estimated for cold brown dwarfs using such a figure. 
%The parameter grouping is
%based on the solar and enhanced-solar metallicity modified-adiabat disequilibrium chemistry models, with an empirical correction to [3.6] $-$ [4.5] determined from the fits to the SEDs  presented in Section 5.2. 

Figure 15 includes all 50 currently known candidate Y dwarfs.
%\citep[no resolved $5~\mu$m photometry is published for the close binary system WISE 014656.66$+$423410.0AB,][]{Dupuy_2015}.
Seven of these  do not have a measurement of $J$.
A lower limit of $J \gtrsim 24.6$ was determined for WISEA J083011.95$+$283716.0 by transforming the F125W limit given by \citet{Bardalez_2020}  using transformations from \citet{Leggett_2017}.
Lower limits on $J$ were taken from \citet{Meisner_2020a, Meisner_2020b} for 
CWISEP J104756.81$+$545741.6 ($J \gtrsim 19.8$),
 and
CWISEP J201146.45$-$481259.7  ($J \gtrsim 20.1$).
For three other  objects we determined limits from the UKIDSS and VISTA surveys' imaging data: 
CWISEP J023842.60$-$133210.7 ($J \gtrsim 23.0$),
%CWISEP J040235.55$-$265145.4 ($J \gtrsim 22.0$),
CWISEP J063428.10$+$504925.9 ($J \gtrsim 20.0$), and
CWISEP J135937.65$-$435226.9 ($J \gtrsim 20.5$).
No constraint on $J$ is currently available for WISEA J235120.62$-$700025.8.

Table 6 lists the 50 Y dwarfs (or Y dwarf candidates) along with spectral type, $T_{\rm eff}$ and (where there is sufficient information) [m/H]. 
For six of the Y dwarfs, identified in the Table, we carried out a detailed atmospheric analysis in Section 5.2, and those values of $T_{\rm eff}$ and  [m/H] are given in the Table, as well as in Table 3. The parameters for the other Y dwarfs are based on one to three colors, using the relationships given in Table 4. $T_{\rm eff}$ is determined from 
[3.6] $-$ [4.5], $J -$ [4.5] and $M_{[4.5]}$, with the $T_{\rm eff}$ values rounded to 5~K. 
The average of the color-implied $T_{\rm eff}$ value is adopted, unless all three estimates are available and the $J -$ [4.5] color is discrepant (suggesting a non-solar metallicity, Figures 14 and 15), in which case the two other values are averaged.

\medskip
\subsection{Super-Luminous Y Dwarfs and Binarity}

There are four Y dwarfs for which the luminosity-implied $T_{\rm eff}$ is only consistent with the SED- or color-implied value if the dwarf is an unresolved multiple system: 
CWISEP  J021243.55+053147.2
WISE J053516.80$-$750024.9, WISEPA J182831.08$+$265037.8, and CWISEP J193518.59$-$154620.3 (see Figures 3, 13, and 15). 

If these four objects are approximately-equal-mass binaries, the sample of 50 Y dwarf systems then includes the secondaries of three known resolved systems, plus these four unresolved binaries, for a notional binary fraction of 14\%. Table 7 summarises the properties of these confirmed and candidate binaries.   
The number of candidate binary systems is consistent with studies of the binary fraction of substellar objects --- for example \citet{Fontanive_2018} find a binary fraction of $8 \pm 6\%$  for T5 -- Y0 brown dwarfs at separations of 1.0 -- 1000 AU, with a mass ratio distribution peaking around unity.

\setlength\tabcolsep{7pt}
\begin{deluxetable*}{lccccc}[t]
\tabletypesize{\normalsize}
\tablecaption{Known and Candidate Binary Y Dwarfs}
\tablehead{
\colhead{{\it WISE} (Other) Name} &  \multicolumn{3}{c}{Separation}
&  \multicolumn{2}{c}{Mass Ratio}\\
  & \colhead{arcsecond}  & \colhead{AU\tablenotemark{a}}   & \colhead{Ref.}  & \colhead{Value} & \colhead{Ref.} 
  }
\startdata
014656.66$+$423410.0B & $0\farcs 09$  & 1.7 & Du15 & 0.9 & Du15 \\
021243.55$+$053147.2(AB) &  $< 0\farcs 36$  & $< 9$ & 1 &  1.0 & 1 \\
053516.80$-$750024.9(AB) &  $< 0\farcs 15$  & $< 2.2$ & Op16 & 1.0 & 1 \\
080714$-$661848 (WD 0806$-$661B)  & 130  & 2500 & Lu11 & 0.004\tablenotemark{b} & Lu11 \\
121756.91$+$162640.2B & $0\farcs 76$ & 7.1 & Li12 & 0.7 & Le14 \\
182831.08$+$265037.8(AB) &  $< 0\farcs 05$  & $< 0.4$ & Be13 & 1.0 & 1 \\
193518.59$-$154620.3(AB) &   $< 0\farcs 35$  & $< 5$ & 1 &  1.0 & 1 \\
\enddata
\tablenotetext{a}{Distance in AU calculated using parallaxes from \citet{Kirkpatrick_2020}. For J0212 the upper limit on the parallax is used, which is more consistent with the observed colors (Figures 3, 13, 15).}
\tablenotetext{b}{The mass ratio uses the white dwarf progenitor mass.}
\vskip -0.05in
\tablerefs{
(1) this work; Be13 -- \citet{Beichman_2013};
Du15 -- \citet{Dupuy_2015};
Le14 -- \citet{Leggett_2014}; Li12 -- \citet{Liu_2012};
Lu11 - \citet{Luhman_2011}; Op16 -- \citet{Opitz_2016}.
}
\end{deluxetable*}

\bigskip
\bigskip
\section{Conclusions}

The cold Y dwarfs   are important laboratories for atmospheric dynamics because the regions from which the 1 --10~$\mu$m light emerges span a  range in  pressure  of  3 orders of  magnitude  (Figure 7). They are also rapid rotators \citep{Cushing_2016, Esplin_2016, Leggett_2016b, Tannock_2021}. Under these conditions,  small departures from   standard radiative/convective equilibrium is a natural and stable phenomenon \citep[e.g.][]{Guillot_2005, Augustson_2019, Tremblin_2019, Zhang_2020}. In this work we show that a $\sim 10$\% reduction in the standard adiabat in the upper photosphere of Y dwarfs leads to cooler deeper photospheres. This change  yields significant and comprehensive improvements in the agreement between  modelled and  observed colors and spectra of  brown dwarfs with $T_{\rm eff} < 600$~K (Figures 5, 12, 13). The modified-adiabat models with non-equilibrium chemistry that we outline here 
produce the best fit to date of the 
%are the first models to reproduce almost the entire  
1 -- 20~$\mu$m flux distribution of brown dwarfs cooler than 600~K (Figures 5, 8, 9). A summary of key results follows.

\begin{itemize}
    \item New near-infrared photometry is presented for  4 late-T  and 17 Y dwarfs (Table 1).
    \item New or revised mid-infrared  photometry is presented for one L, 10 T, and 4 Y dwarfs (Table 2).
    \item Spectral type estimates are revised in Section 2 for three brown dwarfs,  using the new photometry:
    \begin{itemize}
        \item CWISEP 021243.55$+$053147.2 from background source to likely binary Y dwarf system
        \item CWISE J092503.20$-$472013. from Y0 to T8
        \item CWISE J112106.36$-$623221.5 from Y0 to T7
    \end{itemize}
    \item We reconfirm that chemical abundances are not in equilibrium, due to vertical mixing (Figures 2, 3, 5). The decrease in NH$_3$ and increase in CO impacts the flux at $H$, $K$, [4.5] and W3 by  30\%  to a factor of two. Of particular importance for {\it JWST}, chemical equilibrium  models will  underestimate  the [4.5] $-$ W3($\sim 14~\mu$m) and [4.5] $-$ W4($\sim 22~\mu$m) colors of T and Y dwarfs by $\sim 1$ magnitude (Figure 4).
    \item Current (2020) atmospheric models generate $J - K$ and [3.6] $-$ [4.5] colors that deviate from observations by a factor of $\sim 3$, for $T_{\rm eff} < 600$~K (Figure 2).
    \item As a first step towards including processes currently missing in all brown dwarf models, we parameterize  the pressure-temperature atmospheric profile  in  the one-dimensional ATMO 2020 disequilibrium chemistry models, and explore fits to the SEDs of 7 brown dwarfs with $260 \lesssim T_{\rm eff}$~K $\lesssim 540$~K (Section 5). A  decrease  in  the  adiabatic  gradient    at  pressures of 10 -- 50  bar and temperatures $\sim 800$~K 
    produces cooler deep atmospheres for a given $T_{\rm eff}$, and effectively reproduces observations at $1 \lesssim \lambda ~\mu$m $\lesssim 20$ (Figures 5, 8, 9). Discrepancies that remain are at the factor of $\sim 2$ level in the $Y$- and [3.6]-band for $T_{\rm eff} \lesssim 400$~K (Figure 12). Note that the discrepancy at [3.6] is reduced by a factor of $\sim 5$ compared to standard-adiabat models. 
    \item Spectroscopy shows that the problems at $Y$ and [3.6] for the  $T_{\rm eff} \lesssim 400$~K Y dwarfs occur at the blue side of the passbands. 
    \begin{itemize}
        \item For $Y$, the issue is most likely to be deficiencies in  modelling the red wing of the K~I resonant line \citep{Phillips_2020}.
        \item For [3.6], it appears that high in the atmosphere, where pressures are $\sim 0.1$~bar and this flux originates, the temperature needs to be higher. The heating could  be caused by  breaking gravity waves, as is likely in the solar system giant planet atmospheres above the 1-bar pressure surface \citep[e.g.][]{Schubert_2003, O'Donoghue_2016}. The  $T_{\rm eff} \lesssim 350$~K Y dwarfs may have an upper atmosphere heated by water condensation (Figure 7).
    \end{itemize}
         \item The fact that the adiabat changes at temperatures around 800~K and pressures of 10 -- 50 bar, for the 6 Y dwarfs studied in detail, may indicate that convection is disrupted in Y  dwarf atmospheres by a change in nitrogen chemistry and/or the condensation of chlorides and sulfides (Figure 7, top left panel).
    \item The atmospheric parameters combined with evolutionary models indicate that the six Y dwarfs  have an age between 0.5 and 3~Gyr and masses of 5 -- 12 Jupiters (Table 3).
    \item We generate a limited grid of modified-adiabat disequilibrium chemistry models and provide relationships between  $T_{\rm eff}$ and the commonly used colors: [3.6] $-$ [4.5], $J -$ [4.5], $M_{[4.5]}$ (Table 4).  The models indicate that there are two  opacities  which  drive  the  pressure  (gravity)  and  metallicity  sensitivity  in  the  models:  the  CO  absorption  at $\lambda \approx 4.6~\mu$m (Figure 6), and collision-induced H$_2$ opacity with  broad peaks at $\lambda \approx$ 1.2, 2.2 and 11.1~$\mu$m \citep[][their Figure 1]{Saumon_2012}.
    \item We show that the $J -$ [4.5] color is particularly sensitive to metallicity (Table 5),
    and that a diagram which plots  [3.6] $-$ [4.5] and $M_{[4.5]}$ as a function of $J -$ [4.5] can be used to estimate   $T_{\rm eff}$ and metallicity (Figure 15). We estimate these parameters for the 50 known candidate Y dwarfs  (Table 6).
    \item We find that there are four super-luminous Y dwarfs which are likely to be unresolved binaries; together with the three known resolved binary Y dwarf components, this suggests a
binary fraction of $\sim 14$\%  for Y dwarfs (Table 7). Such a number is consistent with what is found for L and T dwarfs \citep[e.g.][]{Fontanive_2018}.
\item The Appendix gives examples of temperature-sensitive {\it JWST} colors, tables of colors generated by the modified-adiabat model grid, and a compilation of the photometry used in this work.
\end{itemize}

\acknowledgments

Supported by the international Gemini Observatory, a program of NSF’s NOIRLab, which is managed by the Association of Universities for Research in Astronomy (AURA) under a cooperative agreement with the National Science Foundation, on behalf of the Gemini partnership of Argentina, Brazil, Canada, Chile, the Republic of Korea, and the United States of America.

This work was enabled in part by observations made from the Gemini North telescope, located within the Maunakea Science Reserve and adjacent to the summit of Maunakea. We are grateful for the privilege of observing the Universe from a place that is unique in both its astronomical quality and its cultural significance.

This publication makes use of data products from the Wide-field Infrared Survey Explorer, which is a joint project of the University of California, Los Angeles, and the Jet Propulsion Laboratory/California Institute of Technology, funded by the National Aeronautics and Space Administration.

This work is based in part on observations made with the {\it Spitzer} Space Telescope, obtained from the NASA/ IPAC Infrared Science Archive, both of which are operated by the Jet Propulsion Laboratory, California Institute of Technology under a contract with the National Aeronautics and Space Administration.

C.V.M. acknowledges the support of the National Science Foundation grant number 1910969. 

P.T. acknowledges supports by the European Research Council under Grant Agreement ATMO 757858.

\bigskip
\bigskip
{\bf We dedicate this work to France Allard and Adam Showman, both prematurely lost to astronomy in 2020. They leave a legacy of work that is vital to the understanding of  low mass stars, brown dwarfs, exoplanets and the solar system.} 

\clearpage
%\bigskip

\appendix

\section{JWST Colors}

Figure 16 shows color-color diagrams for  {\it JWST} NIRCam and MIRI, generated by the same small grid of modified-adiabat models, for $250 \leq T_{\rm eff}$~K $\leq 500$.  We chose the color combinations shown in the figure based on sensitivity to $T_{\rm eff}$ and measurability. The latter was determined from the model-calculated brightness of the brown dwarf and the   throughput of the  NIRCam  \footnote{\url{https://jwst-docs.stsci.edu/near-infrared-camera/nircam-observing-modes/nircam-imaging}} and MIRI \footnote{\url{https://jwst-docs.stsci.edu/mid-infrared-instrument/miri-predicted-performance/miri-sensitivity}} filters. We found, for example, that  the best short-wavelength filter in NIRCam for cold brown dwarf work is the F162M. The shorter wavelength filters either sample regions where there is very little signal (F070W, F090W, F140M) or are wide enough to include a significant wavelength region with no signal (F115W, F150W). We give the {\it JWST} magnitudes in Table 9 and the reader can explore other color combinations. The modified-adiabat models we present here indicate that {\it JWST} colors can be used to estimate brown dwarf temperatures, and vice versa; this is especially true at shorter wavelengths, as can also be seen in the top left panel of Figure 6.

\begin{figure}[b]
%\vskip -0.3in
%\plotone{JWST_color.pdf}
\plotone{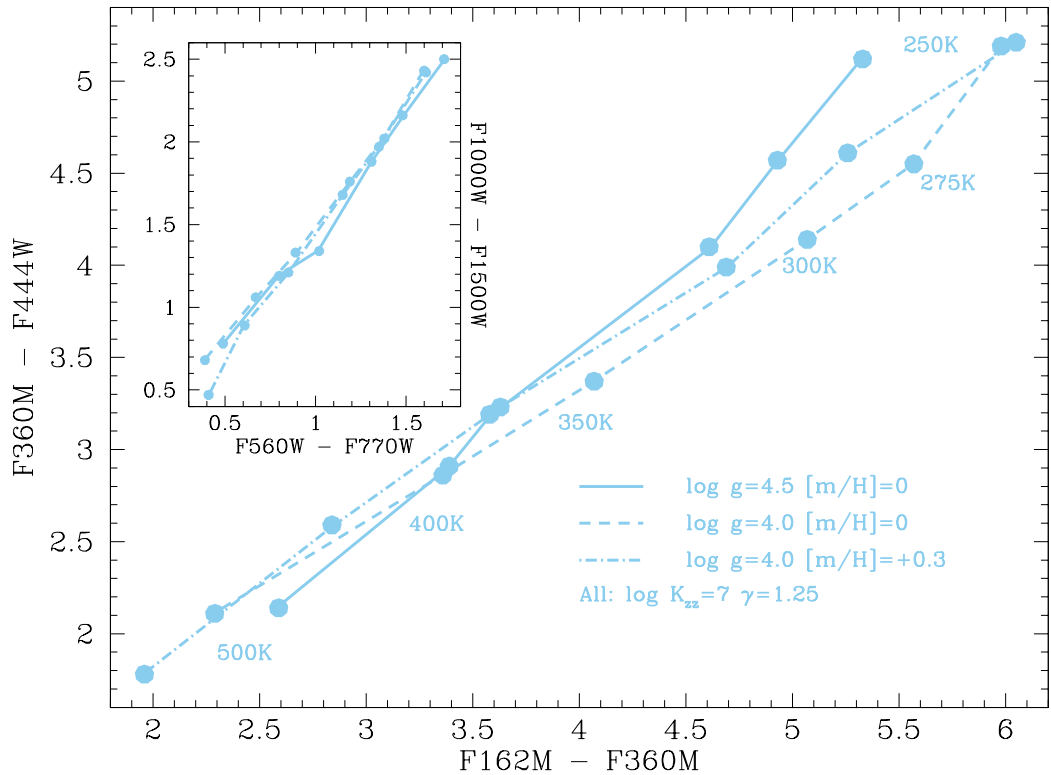}
\vskip -0.2in
\caption{Blue lines are color-color sequences generated by the grid of modified-adiabat models  for {\it JWST} filters. The atmospheric parameters are given in the legend. Dots along each sequence indicate  where $T_{\rm eff} =$ 500, 400, 350, 300, 275, and 250~K, from left to right.   The large diagram is for NIRCam  and the inset for MIRI. The colors were chosen for sensitivity to $T_{\rm eff}$ (based on the models) and measurability (based on the brightness of the brown dwarf and the throughput of the filters). 
}
\end{figure}

\bigskip
\section{Model Grid}

Tables 8 and 9 give colors generated by the modified-adiabat ATMO 2020 disequilibrium chemistry model atmospheres. The models have  $K_{zz} = 10^7$~cm$^2$s$^{-1}$, $\gamma = 1.25$ and $P_{\gamma -max} = 15$~bar. Table 8 gives  magnitudes on the MKO near-infrared system, as well as {\it Spitzer} [3.6] and [4.5], and {\it WISE} W3 and W4. Table 9 gives magnitudes for a subset of the {\it JWST} filters, those which are  likely to be used for observations of  brown dwarfs.

%\smallskip
\setlength\tabcolsep{7pt}
\begin{deluxetable*}{rrcrrrrrrrr}[b]
\tabletypesize{\normalsize}
\tablecaption{ATMO 2020 Grid  with Modified $P-T$ Profile: MKO, {\it Spitzer} and {\it WISE} Filters}
\tablewidth{0pt}
\tablehead{
\colhead{$T_{\rm  eff}$ K} &   \colhead{$\log g$} &
\colhead{[m/H]} & \colhead{$Y$} &  \colhead{$J$} &  \colhead{$H$} &   \colhead{$K$} &
  \colhead{[3.6]} &   \colhead{[4.5]} &  \colhead{W3} & \colhead{W4} 
}
\startdata
250	&	4.0	&	0.0	&	32.02	&	30.61	&	29.21	&	31.27	&	22.60	&	17.26	&	15.05	&	12.95	\\
275	&	4.0	&	0.0	&	29.79	&	28.60	&	27.63	&	28.94	&	21.48	&	16.75	&	14.56	&	12.70	\\
300	&	4.0	&	0.0	&	27.94	&	26.83	&	26.15	&	27.10	&	20.53	&	16.21	&	14.08	&	12.42	\\
350	&	4.0	&	0.0	&	24.86	&	23.85	&	23.67	&	24.16	&	19.10	&	15.55	&	13.39	&	12.08	\\
400	&	4.0	&	0.0	&	22.78	&	21.72	&	21.81	&	21.97	&	18.03	&	15.00	&	12.84	&	11.80	\\
500	&	4.0	&	0.0	&	19.77	&	18.68	&	19.05	&	18.92	&	16.49	&	14.28	&	12.04	&	11.38	\\
250	&	4.5	&	0.0	&	30.66	&	30.09	&	28.54	&	31.85	&	22.58	&	17.31	&	15.24	&	13.11	\\
275	&	4.5	&	0.0	&	28.71	&	28.11	&	27.03	&	29.61	&	21.51	&	16.78	&	14.80	&	12.87	\\
300	&	4.5	&	0.0	&	27.17	&	26.54	&	25.80	&	27.75	&	20.63	&	16.35	&	14.39	&	12.66	\\
350	&	4.5	&	0.0	&	24.77	&	23.57	&	23.28	&	24.80	&	19.20	&	15.86	&	13.63	&	12.31	\\
400	&	4.5	&	0.0	&	22.80	&	21.95	&	22.10	&	22.90	&	18.28	&	15.18	&	13.25	&	12.09	\\
500	&	4.5	&	0.0	&	20.16	&	19.09	&	19.55	&	19.73	&	16.68	&	14.38	&	12.41	&	11.66	\\
250	&	4.0	&	0.3	&	31.50	&	30.08	&	29.42	&	29.89	&	22.73	&	17.37	&	14.87	&	12.90	\\
275	&	4.0	&	0.3	&	29.14	&	27.85	&	27.50	&	27.74	&	21.63	&	16.87	&	14.40	&	12.69	\\
300	&	4.0	&	0.3	&	27.12	&	25.88	&	25.83	&	25.75	&	20.56	&	16.41	&	13.90	&	12.38	\\
350	&	4.0	&	0.3	&	24.22	&	23.11	&	23.31	&	23.02	&	19.17	&	15.80	&	13.20	&	12.06	\\
400	&	4.0	&	0.3	&	22.07	&	20.93	&	21.27	&	20.82	&	18.01	&	15.35	&	12.63	&	11.77	\\
500	&	4.0	&	0.3	&	19.48	&	18.36	&	18.80	&	18.27	&	16.57	&	14.82	&	11.94	&	11.38	
\enddata
\tablecomments{All models have $\gamma = 1.25$ and $\log K_{zz} = 7.0$ Magnitudes are for a distance of 10~pc  and are on the Vega system. 
}
\end{deluxetable*}

%\smallskip
\setlength\tabcolsep{2pt}
\begin{deluxetable*}{rrcrrrrrrrrrrrrrrrrr}[b]
\tabletypesize{\scriptsize}
\tablecaption{ATMO 2020 Grid  with Modified $P-T$ Profile: {\it JWST}  Filters}
%\tablewidth{0pt}
\tablehead{
\colhead{$T_{\rm  eff}$ K} &   \colhead{$\log g$} & \colhead{[m/H]} & 
\multicolumn{9}{c}{NIRCam F} & \multicolumn{8}{c}{MIRI F} \\
  &   &   & 
\colhead{115W} & \colhead{150W} &
\colhead{162M} & \colhead{200W} &
\colhead{210M} & \colhead{300M} & \colhead{360M} & \colhead{410M} & \colhead{444W} &
\colhead{560W} & \colhead{770W} & \colhead{1000W} & \colhead{1280W} 
& \colhead{1500W}
& \colhead{1800W} 
& \colhead{2100W}
& \colhead{2550W}  
}
\startdata
250	&	4.0	&	0.0	&	31.16	&	29.56	&	28.65	&	31.72	&	30.81	&	26.36	&	22.67	&	18.61	&	17.48	&	19.21	&	17.61	&	16.22	&	14.57	&	13.79	&	13.32	&	13.01	&	12.86	\\
275	&	4.0	&	0.0	&	29.12	&	27.98	&	27.08	&	29.38	&	28.47	&	24.64	&	21.51	&	17.89	&	16.96	&	18.39	&	17.02	&	15.43	&	14.10	&	13.41	&	12.99	&	12.75	&	12.64	\\
300	&	4.0	&	0.0	&	27.34	&	26.51	&	25.61	&	27.52	&	26.62	&	23.25	&	20.54	&	17.19	&	16.41	&	17.67	&	16.48	&	14.77	&	13.63	&	13.01	&	12.63	&	12.45	&	12.36	\\
350	&	4.0	&	0.0	&	24.34	&	24.03	&	23.15	&	24.54	&	23.66	&	21.03	&	19.08	&	16.24	&	15.71	&	16.57	&	15.68	&	13.80	&	12.97	&	12.47	&	12.18	&	12.09	&	12.04	\\
400	&	4.0	&	0.0	&	22.23	&	22.17	&	21.34	&	22.31	&	21.45	&	19.47	&	17.98	&	15.45	&	15.12	&	15.71	&	15.03	&	13.10	&	12.42	&	12.04	&	11.83	&	11.80	&	11.76	\\
500	&	4.0	&	0.3	&	19.21	&	19.42	&	18.71	&	19.19	&	18.37	&	17.37	&	16.42	&	14.31	&	14.30	&	14.45	&	14.06	&	12.12	&	11.60	&	11.44	&	11.32	&	11.36	&	11.34	\\
250	&	4.5	&	0.0	&	30.61	&	28.89	&	27.98	&	32.23	&	31.40	&	26.52	&	22.65	&	18.52	&	17.53	&	19.47	&	17.76	&	16.46	&	14.76	&	13.96	&	13.49	&	13.18	&	13.02	\\
275	&	4.5	&	0.0	&	28.61	&	27.39	&	26.48	&	29.99	&	29.15	&	24.99	&	21.55	&	17.85	&	16.98	&	18.68	&	17.20	&	15.76	&	14.33	&	13.60	&	13.17	&	12.92	&	12.79	\\
300	&	4.5	&	0.0	&	27.01	&	26.16	&	25.25	&	28.15	&	27.29	&	23.69	&	20.64	&	17.28	&	16.54	&	18.00	&	16.69	&	15.16	&	13.95	&	13.28	&	12.89	&	12.69	&	12.59	\\
350	&	4.5	&	0.0	&	24.16	&	23.64	&	22.76	&	25.14	&	24.31	&	21.25	&	19.18	&	16.35	&	15.99	&	16.87	&	15.85	&	14.05	&	13.21	&	12.71	&	12.43	&	12.33	&	12.26	\\
400	&	4.5	&	0.0	&	22.44	&	22.47	&	21.61	&	23.25	&	22.41	&	20.12	&	18.22	&	15.67	&	15.31	&	16.13	&	15.33	&	13.58	&	12.84	&	12.39	&	12.15	&	12.09	&	12.05	\\
500	&	4.5	&	0.0	&	19.63	&	19.93	&	19.17	&	20.03	&	19.22	&	17.88	&	16.58	&	14.51	&	14.44	&	14.80	&	14.30	&	12.54	&	12.00	&	11.76	&	11.62	&	11.64	&	11.61	\\
250	&	4.0	&	0.3	&	30.56	&	29.77	&	28.86	&	30.41	&	29.44	&	26.16	&	22.81	&	18.62	&	17.59	&	19.19	&	17.58	&	16.01	&	14.40	&	13.59	&	13.17	&	12.94	&	12.85	\\
275	&	4.0	&	0.3	&	28.31	&	27.86	&	26.95	&	28.23	&	27.27	&	24.42	&	21.69	&	17.92	&	17.08	&	18.37	&	17.03	&	15.21	&	13.93	&	13.24	&	12.87	&	12.71	&	12.65	\\
300	&	4.0	&	0.3	&	26.35	&	26.19	&	25.29	&	26.23	&	25.27	&	22.95	&	20.59	&	17.45	&	16.60	&	17.58	&	16.44	&	14.52	&	13.46	&	12.84	&	12.50	&	12.39	&	12.35	\\
350	&	4.0	&	0.3	&	23.57	&	23.67	&	22.80	&	23.44	&	22.51	&	20.70	&	19.17	&	16.31	&	15.93	&	16.51	&	15.66	&	13.53	&	12.76	&	12.32	&	12.07	&	12.05	&	12.04	\\
400	&	4.0	&	0.3	&	21.42	&	21.63	&	20.83	&	21.19	&	20.28	&	19.11	&	17.99	&	15.51	&	15.40	&	15.60	&	15.00	&	12.80	&	12.18	&	11.91	&	11.73	&	11.75	&	11.75	\\
500	&	4.0	&	0.3	&	18.87	&	19.15	&	18.47	&	18.58	&	17.71	&	17.25	&	16.51	&	14.50	&	14.73	&	14.48	&	14.07	&	11.94	&	11.47	&	11.47	&	11.29	&	11.35	&	11.34	\\
\enddata
\tablecomments{All models have $\gamma = 1.25$ and $\log K_{zz} = 7.0$. Magnitudes are for a distance of 10~pc  and are on the Vega system. 
}
\end{deluxetable*}

%\clearpage
\bigskip
\section{Photometry Compilation}

Table 10 presents a compilation of the photometry used in this work.

\setlength\tabcolsep{1pt}
%\movetabledown=2.8in
\movetabledown=2.7in
\begin{rotatetable}
\begin{deluxetable*}{lcccrrrrrrrrrrrrrrrrrrrrrrccccc}
\tabletypesize{\tiny}
\tablecaption{Compilation of Measurements for T6 and Later Brown Dwarfs}
\tablehead{
\colhead{Survey} & 
\multicolumn{2}{c}{Discovery RA Decl.}  & 
%\colhead{Other} & 
\colhead{Sp.} & 
\colhead{$M-m$} & 
\colhead{$Y$} & \colhead{$J$} & \colhead{$H$} & \colhead{$K$} & \colhead{$L^{\prime}$} &
% \colhead{$M^{\prime}$} & 
%\colhead{IR Note} &
\colhead{[3.6]} & \colhead{[4.5]} &
% \colhead{[5.8]} & \colhead{[8.0]} & 
\colhead{W1} & \colhead{W2} & \colhead{W3} & 
%\colhead{W4} & 
%\colhead{WISE Note} &
\colhead{$e_{Mm}$} & 
\colhead{$e_Y$} & \colhead{$e_J$} & \colhead{$e_H$} & \colhead{$e_K$} & \colhead{$e_{L^{\prime}}$} & %\colhead{$e_{M^{\prime}}$} & 
\colhead{$e_{3.6}$} & \colhead{$e_{4.5}$} & 
%\colhead{$e_{[5.8]}$} & \colhead{$e_{[8.0]}$} &
\colhead{$e_{W1}$} & \colhead{$e_{W2}$} & \colhead{$e_{W3}$} & 
%\colhead{$e_{W4}$} & 
\multicolumn{5}{c}{References}\\
\colhead{Name} &
\colhead{hhmmss.ss} & \colhead{$\pm$ddmmss.s} & 
%\colhead{Names} & 
\colhead{Type} & 
\multicolumn{22}{c}{mag} & 
\colhead{Discovery} & \colhead{Sp. Type} & \colhead{Parallax} & \colhead{Near-IR} & \colhead{$Spitzer$} 
}
\startdata
CWISEP &   000229.93 & $+$635217.0     &  7.5    &      &         &      &        &       &      &  17.35  & 15.69 &       &      &      &                                                                                                  &        &      &       &        &       &  0.25   & 0.06     &     &       &  &                                            Meisner{\_}2020b  &  Meisner{\_}2020b &  & & Meisner{\_}2020b  \\            
WISE   & 000517.48 & $+$373720.5       &   9.0  & 0.52  &  18.48  & 17.59 & 17.98 & 17.99 & 14.43 & 15.43  & 13.28 & 16.76 & 13.29 & 11.79 &
                                      0.04  & 0.02  & 0.02  & 0.03  & 0.03  & 0.10  & 0.04  & 0.04   &  0.09  & 0.03  & 0.24   &      
    Mace{\_}2013a   &    Mace{\_}2013a     &      Kirkpatrick{\_}2019    &        Leggett{\_}2015    &           Kirkpatrick{\_}2019 \\               
CWISEP   &    001146.07 & $-$471306.8   &    8.5  &   &    & 19.28  &  19.69    &     &         &    17.74  & 15.81  &  18.94 & 15.99 &   &
                               &  &        0.07 & 0.20   & & &  0.09  &  0.02   & 0.27 &  0.06  &      &        
    Meisner{\_}2020a                 &            Meisner{\_}2020a        &    &   VISTA{\_}VHS     &                Meisner{\_}2020a   \\                
WISE   &      001354.39 & $+$063448.2   &   8.0   &    & 20.56  & 19.54  & 19.98  & 20.79    &      &    17.15 &  15.16  &  &   15.23 & &
                                             & 0.04 & 0.03 & 0.04 & 0.10 &   &           0.03  &   0.03   &  &       0.09 & &
    Pinfield{\_}2014a            &                 Pinfield{\_}2014a                   &                     &             Leggett{\_}2015       &   Pinfield{\_}2014a    \\              
WISEA   &     001449.96 & $+$795116.1    &   8.0  &     &   20.32  & 19.36   &       &     &       &     17.76 &  15.88 &  18.72 & 16.00 &  13.69  &     
                     &   0.10 & 0.10 &   &   &   &   0.04  &    0.02 &    0.28  & 0.06 &  0.40 &
                     Bardalez{\_}2020        &                    Bardalez{\_}2020             &    &     this{\_}work      &        Bardalez{\_}2020                   \\
\enddata
\tablecomments{Table 1 is published in its entirety in the machine-readable format.
      A portion is shown here for guidance regarding its form and content.}
\vskip -0.05in
\tablerefs{(1) this work; 
Albert{\_}2011 -- \citet{Albert_2011};
Artigau{\_}2010 -- \citet{Artigau_2010};
Bardalez{\_}2020 -- \citet{Bardalez_2020};
Best{\_}2015 -- \citet{Best_2015};
Best{\_}2020 -- \citet{Best_2020};
Burgasser{\_}1999 -- \citet{Burgasser_1999};
Burgasser{\_}2000 -- \citet{Burgasser_2000};
Burgasser{\_}2002-- \citet{Burgasser_2002};
Burgasser{\_}2003 -- \citet{Burgasser_2003};
Burgasser{\_}2004-- \citet{Burgasser_2004};
Burgasser{\_}2006 -- \citet{Burgasser_2006};
Burgasser{\_}2008-- \citet{Burgasser_2008};
Burningham{\_}2008 -- \citet{Burningham_2008};
Burningham{\_}2009 -- \citet{Burningham_2009};
Burningham{\_}2010a -- \citet{Burningham_2010a};
Burningham{\_}2010b -- \citet{Burningham_2010b};
Burningham{\_}2011 -- \citet{Burningham_2011};
Burningham{\_}2013 -- \citet{Burningham_2013};
Chiu{\_}2006 -- \cite{Chiu_2006};
Cushing{\_}2011 -- \citet{Cushing_2011};
Cushing{\_}2014 -- \citet{Cushing_2014};
Cushing{\_}2016 -- \citet{Cushing_2016};
Delorme{\_}2008 -- \citet{Delorme_2008};
Delorme{\_}2010 -- \citet{Delorme_2010};
Dupuy{\_}2012 -- \citet{Dupuy_2012};
Dupuy{\_}2012 -- \citet{Dupuy_2013};
Dupuy{\_}2015 -- \citet{Dupuy_2015};
Faherty{\_}2012 -- \citet{Faherty_2012};
Faherty{\_}2020 -- \citet{Faherty_2020};
GAIA -- \citet{GAIA};
Gelino{\_}2011 -- \citet{Gelino_2011};
Goldman{\_}2010 -- \citet{Goldman_2010};
Greco{\_}2019 -- \citet{Greco_2019};
Griffith{\_}2012 -- \citet{Griffith_2012};
Kirkpatrick{\_}2011 -- \citet{Kirkpatrick_2011};
Kirkpatrick{\_}2012 -- \citet{Kirkpatrick_2012};
Kirkpatrick{\_}2013  -- \citet{Kirkpatrick_2013};
Kirkpatrick{\_}2019  -- \citet{Kirkpatrick_2019};
Kirkpatrick{\_}2020  -- \citet{Kirkpatrick_2020};
Knapp{\_}2004 -- \citet{Knapp_2004};
Leggett{\_}2002 - \citet{Leggett_2002};
Leggett{\_}2009 - \citet{Leggett_2009};
Leggett{\_}2010 - \citet{Leggett_2010};
Leggett{\_}2012 - \citet{Leggett_2012};
Leggett{\_}2014 - \citet{Leggett_2014};
Leggett{\_}2015 - \citet{Leggett_2015};
Leggett{\_}2017 - \citet{Leggett_2017};
Leggett{\_}2019 - \citet{Leggett_2019};
Liu{\_}2011 - \citet{Liu_2011};
Liu{\_}2012 - \citet{Liu_2012};
Lodieu{\_}2007 -- \citet{Lodieu_2007};
Lodieu{\_}2009 -- \citet{Lodieu_2009};
Lodieu{\_}2012 -- \citet{Lodieu_2012};
Looper{\_}2007 -- \citet{Looper_2007};
Lucas{\_}2010 -- \citet{Lucas_2010};
Luhman{\_}2011 -- \citet{Luhman_2011};
Luhman{\_}2012 -- \citet{Luhman_2012};
Luhman{\_}2014 -- \citet{Luhman_2014};
Mace{\_}2013a -- \citet{Mace_2013a}; 
Mace{\_}2013b -- \citet{Mace_2013b}; 
Mainzer{\_}2011 -- \citet{Mainzer_2011};
Manjavacas{\_}2013 -- \citet{Manjavacas_2013};
Marocco{\_}2010 -- \cite{Marocco_2010};
Marocco{\_}2020 -- \cite{Marocco_2020};
Martin{\_}2018 -- \citet{Martin_2018};
Meisner{\_}2020a -- \citet{Meisner_2020a};
Meisner{\_}2020b -- \citet{Meisner_2020b};
Murray{\_}2011 -- \citet{Murray_2011};
Patten{\_}2006 -- \citet{Patten_2006};
Pinfield{\_}Gromadzki{\_}2014 -- \rm{Pinfield, P. and Gromadzki, M. private communication 2014;}
Pinfield{\_}2008 -- \citet{Pinfield_2008};
Pinfield{\_}2012 -- \citet{Pinfield_2012};
Pinfield{\_}2014a -- \citet{Pinfield_2014a};
Pinfield{\_}2014b -- \citet{Pinfield_2014b};
Scholz{\_}2010a -- \citet{Scholz_2010a};
Scholz{\_}2010b -- \citet{Scholz_2010b};
Scholz{\_}2011 -- \citet{Scholz_2011};
Smart{\_}2010 -- \citet{Smart_2010};
Strauss{\_}2099 -- \cite{Strauss_1999};
Stephens{\_}2004 -- \citet{Stephens_2004};
Subasavage{\_}2009 -- \citet{Subasavage_2009};
Thompson{\_}2013 -- \citet{Thompson_2013};
Tinney{\_}2003 -- \citet{Tinney_2003};
Tinney{\_}2005 -- \citet{Tinney_2005};
Tinney{\_}2012 -- \citet{Tinney_2012};
Tinney{\_}2014 -- \citet{Tinney_2014};
Tinney{\_}2018 -- \citet{Tinney_2018};
Tsvetanov{\_}2000 -- \citet{Tsvetanov_2000};
Vrba{\_}2004 -- \citet{Vrba_2004};
Warren{\_}2007 -- \citet{Warren_2007};
Wright{\_}2013 -- \citet{Wright_2013}
}      
\end{deluxetable*}
\end{rotatetable}

\clearpage
\bibliography{Leggett_2021_preprint}{}
\bibliographystyle{aasjournal}

\end{document}